\documentclass[aps,prb,superscriptaddress,twocolumn,showpacs,floatfix,longbibliography]{revtex4-2}

\usepackage{amsmath,amssymb}
\usepackage{graphicx}
\usepackage{dcolumn}
\usepackage{bm}
\usepackage[colorlinks=true,linkcolor=blue,citecolor=blue,urlcolor=blue]{hyperref}
\usepackage{placeins}

\begin{document}

\title{A projection operator approach for ultracold bosons coupled to a cavity}

\author{Sarbarish Sen}
\email{iamsarbarishsen@gmail.com}
\affiliation{School of Physical Sciences, Indian Association for the Cultivation of Science, Kolkata, India}

\author{Bimalendu Deb}
\email{bimalendu.deb@tcgcrest.org}
\affiliation{School of Physical Sciences, Indian Association for the Cultivation of Science, Kolkata, India}
\affiliation{CQuERE, TCG Centres of Research and Education in Science and Technology, Sector V, Salt Lake, Kolkata}

\author{K. Sengupta}
\email{ksengupta1@gmail.com}
\affiliation{School of Physical Sciences, Indian Association for the Cultivation of Science, Kolkata, India}

\date{\today}

\begin{abstract}

We employ a projection operator technique to study the equilibrium phases and the non-equilibrium quench and ramp dynamics of ultracold bosons described by a two-dimensional (2D) Bose-Hubbard model coupled to a high-finesse cavity with competing local short-range and cavity-mediated long-range interactions.  Our analysis relies on a systematic elimination of high-energy virtual hopping processes; this enables us to capture the effects of short-range quantum fluctuations on the phase diagram of the model while treating the cavity-induced long-range interaction within mean-field theory. We find that the phase boundaries between the superfluid (SF), supersolid (SS), charge-density-wave (CDW) and Mott insulating (MI) phases are significantly modified by the presence of these fluctuations. We use this approach to study the dynamics of bosons following a quench/ramp from the CDW to the SS and SF phases; our analysis indicates the presence of oscillatory dynamics characterized by a single frequency due to a quench which takes the system near the critical point. In contrast, for a quench deep inside the SF phase, the dynamics involves multiple frequencies. An explicit comparison with Gutzwiller mean-field theory shows the importance of quantum fluctuations in shaping such post-quench dynamics; this feature may be tested experimentally as we discuss. We also study the ramp dynamics of these bosons and identify a bound on the ramp rate above which the present method is expected to produce accurate results. 

\end{abstract}

\maketitle

\section{Introduction}

Ultracold atoms in optical lattices have emerged as a key emulator for strongly-interacting quantum many-body models\cite{Bloch2008ManyBody, Lewenstein2007Quantum,Greiner2002Quantum}. One such model is the Bose-Hubbard model; the equilibrium phase diagram and the non-equilibrium dynamics of this model have been well-studied in different contexts \cite{Fisher1989Boson,Sheshadri1993Superfluid, trivedi1, Jaksch1998Cold, dupuis1, Dutta_2015}. Currently, theoretical and experimental efforts are being devoted to engineering more complex systems that go beyond purely local on-site interactions of the Bose-Hubbard model. Inclusion of long-range or structured interactions allows for the simulation of novel quantum phases of matter that lack direct analogues in conventional solid-state physics~\cite{Lahaye2009Physics, Defenu2023LongRange}. 

A particularly fascinating frontier in this regard involves coupling the quantum gases of atoms to a high-finesse optical cavity~\cite{Maschler2008, revcav}. In such configurations, the collective coherent scattering of photons due to a pump laser and the cavity field generates global, infinite-range interactions among the bosons~\cite{Baumann2010Mott, Chanda_2025,ergodicity,GS1,GS2,metastable1,metastable2,metastable3,MBL1,MBL2,DPT1,dipbose}. This cavity-mediated global interaction acts alongside the usual short-range on-site repulsion $U$; this allows for a realization of an extended Bose-Hubbard model (EBHM) that exhibits coexisting quantum orders~\cite{Landig2015QuantumPF, Chen2016QuantumPT, PhysRevA.99.043633, Sinha_2025}.

In two-dimensional (2D) square lattices, the infinite-range interactions mediated by a single-mode cavity operate as an effective all-to-all checkerboard potential~\cite{PhysRevA.94.023632, Sundar2016LatticeBW}. This potential self-consistently divides the lattice into two interpenetrating sublattices, conventionally denoted as the even ($A$) and odd ($B$) sublattices. In the strong on-site interaction regime, where the nearest-neighbor hopping amplitude $J$ is small compared to the on-site repulsion $U$ ($J/U \ll 1$), the competing energy scales yield an extensive equilibrium phase diagram~\cite{Li2012LatticesupersolidPO, mft4, mft1,PD1,mft3,mft5}. 

In the atomic limit ($J=0$), the global interaction manifests as an effective chemical potential imbalance between the sublattices. This imbalance stabilizes a translationally invariant MI phase alongside configurations with spontaneously broken discrete translational symmetry, such as partially or fully polarized charge-density-wave (CDW) states~\cite{Larson2008Mott, Chen2020}. When tunneling is introduced ($J/U > 0$), the system can exhibit quantum phase transitions to compressible states. Crucially, the combination of off-site interaction and phase coherence triggers the emergence of a supersolid (SS) state~\cite{Leonard2017Supersolid, Li2017A_Supersolid}, characterized by the simultaneous existence of diagonal crystalline order (spatial density modulations) and off-diagonal long-range order (superfluidity). These systems have been studied using a host of theoretical methods such as Gutzwiller mean-field theory \cite{mft1,PhysRevA.94.023632,mft3,mft4}, slave-rotor formalism \cite{rotor1}, quantum Monte Carlo technique \cite{qmc1} and bosonic dynamical mean-field theory \cite{bdmft1,bdmft2}. However, a systematic semi-analytic method for studying the effect of short-range fluctuations arising from virtual hopping processes on the phase diagram and dynamics of such bosons is, to the best of our knowledge, still lacking.

In this work, we address this problem by applying a projection operator technique originally developed for the Bose-Hubbard model in the absence of any long-range interaction~\cite{PhysRevB.86.085140, PhysRevLett.106.095702}. We study both the equilibrium phase diagram and the non-equilibrium dynamics (quench and ramp dynamics) of the bosons coupled to the cavity at the same footing. The key point of our work is to develop a systematic Schrieffer-Wolff (SW) transformation~\cite{Schrieffer1966Relation} which allows one to obtain an effective low-energy Hamiltonian for the cavity-generated EBHM in two-dimension (2D). Our approach distinguishes carefully between boson hopping which keeps the system within the ground-state manifold and those which takes it outside this manifold. By eliminating the the latter hopping systematically using a Schrieffer-Wolff (SW) transformation, we obtain an effective low-energy Hamiltonian and thereby study the effects of short range quantum fluctuation on the phase diagram. We also analyze the effects of the fluctuations on the dynamics of these bosons. Our analysis here constitutes a generalization of the projection operator technique to address models which may have phases with broken translational symmetry; in particular, it enables us to investigate the effects of short-range quantum fluctuations of bosons coupled to the cavity in their CDW and SS phases. 

The main results of our work are as follows. First, using a variational wavefunction which takes into account the bipartite nature of the boson system, we determine the equilibrium ground-state phases and phase boundaries of the effective low-energy Hamiltonian. This procedure allows one to keep track of short-range quantum fluctuations over the mean-field theory \cite{PhysRevLett.106.095702}; we find that the phase diagram of the bosons receives important contribution from these fluctuations which can not be taken into account within a simple mean-field theory. The incorporation of quantum fluctuations visibly extends the Mott lobes, enlarging the regions of stability for the insulating phases. Moreover, it leads to an extended regime (compared to the mean-field result) of the supersolid (SS) phase. In particular, these fluctuations tend to push the phase boundary towards the QMC results showing the vital role of short-range quantum fluctuations in determining the phase diagram \cite{qmc1,haldaneWorm}. 

Second, we use this projection operator technique to address the non-equilibrium dynamics of these bosons; our work generalizes the framework introduced in Ref.~\cite{PhysRevLett.106.095702} and allows us to study the dynamics of bosonic systems with broken translational symmetry. We use a time-dependent SW transformation to identify an instantaneous low-energy subspace which provides key contribution to their dynamics. This leads to a set of equations of motion that describe the dynamics of the model keeping the effects of short-range correlations. Thus our method provides an improvement over results obtained using standard mean-field theory. 

Third, we use these equations to study sudden quenches of the hopping parameter across different phase boundaries. This allows us to explicitly calculate the time evolution of the superfluid order parameters, sublattice population imbalances, residual energy, and fidelity of the driven state. We show that the dynamics of these bosons following quench from the CDW phase to the SS phase near the critical point leads to a single-frequency oscillations of both the SF order parameter amplitude and the boson density imbalance between the sublattices. The time period of these oscillations shows a divergence at criticality demonstrating a critical slowing down of the system at this point. This behavior also influences the fidelity and the residual energy following the quench. In contrast, an analogous quench which takes the system deep inside the SF phase leads to more complicated dynamics involving multiple frequencies. We note the importance of the short-range quantum fluctuation in determining the properties of such dynamics; this aspect, to the best of our knowledge, has not been addressed in the literature for dynamics across a CDW-SS transition. 

Fourth, we also study the ramp dynamics of such systems by ramping down the hopping amplitude which takes the bosons from the SS to the CDW phase using a power-law protocol. Our analysis of the fidelity and the residual at the end of the ramp shows that both of these quantities reach a plateau at slow ramp rate; for fast and intermediate ramp rate they show a monotonic behavior as a function of the ramp rate which we analyze. We note that the projection-operator approach captures only short-range fluctuations over mean-field theory; consequently, it does not capture the Kibble-Zureck scaling in ramps across critical points studied in Ref.\ \cite{rotor1}. However, we expect it to capture the dynamics accurately for fast and intermediate ramp rates where such long-wavelength fluctuation do not play a significant role; we provide a qualitative estimate for ramp rates where this approach is expected to be accurate. Finally, we discuss possible experiments which can test the predictions of our work. 

The rest of the paper is organized as follows. In Sec.~\ref{secham}, we introduce the model and discuss the phase diagram in the atomic limit. This is followed by Sec.\ \ref{secform} where we present our theoretical framework for deriving the effective Hamiltonian and use it to obtain the ground state phase diagram. Next, in Sec.\ \ref{secdyn}, we analyze the non-equilibrium dynamics of the driven bosons following a quench or at the end of a ramp. Finally, we present a summary of our findings and discuss experiments which can test our theory in Sec.\ \ref{dissc}. Some details of our calculations are presented in the appendices. 

\section{Hamiltonian and its phase in the atomic limit}
\label{secham} 

In an ensemble of ultracold atoms inside a cavity, atoms interact with each other via an effective long-range potential. When an external laser drives the atoms, the light scattered by the atoms gets trapped between two highly reflective mirrors and bounces back and forth so as to reach every single atom regardless of the distance between them. This trapped light and the laser form an effective optical lattice. The behavior of the atoms in such a lattice is governed by three processes, namely, the motion of the atoms between the lattice sites, their on-site interaction, and the effective long-range interaction mediated by the cavity. A control over the laser parameters in such a setup may lead to formation of translationally broken states (such as the CDW and the SS) of these bosons \cite{Landig2015QuantumPF, Maschler2008, Sinha_2025, D.L.Kovrizhin_2005, Leonard2017Supersolid}

In what follows, we shall analyze an effective EBHM which describes bosons coupled to a cavity. We shall consider a system of $N$ bosons in a two-dimensional (2D) optical lattice with $N_0= L\times L$ sites (with lattice spacing set to unity). These bosons describe a cavity-mediated Bose-Hubbard model whose Hamiltonian is given by \cite{revcav}
\begin{eqnarray}
    \hat{H} &=& - J\sum_{\langle i,j \rangle} (\hat{b}_i^{\dagger} \hat{b}_j + h.c.) + \frac{U}{2} \sum_{i} \hat{n}_i(\hat{n}_i-1) \nonumber\\
    && - \mu \sum_i \hat{n}_i - N_0 V\hat{\Phi}^2. \label{s1}
\end{eqnarray}
 Here $\hat{b_i}$ is the annihilation operator of the boson at site $i=(i_x,i_y)$ which obeys the standard bosonic commutation relation $[\hat{b_i}, \hat{b_j}^\dagger] = \delta_{ij}$ and $n_i=\hat{b_i}^\dagger \hat{b_i}$ is the boson number operator at site $i$, $\langle ij\rangle$ indicates that $j$ is one of the nearest neighboring site of $i$ on the square lattice, $\mu$ is the chemical potential, and $J$ denotes the nearest-neighbor hopping amplitude of the boson. The repulsive interaction between the bosons is approximated to be on-site with strength $U$; in addition, these bosons also experience a long-range attraction with strength $V$. For this setup \cite{PhysRevA.94.023632, PhysRevA.99.043633}, the wavelengths of the single-mode cavity and the optical lattice are equal, the operator $\hat{\Phi}$ takes the form 
 \begin{equation}
 \label{s2}
     \hat{\Phi} = \sum_i (-1)^{i_x+i_y} \, \hat{n}_{i}/N_0 \
 \end{equation}
Note that the presence of this interaction splits the 2D square lattice into two inter-penetrating sub-lattices with opposite sign of $\hat \Phi$. 

To obtain the phase diagram of this model, we first decouple off-site long-range interactions. Defining the long-range fluctuations $\delta\hat{\Phi}^2=(\hat{\Phi}-\langle\hat{\Phi}\rangle)^2$, we write  
\begin{equation}
\label{s3}
    \hat{\Phi}^2 \simeq \delta n\hat{\Phi}-\frac{\delta n^2 }{2}
\end{equation}
where we have ignored the fluctuations $\delta \Phi^2$ term and 
\begin{eqnarray} 
\delta n &=& \frac{1}{N_0} \langle \hat{n}_{2j} - \hat{n}_{2j+1} \rangle= n_A-n_B = 2 \langle \Phi \rangle,
\end{eqnarray}
where $A(B)$ denotes sublattices for which sign of $\hat \Phi$ are opposite. This allows us to write the Hamiltonian in Eq.(\ref{s1}) as
\begin{eqnarray}
    \hat{H_1} &=& -\sum_{<i,j>} J_{ij}(\hat{b}_i^\dagger \hat{b}_j + H.c.) + \frac{U}{2} \sum_{i} \hat n_i(\hat n_i-1) \nonumber\\
    && - \sum_i  \mu_i \hat n_i + V\frac{\delta n^2}{2} \label{s4}
\end{eqnarray}
where $\mu_i=\mu + (-1)^{i_x+i_y} V\delta n$. We note that $\delta n$ is a measure of the charge imbalance of the system; it's finite value represents broken translational symmetry. 

\begin{figure}
    \centering
    \includegraphics[width=\linewidth]{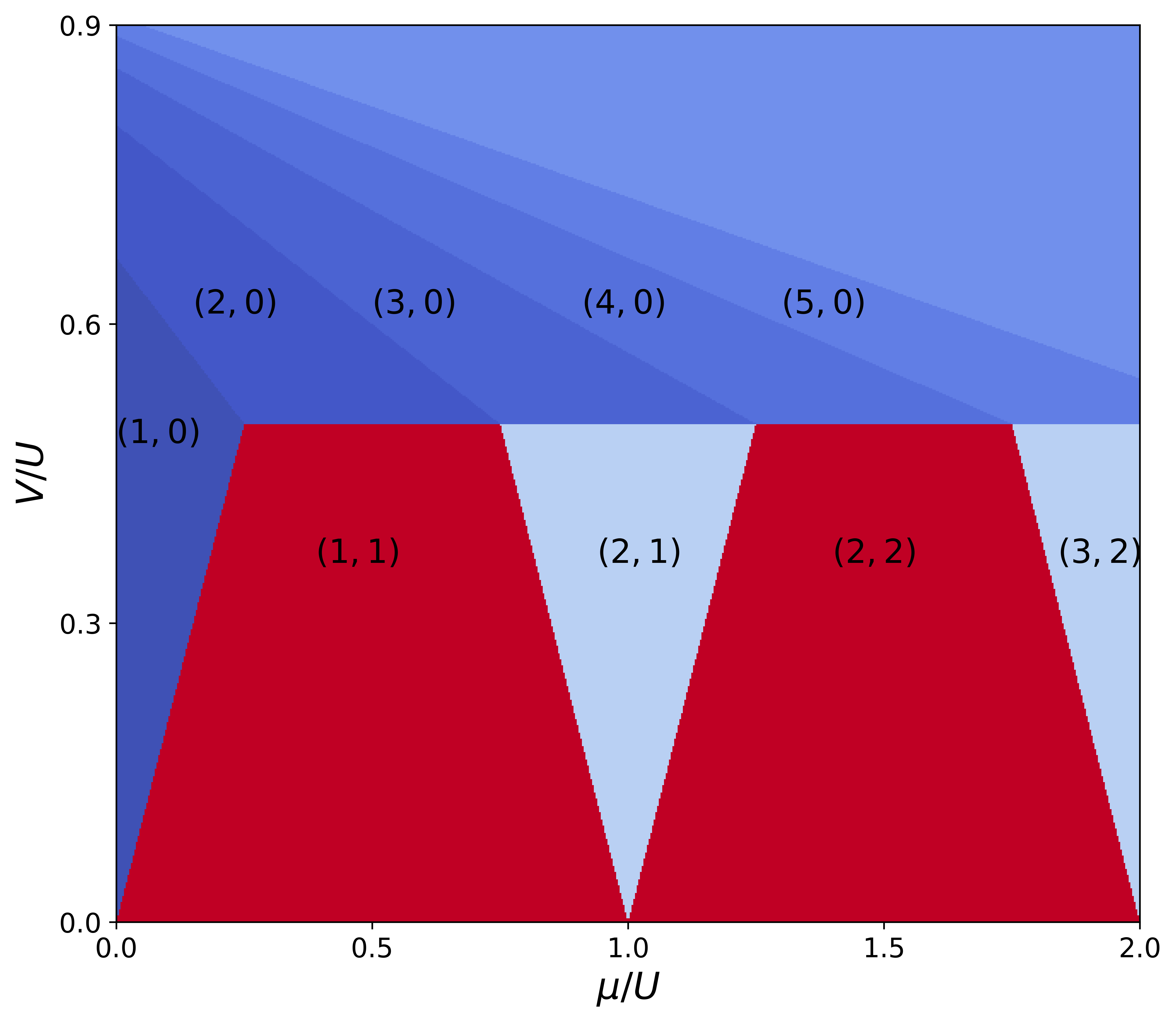}
    \caption{Ground state phase diagram of the model in the atomic limit $(J/U=0)$ as a function of the chemical potential $\mu/U$ and the long-range interaction $V/U$. For $V/U \le 1/2$, one finds translationally invariant MI phases with $n_A=n_B$ (red) separated by partially polarized CDW phase with $n_A-n_B=1$ sandwiched between them (light blue). For $V/U> 0.5$, one finds several fully polarized CDW phases with $n_B=0$ (dark blue); for these phases $\delta n$ incoreses with $\mu/U$ as shown. The white region indicates instability of the system due to large $V/U$; see text for details. }
    \label{fig1}
\end{figure}

First, we analyze this model in the atomic limit where $J/U =0$. In this case $n_A$ and $n_B$ are good quantum numbers; consequently one can choose a variational wavefunction $|\psi\rangle = \prod_{i \in A} \prod_{j \in B} |n_A,n_B\rangle$. The on-site energy is then computed using $E[n_A, n_B]= \langle \psi|H_1[J=0]| \psi\rangle$
and is given by 
\begin{eqnarray}
E[n_A,n_B] &=&  (-\mu +V\delta n)n_B - (\mu+V \delta n)n_A  \label{s10}\\
&& + \frac{U}{2} [n_A(n_A-1) + n_B(n_B-1)] + \frac{V}{2} \delta n^2 \nonumber
\end{eqnarray}
The ground state is found by the configuration that minimizes the energy $E[n_A,n_B]$ and is a function of the dimensionless parameters $V/U$ and $\mu/U$. We note that $V=0$ corresponds to the standard Bose-Hubbard model in the atomic limit whose phase diagram is well-known; also, for $V/U \gg \mu/U$, we expect $\delta n \ne 0$ leading to translational symmetry broken CDW phase. The phase diagram, shown in Fig.\ \ref{fig1}, 

The analytic expression of the phase boundaries as shown in Fig.\ \ref{fig1} are calculated by comparing $E[n_A,n_B]$ for different $n_A$ and $n_B$. For $V=0$, it is well-known that the translationally symmetric MI phase with $n$ bosons per site is realized for $(n-1) \le \mu/U \le n$. Upon increasing the value $V/U$, the MI phase with commensurate density $n$ remains stable provided $V/U \le 1/2$ and for 
\begin{equation}
    (n-1)+\frac{V}{2U} \le  \frac{\mu}{U} \le n-\frac{V}{2U} \label{pcond1}
\end{equation}
 At the boundaries where the equalities in Eq.\ \ref{pcond1} hold, there is an abrupt jump from the MI to the partially polarized CDW phase $|\delta n|=1$. This signifies a first order transition and leads to a doubly degenerate CDW state with $n=\pm 1$; this degeneracy is spontaneously broken by the ground state. 
 
 At $V/U>1/2$, the ground state becomes fully polarized with $n_A=n$ and $n_B=0$. Such a state remain stable till $V/U\le 1$ and for  
 \begin{equation}
     (1-V/U)(\delta n-1)-\frac{V}{2U} \le \mu/U \le (1-V/U)\delta n-\frac{V}{2U}
 \end{equation}
For $V/U>1$, the on-site energy is not bounded from below. In this regime an analysis based on grand canonical ensemble can not be performed and we are not going to address this regime in our work. 

\section{Formalism and Equilibrium Phase Diagram}
\label{seceq}
In this section, we provide a detailed description of the projection operator formalism. In Sec.\ \ref{secform}, we use this formalism to derive an effective low-energy Hamiltonian for the extended Bose-Hubbard model while in Sec.\ \ref{secph}, we compute the variational energy and operator expectation values using this effective Hamiltonian. Finally, in Sec.\ \ref{secnumeq}, we present our numerical results on the equilibrium phase diagram of the model. 

\subsection{Formalism}
\label{secform}
In this section, we develop the projection operator formalism which allows one to incorporate the effect of short-range quantum fluctuations. We note at the outset that in our scheme the all-to-all long-range interaction mediated by $\hat \Phi$ (Eq.\ \ref{s2}) will be treated within mean-field approximation. The rationale behind this approach is that quantum fluctuations arising from such all-to-all long-range interaction terms do not depend on dimensions; moreover, they are usually small in the regime of interest.  

To this end, we divide the Hamiltonian in Eq.\ \ref{s1} into two parts $H= H_0 +H_1$ where $H_1= -J \sum_{\langle ij \rangle } (b_i^{\dagger} b_j +{\rm h.c.})$ and $H_0$, which is the collection of all on-site terms, can be written as  
       \begin{eqnarray}
            \hat{H}_0 &=& -\sum_j[(\mu+V\delta n) \hat{n}_{2j} + (\mu -V\delta n)  \hat{n}_{2j+1}] \label{h0eq}\\
            && + \frac{U}{2}  \sum_j[\hat{n}_{2j} (\hat{n}_{2j} - 1) + \hat{n}_{2j+1} (\hat{n}_{2j+1} - 1)] + \frac{V}{2}\delta n^2 \nonumber
        \end{eqnarray}
where we have distinguished between even and odd sites.

The first step for implementing the projection operators to be defined is to distinguish between high- and low-energy hopping processes in the Hamiltonian. The former costs an energy $O(U)$ while the latter does not. Now we define the projection operators
 \begin{eqnarray}
   \hat{T}_1 &=&  -J\sum_{n_A,n_B,j} \sqrt{n_A(n_B+1)} |n_A-1,n_B+1\rangle \langle n_A,n_B| \nonumber\\
   &=&  \sum_{n_A,n_B,j} T_1^{n_A,n_B}  \label{t1eq} \nonumber\\
    \hat{T}_2 &=& -J\sum_{n_A,n_B,j} \sqrt{n_B(n_A+1)} |n_A+1,n_B-1\rangle \langle n_A,n_B| \nonumber\\
    &=&  \sum_{n_A,n_B,j} T_2^{n_A,n_B} \label{t2eq} 
\end{eqnarray}
which express hopping between the neighboring sites in terms of the change of occupations $n_A$ and $n_B$. Here and in rest of this work, we shall choose $n_{A(B)}$ to be occupation of sites in the even(odd) sublattice. In terms of these operators, one can write 
\begin{eqnarray}
\hat{H}_1 &=& \hat{T}_1 + \hat{T}_2  \label{hopeq}   
\end{eqnarray}
 The crucial point of this decomposition is that it distinguishes between low and high-energy tunneling processes. We note here that in this scheme the low-energy subspace constitutes a set of states which are separated by $O(J)$ energy from the ground state; this scheme is therefore useful for $J/U \ll 1$.

To identify the high energy hopping process, we compare the energy of an arbitrary Fock state labeled by $|n_A, n_B\rangle$ (note that these are eigenstates of $H_0$) before and after a hopping process on a given link $\ell$ between adjacent sites $i$ and $j$. This leads to the operators identities involving $H_0 $ and $T_{1(2)}$ which reads 
\begin{eqnarray}
[H_0,T_{\alpha}^{n_A,n_B}] &=& \Delta E_{\alpha} T_{\alpha}^{n_A,n_B}, \quad \alpha=1,2, \nonumber\\
\Delta E_{\alpha} &=& (U - 2V) (1+(-1)^{\alpha}\delta n). \label{eqcomm}     
\end{eqnarray}
Thus as long as $(U-2V) \gg J$ which is the regime we shall be addressing, we find that for a hopping process due to $T_{\alpha}$ does not change the energy of the state 
when $n_{A(B)}= {\bar n_{A(B)}^{\alpha}}$ such that 
\begin{eqnarray} 
1+(-1)^{\alpha} ({\bar n_{A}^{\alpha}}-{\bar n_{B}^{\alpha}})=0. \label{eqcond1} 
\end{eqnarray} 
When the condition given by Eq.\ \ref{eqcond1} is not satisfied, the hopping process takes the system out of the low-energy manifold. 

In what follows, we eliminate these high-energy processes perturbatively using a Schrieffer-Wolff (SW) transformation. To this end, we first rearrange the hopping terms and write 
$H_1 = \sum_{\alpha=1,2} (T_{\alpha}+\bar{T}_{\alpha})$ where
 \begin{eqnarray}
     \bar{T}_{\alpha} &=& \sum_{n_A,n_B} T_{\alpha}^{n_A,n_B} \delta_{n_A,\bar{n}_A^{\alpha}} \delta_{n_B,\bar{n}_B^{\alpha}}= T_{\alpha}^{\bar{n}_A^{\alpha}\bar{n}_B^{\alpha}} \nonumber\\
     T_{\alpha} &=& \sum_{n_A,n_B} T_{\alpha}^{n_A,n_B} (1 -\delta_{n_A,\bar{n}_A^{\alpha}} \delta_{n_B,\bar{n}_B^{\alpha}}). \label{eqhop2} 
 \end{eqnarray}
Next we implement a SW transformation via an operator $S$ so that $H_{\rm eff} = e^{i S} H e^{-iS}$ does not contain any high-energy hopping process to $O(J/U)$. 
A straightforward perturbation expansion in $S$ leads to 
\begin{eqnarray}
    H_{\rm eff} &=& H_0 + H_1 +[iS,H_0] + [iS,H_1] \nonumber\\
    && +\frac{1}{2!}[iS,[[iS,H_0]+[iS,H_1]]] + .... \label{eqeffham1}
\end{eqnarray}
where the ellipsis indicates terms which are ${\rm O}(J^3/U^3)$ or higher. 

From Eq.\ \ref{eqeffham1}, we find that elimination of the high-energy hopping processes to $O(J/U)$ requires that $[iS, H_0] = - \sum_{\alpha} T_{\alpha}$. Using Eq.\ \ref{eqcomm}, this yields the expression of $S$ to be 
\begin{eqnarray}
iS &=& - \sum_{\alpha , n_A,n_B\neq \bar{n}_A^{\alpha},\bar{n}_B^{\alpha} } \frac{T_{\alpha}^{n_A^{\alpha}n_B^{\alpha}}}{\Delta E_{\alpha}} \label{eqsw1} 
\end{eqnarray}

Substituting Eq.\ \ref{eqsw1} in Eq.\ \ref{eqeffham1}, we get, after a few lines of algebra
\begin{widetext}
\begin{eqnarray}
    H_{\rm eff} &=& H_0 + \sum_{\alpha = 1,2} T_{\alpha}^{\bar{n}_A^{\alpha}\bar{n}_B^{\alpha}} +\sum_{\alpha,\beta}\sum_{ n_A,n_B\neq \bar{n}_A^{\alpha},\bar{n}_B^{\alpha} }\frac{[T_{\alpha}^{n_A,n_B},T_{\beta}^{\bar{n}_A^{\beta}\bar{n}_B^{\beta}}]}{\Delta E_{\alpha}} + \sum_{\alpha,\beta}\sum_{n_{A(B)} \neq \bar{n}_{A(B)}^{\alpha}} \sum_{n'_{A(B)}\neq \bar{n}_{A(B)}^\beta}\frac{[T_{\alpha}^{n_A,n_B},T_{\beta}^{n'_A,n'_B,}]}{\Delta E_{\alpha}} + \cdots \nonumber\\
    &=& H_0 + H_a + H_b + H_c + \cdots, \label{eqeffham2}
\end{eqnarray}
\end{widetext}
where the ellipsis indicate higher-order terms. We note that the second-order terms in $H_{\rm eff}$ involve hopping processes between links which involves nearest and next-nearest neighboring sites. Thus, in this perturbative scheme, the range of correlations increase with order of perturbation theory. The present analysis therefore takes into account short-ranged correlation. In the next section, we shall use $H_{\rm eff}$ to estimate the ground state energy and determine the phases of the model.

\subsection{Equilibrium Phases}
\label{secph}

In this section, we compute the variational energy starting from $H_{\rm eff}$ (Eq.\ \ref{eqeffham2}) and using a 
variational ground state wavefunction. For this we note that for any chosen variational wavefunction $|\psi\rangle$
we can write
\begin{eqnarray}
E_{\rm var} &=& \langle \psi|H|\psi \rangle = \langle \psi'|H_{\rm eff} |\psi'\rangle + {\rm O}[(J/U)^3] \nonumber\\
|\psi'\rangle &=& e^{i S} |\psi\rangle  \label{eqvaren1}  
\end{eqnarray}
In what follows we shall use a Gutzwiller form for $|\psi'\rangle$ given by 
\begin{eqnarray} 
|\psi'\rangle &=& \prod_{{\bf r}\in A,{\bf r'} \in B} \sum_{n_A,n_B} f^{n_A}_{\bf r} f^{n_B}_{{\bf r'}} |n_A\rangle_{\bf r} \otimes |n_B\rangle_{\bf r'}  \label{eqvarwav1} 
\end{eqnarray}
We note that such a choice does not mean that $|\psi\rangle$ is of the Gutzwiller form. In fact, spatial correlation is built into $|\psi\rangle$ due to the $\exp[iS]$ factor even when $|\psi'\rangle$ is chosen to be of Gutzwiller product form. 

To compute $E_{\rm var}$, we decompose it into different terms and write $E_{\rm var}= E_0 + E_a +E_b + E_c$ where $E_{\mu} = \langle \psi' | H_{\mu} |\psi'\rangle$, where $\mu= 0, a,b ,c$ and $H_{\mu}$ is given by Eq.\ \ref{eqeffham2}. The first term, $E_0$, is given by
\begin{eqnarray} 
E_0 &=& \sum_{\bf r \in A}  |f^{n_A}_{\bf r}|^2 \epsilon_1(n_A) + \sum_{\bf r \in B} |f^{n_B}_{\bf r}|^2  \epsilon_2(n_B)  \label{eqen0} \\
\epsilon_1(n_A) &=& -(\mu + V  \delta n) n_A + \frac{U}{2} n_A(n_A-1) + \frac{V}{4} \delta n^2  \nonumber\\
\epsilon_2(n_B) &=& -(\mu - V  \delta n) n_B + \frac{U}{2} n_B(n_B-1) + \frac{V}{4} \delta n^2  \nonumber
\end{eqnarray} 
We note that due to the long-range interaction, the on-site energies depend on the occupation of neighboring sites. 

The next term $E_a$ can be computed similarly. Here we note that ${\bar n}_A^{\alpha} - {\bar n}_B^{\alpha} = (-1)^{\alpha+1}$ for $\alpha=1,2$. Using this relation, we obtain
\begin{eqnarray} 
E_a &=& -J \sum_{{\bf r} \in A, {\bf r'} \in B} \Big[ f_{{\bf r}}^{{\bar n}_A^{1}-1 \ast} f_{{\bf r}}^{{\bar n}_A^{1}} f_{{\bf r'}}^{{\bar n}_B^{1}+1 \ast} f_{{\bf r'}}^{{\bar n}_B^{1}} \sqrt{{\bar n}_A^{1}({\bar n}_B^{1}+1)} \nonumber\\
&& + f_{{\bf r}}^{{\bar n}_A^{2}+1 \ast} f_{{\bf r}}^{{\bar n}_A^{2}} f_{{\bf r'}}^{{\bar n}_B^{2}-1 \ast} f_{{\bf r'}}^{{\bar n}_B^{2}} \sqrt{{\bar n}_B^{2}({\bar n}_A^{2}+1)} \Big]\label{eqena}
\end{eqnarray}
We note that such terms involve a single link $\ell$ separating the neighboring sites ${\bf r }\in A$ and ${\bf r'} \in B$. 

Next, we compute $E_b$ and $E_c$ which involves either a single link $\ell_1$ between two neighboring sites or two adjacent links 
$\ell_1$ and $\ell_2$ connecting three sites. For this purpose, for the first class of terms, we define 
\begin{eqnarray} 
\langle \psi'| T_{\alpha \ell_1}^{n_A,n_B} T_{\beta \ell_1}^{n'_A,n'_B}|\psi'\rangle &=& J^2 S_{\alpha \beta;\ell_1}(n_A, n_B; n'_A, n'_B) \nonumber\\ \label{eqsdef}
\end{eqnarray} 
where $T_{\ell_1} \equiv T_{\bf r,\bf r'}$ indicates hopping on link $\ell_1$. The expression of $S_{\alpha \beta:\ell_1} (n_A, n_B; n'_A, n'_B) $ can be explicitly obtained but is cumbersome; this is given in Appendix \ \ref{appa}. In terms of $S_{\alpha \beta;\ell_1}(n_A, n_B; n'_A, n'_B)$, we obtain 
\begin{widetext}
\begin{eqnarray}
E_b &=& \sum_{\ell_1} \sum_{\alpha,\beta=1,2} \sum_{n_A,n_B,n'_A,n'_B} \frac{J^2}{\Delta E_{\alpha}} \Big[S_{\alpha \beta; \ell_1}(n_A,n_B; n'_A,n'_B) - S_{\beta \alpha;\ell_1}(n'_A,n'_B; n_A,n_B)\Big] \nonumber\\
&& \times (1-\delta_{n_A, {\bar n}_A^{\alpha}} \delta_{n_B, {\bar n}_B^{\alpha)}}) \delta_{n'_A, {\bar n}_A^{\beta}} \delta_{n'_B, {\bar n}_B^{\beta}} \label{eqenb}
\end{eqnarray}
\end{widetext}
The terms in $E_c$ can be computed similarly. We divide these terms into two parts $E_c = E_c^{(1)} +E_c^{(2)}$ such that $E_c^{(1)}$ 
involves contribution of terms from a single link while $E_c^{(2)}$ contains contribution of terms from two adjacent links. The first of these terms, $E_c^{(1)}$, can be written in terms of $S_{\alpha \beta;\ell_1}(n_A, n_B; n'_A, n'_B)$ and is given by 
\begin{widetext} 
\begin{eqnarray}
E_c^{(1)} &=& \sum_{\ell_1}\sum_{\alpha,\beta=1,2} \sum_{n_A,n_B,n'_A,n'_B} \frac{J^2}{2\Delta E_{\alpha}} \Big[S_{\alpha \beta; \ell_1}(n_A,n_B; n'_A,n'_B) - S_{\beta \alpha;\ell_1}(n'_A,n'_B; n_A,n_B) \Big]\nonumber\\
&& \times (1-\delta_{n_A, {\bar n}_A^{\alpha}} \delta_{n_B, {\bar n}_B^{\alpha)}})(1- \delta_{n'_A, {\bar n}_A^{\beta}} \delta_{n'_B, {\bar n}_B^{\beta}})  \label{eqenc1}
\end{eqnarray}
\end{widetext} 

Next, we consider the terms in $E_c$ which involves two adjacent links $\ell_1$ and $\ell_2$. Such terms involve two distinct cases; the first is when the site which is common to both the links belongs to sublattice $A$ and the second when it belongs to sublattice $B$. We denote these cases by an index $a=1,2$, respectively. We compute the matrix elements where the first hopping occurs on link $\ell_2$ and the second on $\ell_1$. Since we sum over $\ell_1$ and $\ell_2$, we can always choose $\ell_2$ to be to the right of $\ell_1$. This allows us to define
\begin{eqnarray}
    \langle\psi'|T_{\alpha;\ell_1}^{n_A,n_B},T_{\beta ;\ell_2}^{n'_A,n'_B}|\psi'\rangle= J^2 R_{\alpha\beta ;\ell_1,\ell_2}^{(a)}(n_A,n_B;n'_A,n'_B) \nonumber\\  \label{eqrdef}
\end{eqnarray}
Explicit expressions for different components of $R_{\alpha\beta ;\ell_1,\ell_2}^{(a)}(n_A,n_B;n'_A,n'_B)$ in terms of the coefficients $f_{\bf r}$ is charted out in Appendix\ \ref{appa}. The terms in $E_c^{(2)}$ involving two different links can be expressed in terms of $R_{\alpha\beta ;\ell_1,\ell_2}^{(a)}(n_A,n_B;n'_A,n'_B)$ and are given by $E_c^{(2)} = E_2+E_3$ where 
\begin{widetext} 
\begin{eqnarray}
    E_{2} &=& \sum_{a=1,2}\sum_{\alpha,\beta=1,2}\sum_{\ell_1,\ell_2} \sum_{n_A,n_B,n'_A,,n'_B} \frac{J^2}{\Delta E_{\alpha}} \Big[R_{\alpha\beta ;\ell_1,\ell_2}^{(a)}(n_A,n_B;\bar{n}_A^{\beta},\bar{n}_B^{\beta}) - R_{\beta\alpha ;\ell_1,\ell_2}^{(a)}(\bar{n}_A^{\beta},\bar{n}_B^{\beta};n_A,n_B)\Big] \nonumber\\
    && \times (1-\delta_{n_A,\bar{n}_A^{\alpha}}\delta_{n_B,\bar{n}_B^{\alpha}})  \nonumber\\ 
E_3 &=& \sum_{a=1,2}\sum_{\alpha,\beta=1,2}\sum_{\ell_1,\ell_2} \sum_{n_A,n_B,n'_A,,n'_B,} \frac{J^2}{\Delta E_{\alpha}}\Big[R_{\alpha\beta ;\ell_1,\ell_2}^a(n_A,n_B;n'_A,n'_B) - R_{\beta\alpha ;\ell_1,\ell_2}^a(n'_A,n'_B;n_A,n_B)\Big] \nonumber\\
&& \times (1-\delta_{n_A,\bar{n}_A^{\alpha}}\delta_{n_B,\bar{n}_B^{\alpha}}) (1-\delta_{n'_A,\bar{n}_A^{\beta}}\delta_{n'_B,\bar{n}_B^{\beta}}) \label{eqenc2}
\end{eqnarray}
\end{widetext} 
Eqs.\ \ref{eqen0}, \ref{eqena}, \ref{eqenb}, \ref{eqenc1} and \ref{eqenc2} provide an expressions of the variational energy $E_{\rm var}$ in terms of the coefficients $f_{\bf r}^{A(B)}$. A numerical minimization of $E_{\rm var}$ therefore leads to the wavefunction $|\psi'\rangle$ in terms of the coefficients $f_{\bf r}^{n_A(n_B)}$.

Having obtained the ground state wavefunction $|\psi\rangle$ from $|\psi'\rangle$ using Eq.\ \ref{eqvaren1}, one can compute expectation values of several physical observables. 
We note that for any operator $\hat O$, one can write 
\begin{eqnarray}
\langle \psi|\hat{O}|\psi\rangle &=& \langle \psi'|e^{is}\hat{O}e^{-is}|\psi'\rangle =O_1+O_2+O_3/2 \nonumber\\
O_1 &=&\langle \psi'|\hat{O}|\psi'\rangle, \quad  O_2=\langle \psi'|[iS,\hat{O}]|\psi'\rangle \nonumber\\
O_3 &=& \langle \psi'|[iS,[iS,\hat{O}]]|\psi'\rangle \label{opex1}
\end{eqnarray}
where we restrict ourselves up to $O(J^2/U^2)$. Using the expression for $S$ in Eq.\ \ref{eqsw1}, we find 
\begin{eqnarray}
    O_2 &=& \sum_{\alpha=1,2}\sum_{n_A,n_B,n'_A,,n'_B} \frac{1}{\epsilon_{\alpha}} \langle \psi'|[T_{\alpha}^{n_A,n_B},\hat{O}]|\psi'\rangle \nonumber\\
    && \times (1-\delta_{n_A,\bar{n}_A^{\alpha}}\delta_{n_B,\bar{n}_B^{\alpha}}) \nonumber\\
    O_3 &=& \sum_{\alpha=1,2}\sum_{n_A,n_B,n'_A,n'_B} \langle \psi'|[T_{\alpha}^{n_A,n_B},[T_{\beta}^{n'_A,n'_B},\hat{O}]]|\psi'\rangle \nonumber\\
    && \times \frac{(1-\delta_{n_A,\bar{n}_A^{\alpha}}\delta_{n_B,\bar{n}_B^{\alpha}})(1-\delta_{n'_A,\bar{n}_A^{\beta}}\delta_{n'_B,,\bar{n}_B^{\beta}})}{\epsilon_{\alpha} \epsilon_{\beta}} \label{opex2}
\end{eqnarray}
We note that a standard variational mean-field approach leads to $\langle \hat O\rangle = O_1$; thus, $O_2$ and $O_3$ represents contribution of fluctuations over the standard mean-field results. 

\begin{figure}
\rotatebox{0} {\includegraphics[width=0.49\linewidth]{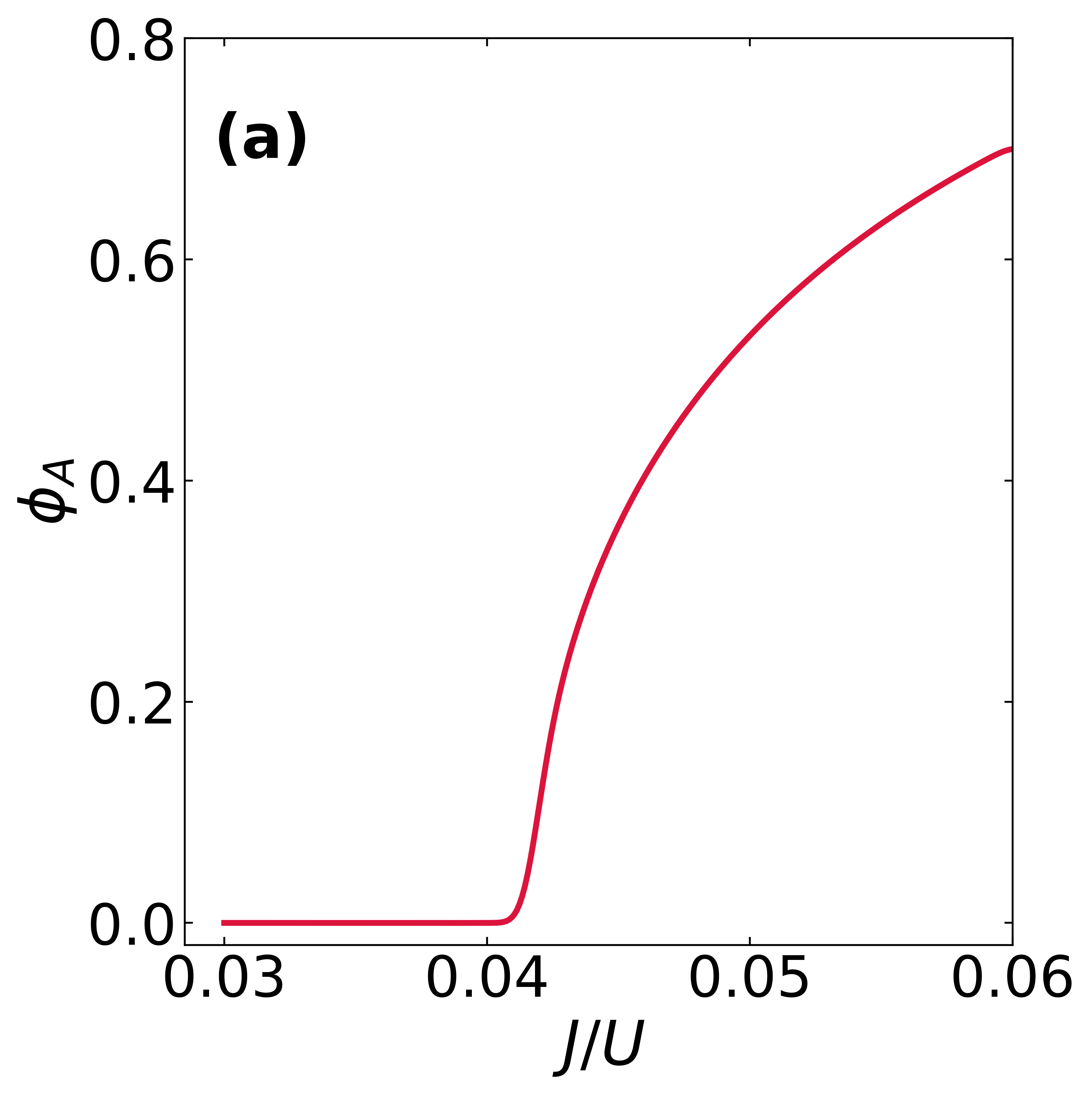}}
\rotatebox{0} {\includegraphics[width=0.49\linewidth]{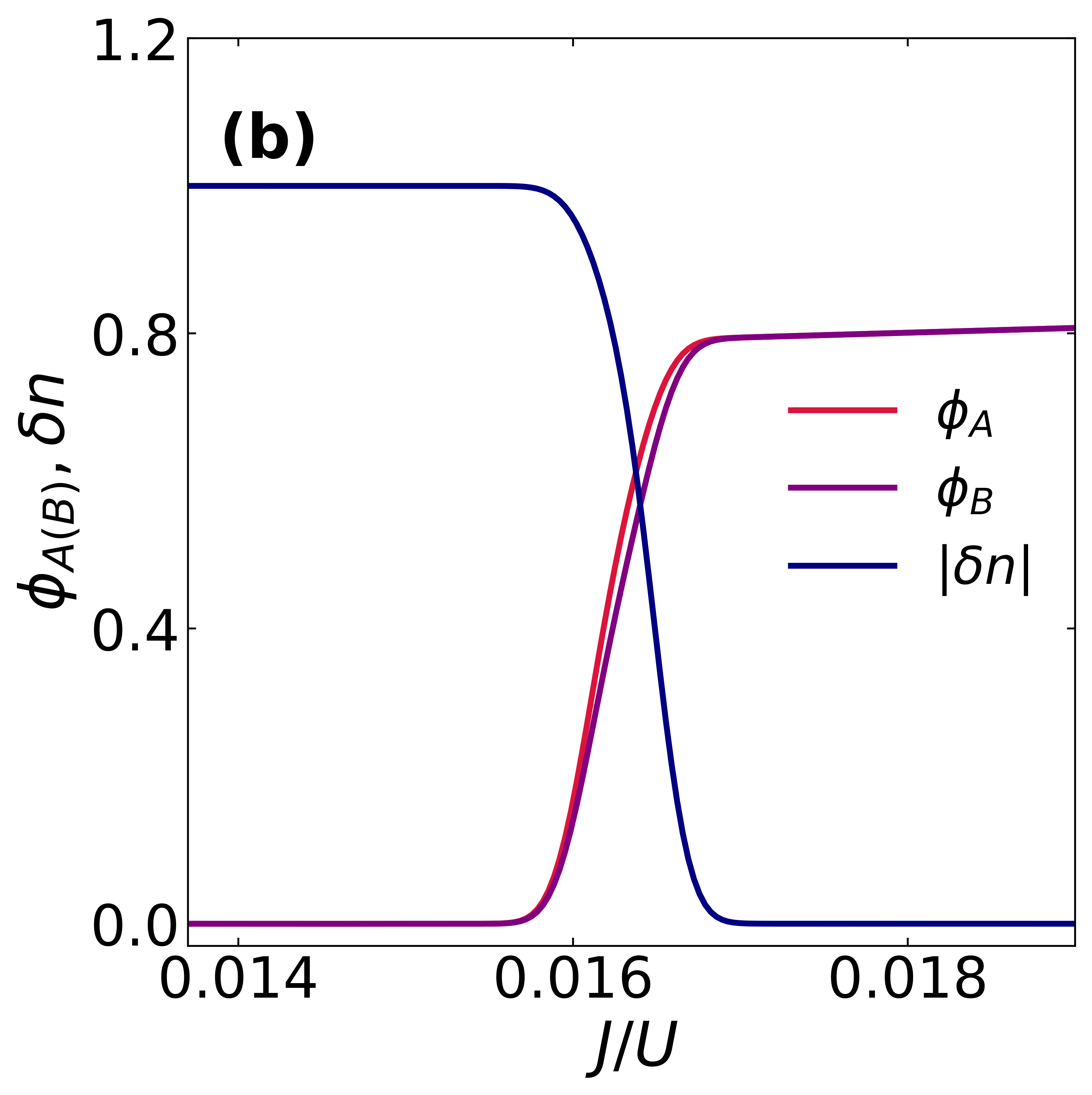}}
\rotatebox{0} {\includegraphics[width=0.49\linewidth]{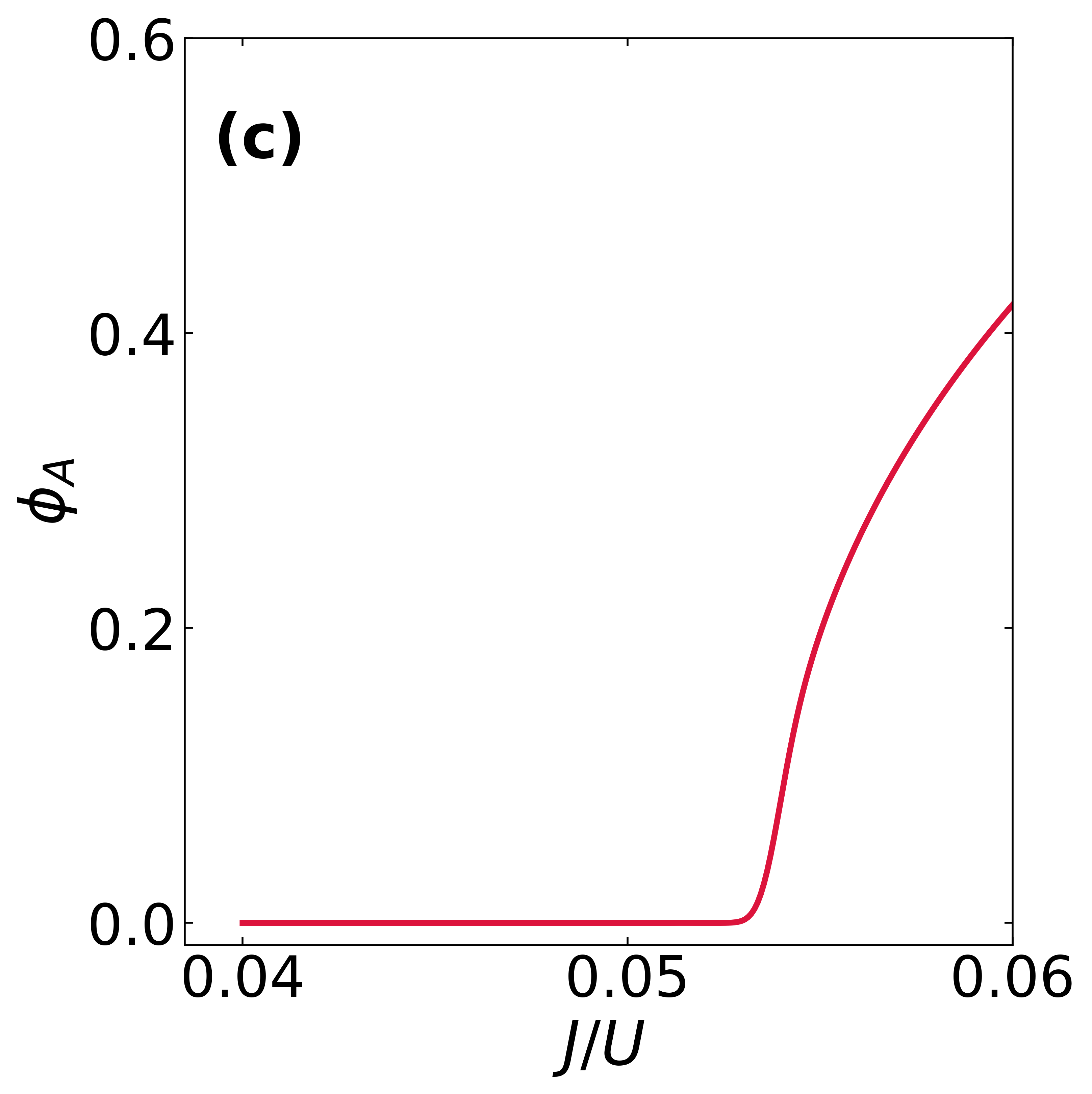}}
\rotatebox{0} {\includegraphics[width=0.49\linewidth]{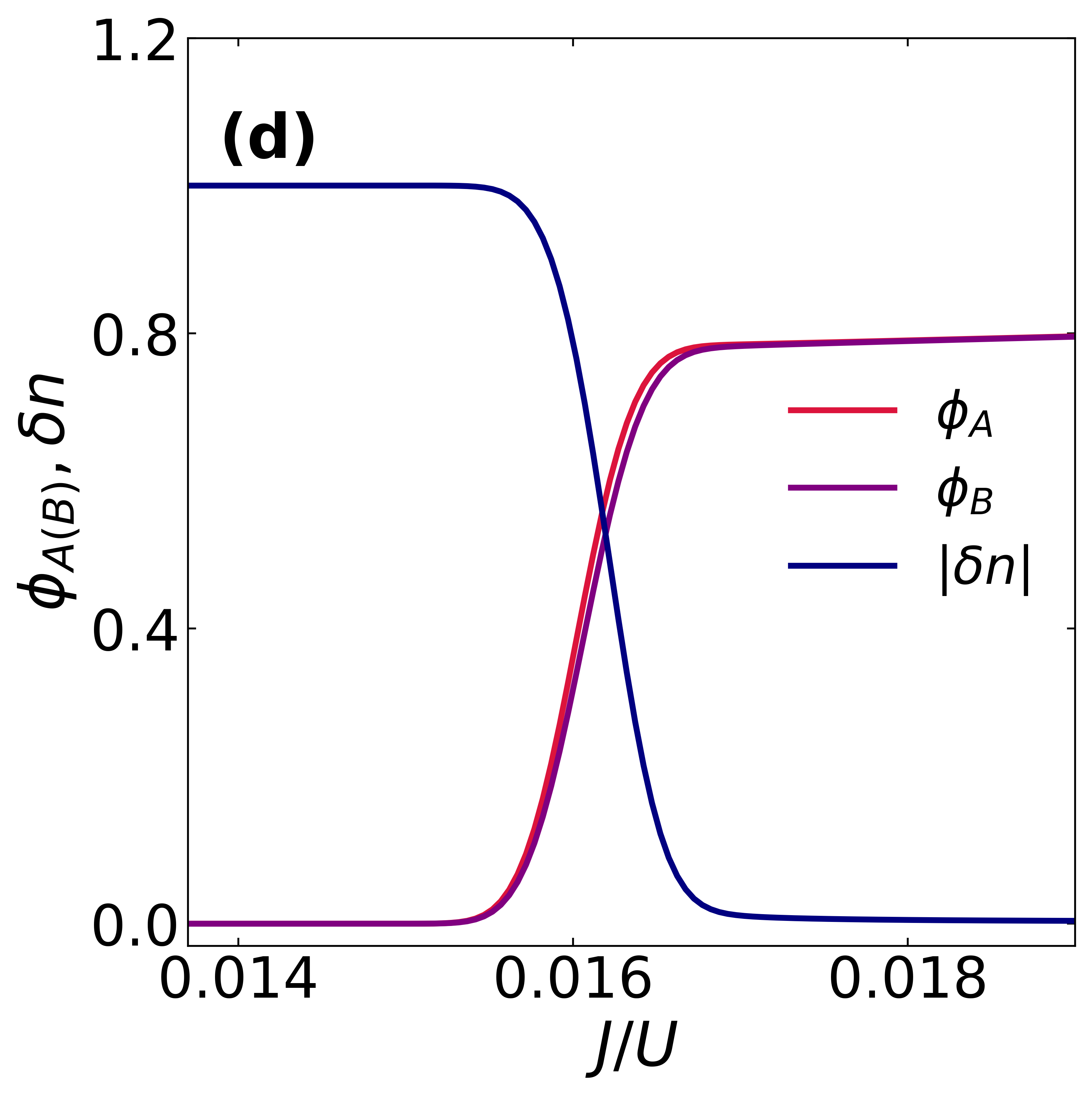}}
    \caption{(a) Plots of the order parameters $\phi_A$, $\phi_B$, and $\delta n =|\langle n_A\rangle-\langle n_B\rangle|$ as a function of $J/U$ for $\mu/U=0.42$ obtained using mean-field theory near the peak of the Mott Insulator (MI) lobe where $\phi_A=\phi_B$ and $\delta n=0$. (b) Similar plot across the tip of the CDW lobe for and $\mu/U=0.98$. Here $\delta n$ decreases from $1$ to $0$ as one moves across the Mott lobe from the Mott to the superfluid phase. In addition, there is a narrow finite region near the tip of the Mott lobe where both $\delta n$ and $\phi_{A,B}$ are finite indicating possibility of a SS phase. (c) and (d), Same as  (a) and (b) but using projection operator technique. These predict a different position of the Mott lobe for the homogeneous MI phase (a); for the CDW lobe, it provides a slightly wider region where $\delta n$ and $\phi_A -\phi_B$ are both non-zero. See text for details.}
    \label{fig2a}
\end{figure}
\begin{figure}
\rotatebox{0}{\includegraphics[width=0.49\linewidth]{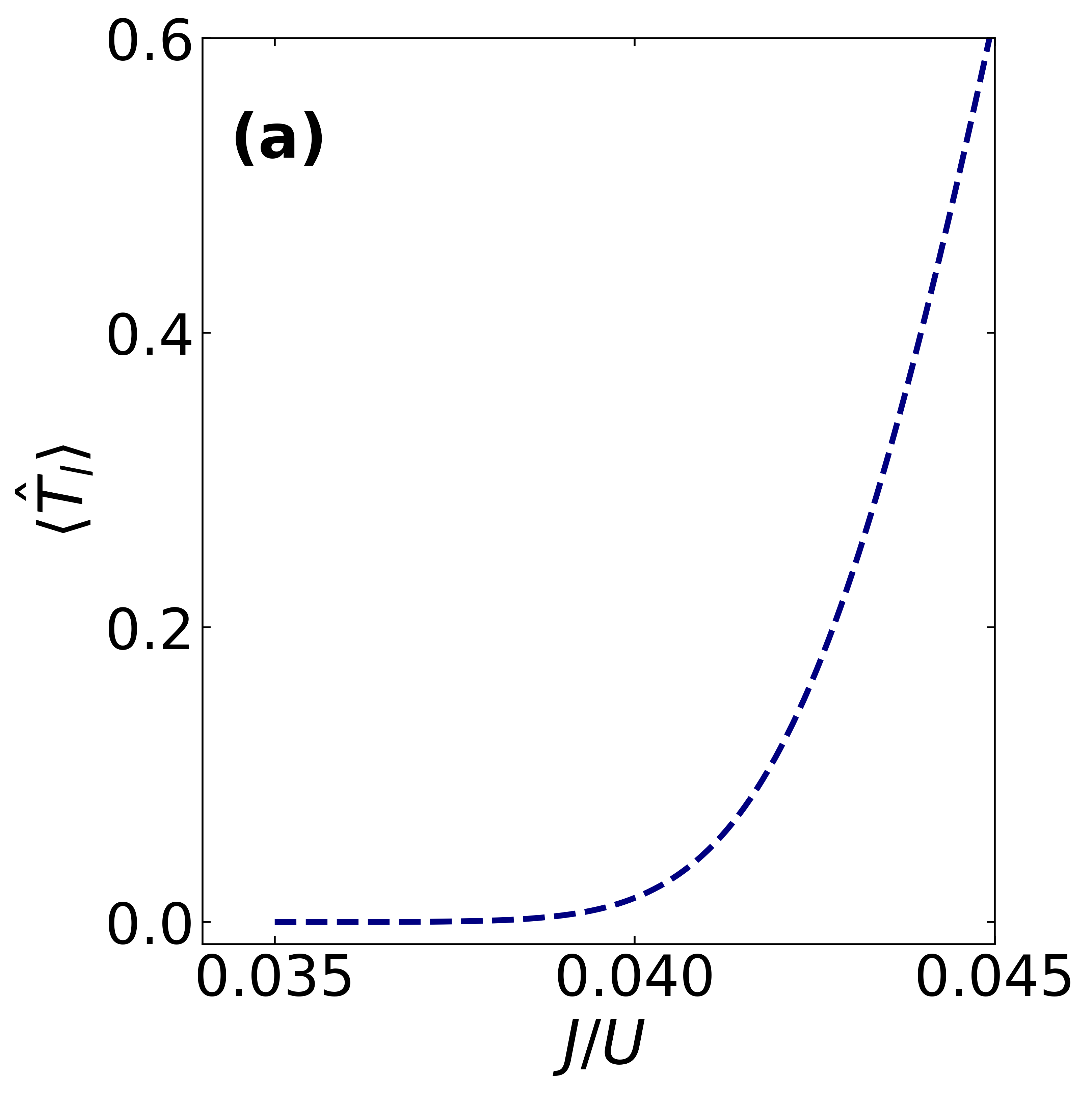}}
\rotatebox{0}{\includegraphics[width=0.49\linewidth]{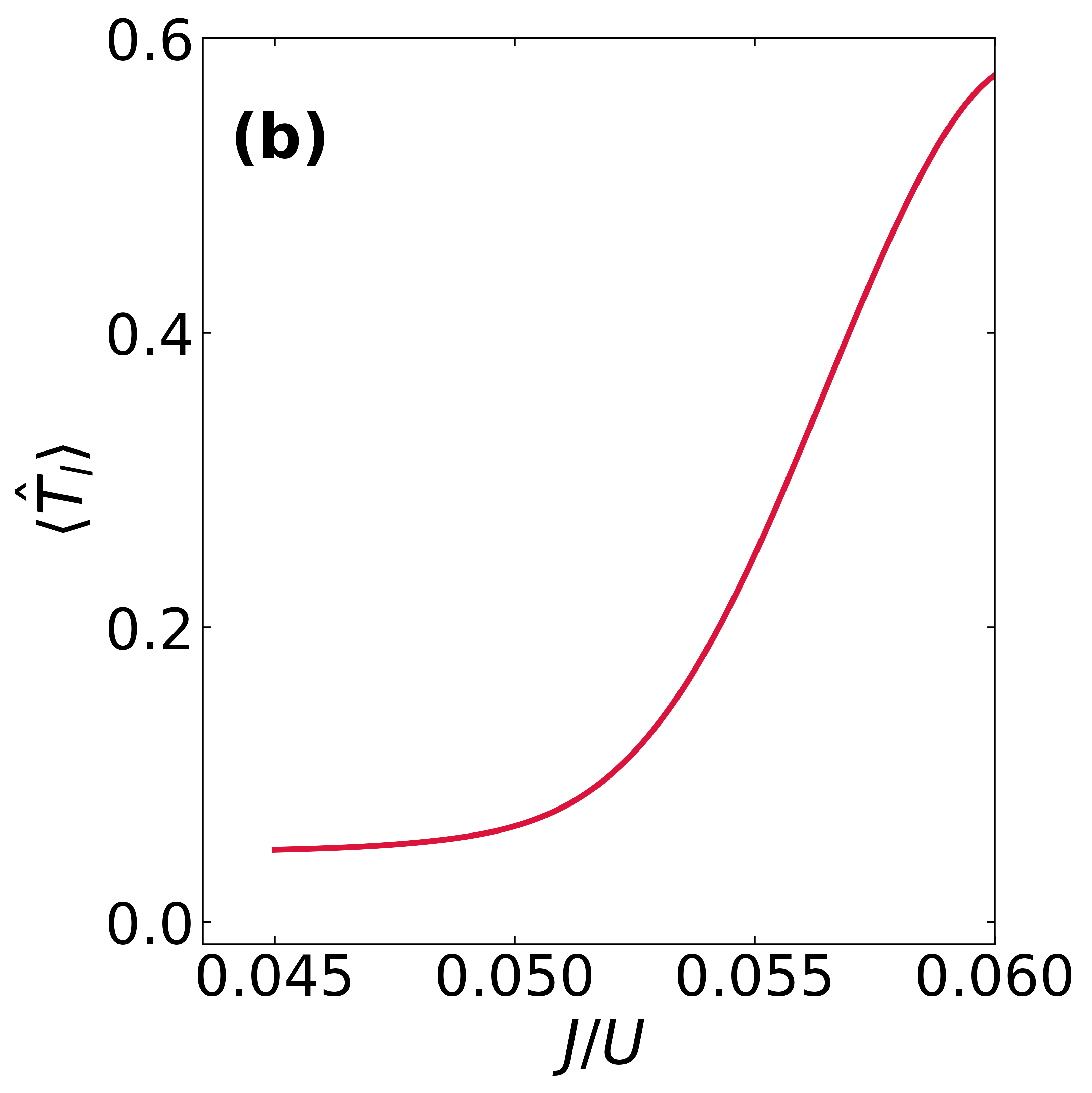}}
\rotatebox{0}{\includegraphics[width=0.49\linewidth]{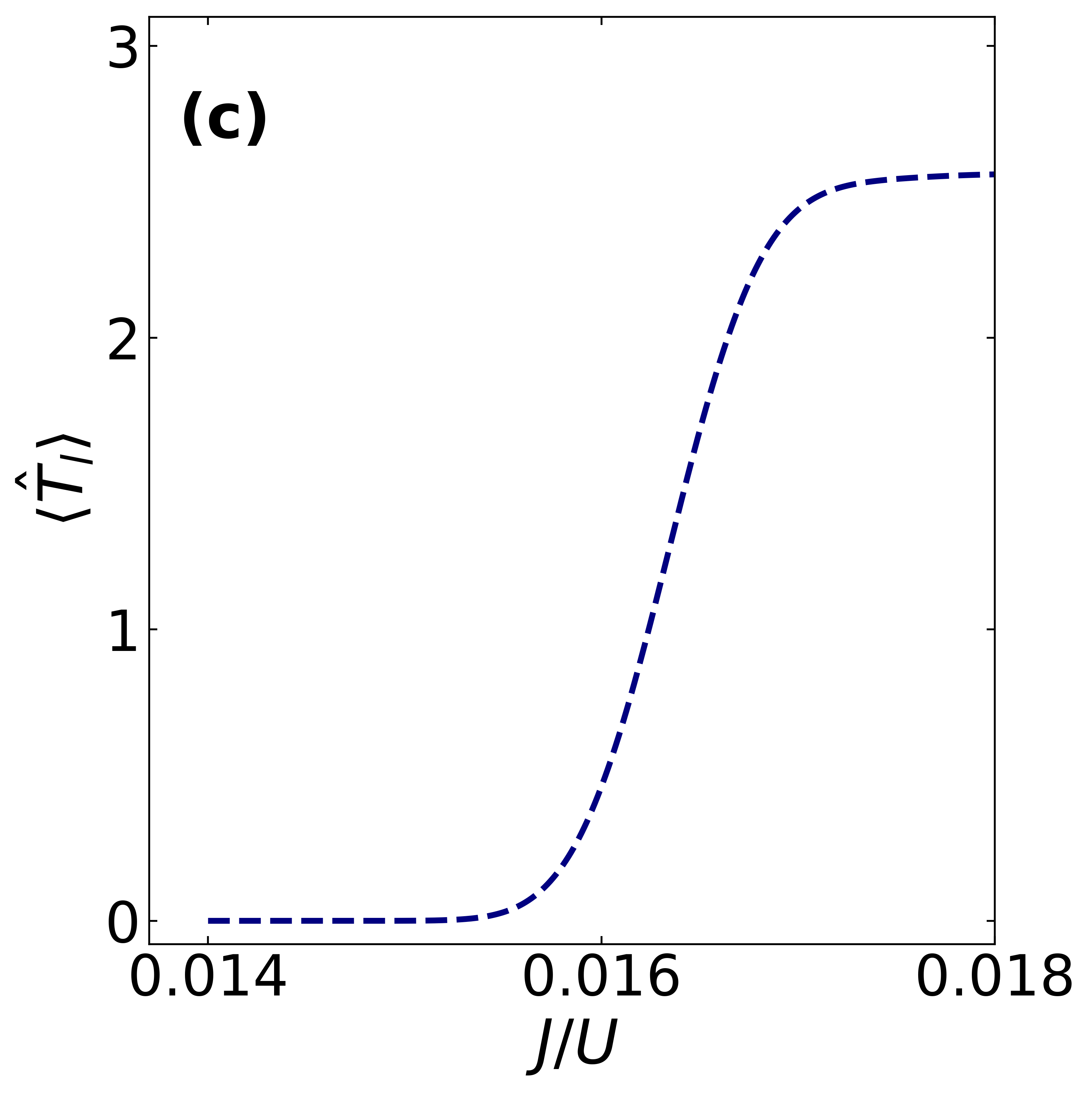}}
\rotatebox{0}{\includegraphics[width=0.49\linewidth]{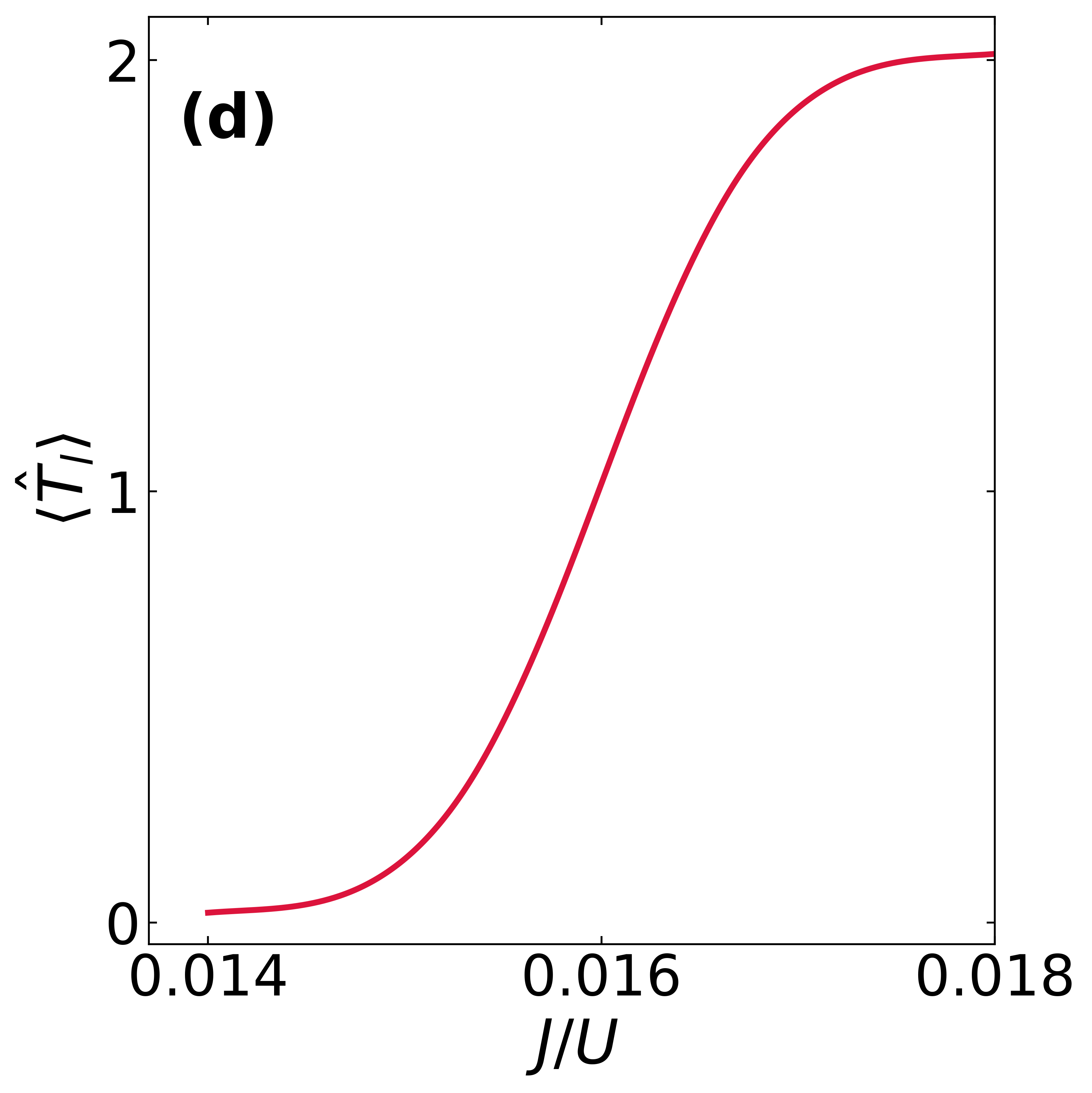}}
    \caption{ Plot for expectation of the kinetic operator $\langle T_l \rangle$ at the MI obtained using mean-field theory (panel (a), dotted lines) and projection operator formalism (panel (b), solid line). (c) and (d) Similar plots for the CDW phase respectively. All other parameters are same in Fig.\ \ref{fig2a}. See text for details.}
    \label{fig2b}
\end{figure}

\subsection{ Numerical results}
\label{secnumeq}
The phase diagram of the extended Bose-Hubbard model can be obtained by minimizing the variational energy $E_{\rm var} = E_0+E_1+E_2+E_3$ whose explicit expressions are given in Eqs.\ \ref{eqen0}, \ref{eqena}, \ref{eqenb}, \ref{eqenc1} and \ref{eqenc2}. In what follows, we carry out this minimization numerically assuming translational invariance within each sublattice; however, translational symmetry breaking between $A$ and $B$ sublattices which is crucial for the realization of the CDW and SS phases are allowed. A numerical minimization of $E$, with a cutoff $n_A, n_B \le 3$ on each site, allows us to obtain the variational parameters $f_{n_A}$ and $f_{n_B}$ on each sublattice leading to $|\psi'\rangle$.

To characterize the different phases, we compute the operator expectation is then numerically computed using Eq.\ \ref{opex1} and \ref{opex2}. In what follows we compute the superfluid and the density wave-order parameters given by
\begin{eqnarray} 
\delta n &=& \langle \psi|\hat \Phi|\psi\rangle, \quad 
\phi_{\alpha} = \langle \psi| \hat b_{\alpha}|\psi\rangle   \label{op1} 
\end{eqnarray} 
where $\alpha=A, B$ is the sublattice index and we have assumed translational invariance within each sublattice. The characterization of the different phases can be achieved using these order-parameters. The MI phase corresponds to $\phi_{\alpha}=0=\delta n$ while the CDW phase corresponds to $\phi_{\alpha}=0, \delta n \ne 0$. The SF phase corresponds to $\delta n=0, \phi_{\alpha} \ne 0$ while an indication of the SS phase is given by $\phi_{\alpha}, \delta n \ne 0$. We also compare our numerics with the standard mean-field phase diagram which is obtained by replacing $|\psi\rangle=|\psi'\rangle$ in our numerics. This allows us to  provide a comparison between the mean-field and the projection operator results.

The result of these computations is shown in Fig.\ \ref{fig2a}. The top panels shows a computation of $\delta n$, $\phi_A$ and $\phi_B$ as a function of $J/U$ using mean-field theory while the bottom panels shown analogous plots obtained using the projection operator technique. For all plots, we have considered a coordination number $z=4$ which corresponds to a 2D square lattice. Moreover, for the left panels, $\mu/U=0.42$ which indicates that varying $J/U$ allows one to exist the MI phase through the tip of the Mott lobe; for the right panels, $\mu/U=0.98$ which corresponds to the tip of the CDW lobe at $J=J_c$. The left panels ((a) and (c)) corresponds to these plots across the uniform Mi-SF phase boundary. Here $\delta n=0$ across the transition and the sublattice symmetry remains unbroken while the superfluid order parameters $\phi_A=\phi_B$ becomes finite at the transition. Importantly, the projection operator method predicts $J_c/U \simeq 0.056$ which is closer to Quantum Monte-Carlo value $J_c^{\rm QMC}/U \sim 0.061$ for the 2D Bose Hubbard model (which we expect to be exact for $\Phi=0$). In contrast, the right panel where $\delta n \ne 0$ for the CDW phase, both the mean-field and the projection operator method predict almost identical values of $J_c/U \sim 0.0155$. Both of these methods predict the presence of a SS phase at the tip of the Mott lobe; as $J$ is increased, one moves to the SF phase from the CDW through this SS phase.  However, the region for which $\delta n$ and $\phi_{A,B}$ coexist (domain of the SS phase) is slightly wider for Fig.\ \ref{fig2a}(d) indicating a wider parameter regime where SS phase can occur. 

A key difference between the mean-field and the projection operator approach is brought out in Fig.\ \ref{fig2b}. Here we plot $\langle T_{\ell} \rangle= \langle b_i^{\dagger} b_j +b_{j}^{\dagger} b_i\rangle$, where $\ell$ is the link between sites $i$ and $j$. The mean-field theory predicts $\langle T_{\ell} \rangle=0$ in the MI phase while the projection operator formalism correctly captures the leading-order fluctuation effects leading to a finite $\langle T_{\ell}\rangle \sim J^2/U$ in the MI phase. This difference is clearly seen from Fig.\ \ref{fig2b} where the left panels (a) and (c) show the mean-field results for the MI and CDW phases. The corresponding result obtained using the projection operator approach, exhibiting a finite value of $\langle T_{\ell}\rangle$ in the insulating phases below the transition, is shown in the right panels (b) and (d) 
  \begin{figure}
    \centering
    \includegraphics[width=\linewidth]{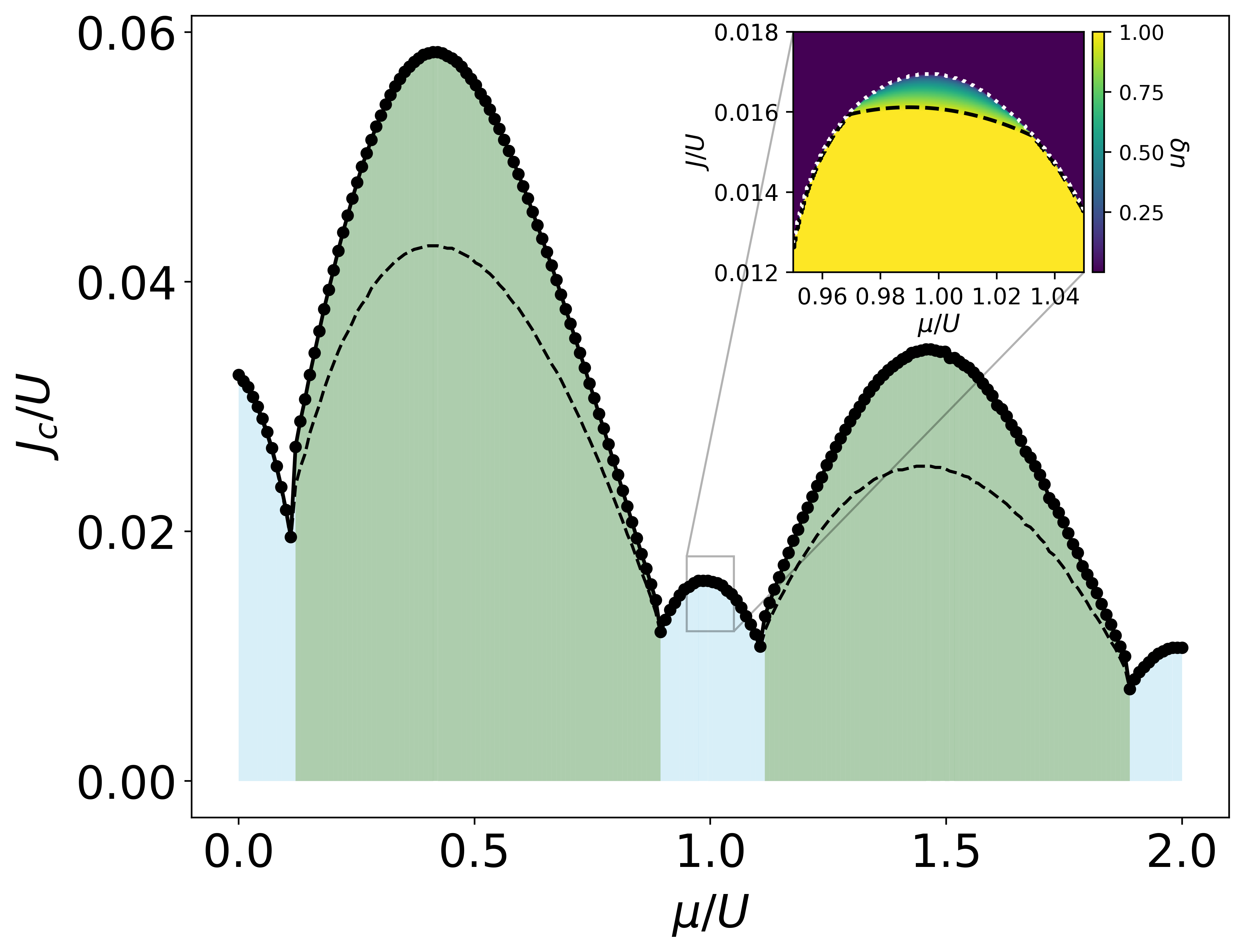}
    \caption{The ground state phase boundary separating the compressible (SF or SS) and incompressible (MI or CDW) for long-range interaction $V/U=0.15$ as a function of chemical potential $\mu/U$ and the critical hopping strength $J/U$. The black dots indicate data obtained using the projection operator method while the dashed lines indicate those from mean-field theory. The small lobes (blue) indicate CDW phases with $n_A-n_B=1$ while the larger lobes (green) corresponds to $n_A=n_B$. The inset shows a close up of the phase diagram obtained using the projection operator method. The black dashed line shows the boundary between the CDW and the SS phase (green region of the inset) where $\delta n$ and $\phi_A$ are both non-zero. The white dotted line shows the border between the SS and SF phases. See text for details. }
    \label{fig4}
\end{figure}

The phase diagram obtained from such computations is shown in Fig.\ \ref{fig4}. The green (blue) regions indicate the MI (CDW) phases while the white region indicates the SF phase. The dots refer to the phase boundary as computed using the projection operator method while the dashed lines indicate the mean-field phase diagram. The inset shows the region near the tip of the CDW lobe showing a small region, marked in blue between the black dashed and the white dotted lines, where the SS phase exists. Here, the black dashed line corresponds to the value of $J/U$ where $\phi_A$ and $\phi_B$ becomes non-zero 
indicating onset of superfluidity while the white dotted line indicates the values of $J/U$ at which $\delta n$ vanishes. Away from the lobe, these two lines merge; this clearly indicates that the bosons, upon lowering $J/U$, undergo a first order transition from the $U(1)$ symmetry broken superfluid ($\phi_A=\phi_B$) to a translational symmetry broken CDW ($\delta n \ne 0$) phase. Our analysis indicates that the SS phase is confined to a narrow regime near the tip of the CDW Mott lobes which is consistent in earlier QMC studies \cite{qmc1}.

\section{Dynamics}
\label{secdyn}

In this section, we shall study non-equilibrium dynamics of the extended Bose-Hubbard model using the projection operator formalism. The details of the formalism for ramp and quench protocol is presented in Sec.\ \ref{dynform} while numerical results for both quench and linear ramp dynamics is presented in Sec.\ \ref{dynres}.

\subsection{Formalism for non-equilibrium dynamics}
\label{dynform} 
In this section, we shall use the projection operator method to study non-equilibrium dynamics of the extended Bose-Hubbard model. In what follows we shall vary the hopping amplitude as a function of time such that its initial and final value both remain small compared to the on-site interaction $U$. This, in turn, guarantees that $J(t)/U \ll 1$ throughout the ramp. 

To begin our analysis, we first note that an exact solution of the non-equilibrium problems requires a solution of the Schrödinger equation 
 \begin{eqnarray}
     i\hbar\partial_t|\psi\rangle = \mathcal{H}[J(t)]|\psi\rangle  \label{sch1}
 \end{eqnarray}
This equation is difficult to solve due to the infinite dimensionality of the boson Hilbert space. However, it is possible to solve this equation without much difficulty for $J(t)/U\ll1$; in this regime, it is possible to work with a finite set of states which are the members of the instantaneous low-energy subspace at any given instant of time. 

To capture the contribution of states belonging to this instantaneous low-energy subspace, we implement a time-dependent SW transformation given by 
\begin{eqnarray} 
|\psi'\rangle=e^{-iS[J(t)]} |\psi\rangle \label{tdsw1} 
\end{eqnarray} 
where $S[J(t)]$ is chosen so as to project out high-energy hopping processes from the instantaneous Hamiltonian $H[J(t)]$. Thus $S[J(t)]$ is given by Eq.\ \ref{eqsw1} with $J \to J(t)$. The Schrodinger equation for $|\psi'(t)\rangle$ can then be easily obtained and is given by 
\begin{eqnarray}
\left(i\hbar\partial_t + \frac{\partial S}{\partial t} \right)|\psi'\rangle &=& H_{\rm eff} [J(t)]|\psi'\rangle \label{sch2}
\end{eqnarray}
We note that the additional term $\frac{\partial S}{\partial t}$ takes into account the possibility of the creation of excitations during the dynamics. The SW transformation has the effect of tracking the dynamics keeping contribution from states which belong to the instantaneous low-energy sector of $H_{\rm eff}[J(t)]$; this provides a faithful description of the dynamics accurate picture for $J(t)/U\ll1$. We also note that this method can treat dynamics with ramp rate $\hbar \dot J/U^2 \ll 1$; for $J/U\ll 1$, this method can therefore treat both slow and fast dynamics at equal footing. 

Next, we choose $|\psi'(t)\rangle$ to have a time-dependent Gutzwiller form given by
\begin{eqnarray} 
|\psi'\rangle = \prod_{{\bf r}\in A} \prod_{{\bf r'}\in B} \sum_{n_A,n_B} f^{n_A}_{\bf r}(t)f^{n_B}_{{\bf r'}}(t) |n_A\rangle \otimes |n_B\rangle  \label{gutztime}
\end{eqnarray} 
We note that this retains instantaneous spatial correlations in $|\psi(t)\rangle$ due to the presence of the $\exp[iS]$ factor. Substituting Eq.\ \ref{gutztime} in Eq.\ \ref{sch2} one gets 
\begin{widetext} 
\begin{eqnarray}
&& i\hbar\frac{\partial}{\partial t}f^{n_A}_{\bf r} +\frac{dJ(t)}{dt} \frac{\delta}{\delta f^{n_A \ast}_{\bf r}} \langle\psi'|\frac{\partial S[J]}{\partial J}|\psi'\rangle = \frac{\delta}{\delta f^{n_A \ast}_{\bf r}}\langle\psi'|H[J(t)]|\psi'\rangle, \nonumber\\ 
&& i\hbar\frac{\partial}{\partial t}f^{n_B}_{\bf r} +\frac{dJ(t)}{dt} \frac{\delta}{\delta f^{n_B \ast}_{\bf r}} \langle\psi'|\frac{\partial S[J]}{\partial J}|\psi'\rangle = \frac{\delta}{\delta f^{n_B \ast}_{\bf r}}\langle\psi'|H[J(t)]|\psi'\rangle \label{feq1} 
\end{eqnarray}
\end{widetext} 

In what follows, we shall systematically express the terms in Eq.\ \ref{feq1} in terms of the Gutzwiller coefficients 
$f_{\bf r}^{A(B)}$. To this end, we begin with the term involving $dJ(t)/dt$. For computing this, we first 
note that $S[J]$ is given by Eq.\ \ref{eqsw1}; defining $S^{(A)[(B)]} ={\delta}/{\delta f^{n_{ A[B]} \ast}_{\bf r}}(\langle\psi'|\partial S/\partial t|\psi'\rangle)$, we find 
\begin{widetext}
\begin{eqnarray}
S^{(A)} &=& iJ \sum_{\langle {\bf r'}\rangle}\Big[\sum_{n_{B}\neq\bar{n}_{B}^1} \frac{1}{\Delta E_{1}} (1-\delta_{n_{A},\bar{n}_{A}^1-1}) f_{\bf r}^{n_{A}+1}f_{{\bf r'}}^{n_B+1 \ast}f_{{\bf r'}}^{n_B} \sqrt{(n_A+1)(n_B +1)} \nonumber\\
&& + \sum_{n_{B} \neq\bar{n}_B^2} \frac{1}{\Delta E_2}(1-\delta_{n_{A},\bar{n}_A^2+1)} f_{\bf r}^{n_A-1}f_{{\bf r'}}^{n_B-1 \ast}f_{{\bf r'}}^{n_B} \sqrt{n_A n_B} \Big]  \nonumber\\
S^{(B)} &=& iJ \sum_{\langle {\bf r'} \rangle}\Big[\sum_{n_A \neq\bar{n}_A^1} \frac{1}{\Delta E_1}(1-\delta_{n_B,\bar{n}_B^1+1}) f_{{\bf r}}^{n_A-1 \ast}f_{{\bf r}}^{n_A}f_{{\bf r}'}^{n_B - 1} \sqrt{n_A n_B} \nonumber\\
&& + \sum_{n_A
    \neq\bar{n}_A^2} \frac{1}{\Delta E_2}(1-\delta_{n_B,\bar{n}_B^2-1)} f_{{\bf r}'}^{n_B+1}f_{{\bf r}}^{n_A+1 \ast}f_{{\bf r}}^{n_A} \sqrt{(n_A + 1)(n_B+1)}\Big] \label{s1s2eq}
\end{eqnarray}
\end{widetext}
where $\langle {\bf r'}\rangle$ indicates that $\bf r'$ is one of the nearest-neighbors of ${\bf r}$ and ${\bf r} \in A(B)$ for $S^{(A)[(B)]}$. Eq.\ \ref{s1s2eq} allows one to express the left-side of Eq.\ \ref{feq1} in terms of $f_{\bf r}^{n_A(n_B)}$.

Next we consider the terms in the right side of Eq.\ \ref{feq1}. To this end, we write 
\begin{eqnarray} 
{\delta}/{\delta f^{n_{A(B)} \ast}_{\bf r}}(\langle\psi^,|H_\mu[J(t)]|\psi^,\rangle)= \sum_{\mu}P^{A(B)}_\mu
\end{eqnarray}  
which can be obtained from ${\delta E_\mu}/{\delta f^{n_{A(B)}\, \ast}_{\bf r}}$ where $\mu=0,a,b,c$ and $E_c= E_{c1} + E_{c2}$. The expressions of $E_{\mu}$ for $\mu=0,a,b,c$ are given in Eqs.\ \ref{eqen0}, \ref{eqena}, \ref{eqenb}, \ref{eqenc1}, and \ref{eqenc2}. 

The first of these terms, corresponds to $\delta E_{0}/\delta f^{n_{A(B)} \, \ast}_{\bf r}$ and are given by 
\begin{eqnarray}
P^{(A)}_{0} &=& \sum_{\bf r \in A} f_{\bf r}^{n_A}\epsilon_1(n_A), \,\, 
P^{(B)}_{0} = \sum_{\bf r \in B} f_{\bf r}^{n_B}\epsilon_2(n_B)  
\end{eqnarray} 
where $\epsilon_1$ and $\epsilon_2$ can be read off from Eq.\ \ref{eqen0}.

The next set of terms are linear in $J$ and originates from ${\delta E_{a}/\delta f^{n_A \ast(n_B \ast)}_{\bf r}}$. These are given by 
\begin{eqnarray}
P^{(A)}_{a} &=& -J\sum_{\langle r'\rangle} \Big[f_{\bf r}^{\bar{n}_A^1}f_{{\bf r}'}^{\bar{n}_B^1+1 \ast}f_{{\bf r}'}^{\bar{n}_B^1}\sqrt{\bar{n}_A^1(\bar{n}_B^1 + 1)} \delta_{n_A,\bar{n}_A^1-1} \nonumber\\
&& + f_{\bf r}^{\bar{n}_A^2}f_{{\bf r}'}^{\bar{n}_B^2-1 \ast}f_{{\bf r}'}^{\bar{n}_B^2}\sqrt{\bar{n}_B^2(\bar{n}_A^2 + 1)} \delta_{n_A,\bar{n}_A^2+1}\Big] 
\nonumber\\
P^{(B)}_{a} &=& -J\sum_{\langle r'\rangle} \Big[f_{\bf r}^{\bar{n}_A^1 - 1 \ast}f_{\bf r}^{\bar{n}_A^1}f_{{\bf r}'}^{\bar{n}_B^1}\sqrt{\bar{n}_A^1(\bar{n}_B^1 + 1)} \delta_{n_B,\bar{n}_B^1+1} \nonumber\\
&& + f_{\bf r}^{\bar{n}_A^2 + 1 \ast}f_{\bf r}^{\bar{n}_A^2}f_{{\bf r}'}^{\bar{n}_B^2}\sqrt{\bar{n}_B^2(\bar{n}_A^2 + 1)} \delta_{n_B,\bar{n}_B^2-1}\Big]
\end{eqnarray}

Next, we consider the contribution from terms ${\rm O}(J^2)$ that involve 
hopping within a single link. These are given by 
\begin{eqnarray} 
P^{(A)[(B)]}_{b} &=&  \delta E_{b}/\delta f^{n_A\ast[n_B\ast]}_{\bf r},\nonumber\\
P^{(A)[(B)]}_{c} &=& \delta E_{c1}/\delta f^{n_A\ast[n_B\ast]}_{\bf r}, \label{pbpcdef}
\end{eqnarray} 
and yield two terms each for a given $\alpha$ and $\beta$. To express these, we define
\begin{eqnarray} 
Q_{\alpha\beta}(n_A,n_B,n_A',n_B') &=& {{\delta }}/{{\delta f^{n_A \ast}_{\bf r}}}(S_{\alpha\beta}(n_A,n_B,n_A',n_B')) \nonumber\\
Y_{\alpha\beta}(n_A,n_B,n_A',n_B') &=& {{\delta }}/{{\delta f^{n_B \ast}_{\bf r}}}(S_{\alpha\beta}(n_A,n_B,n_A',n_B')). \nonumber\\ \label{qydef}
\end{eqnarray} 
where the expression of $S_{\alpha \beta}$ is given in Appendix\ \ref{appa}. The expressions of these terms are given in Appendix\ref{appb}; in terms of these one finds the contribution from the term $P^{(A)[(B)]}_{b}$ to be 
\begin{widetext}
\begin{eqnarray}
 P^{(A)}_{b} &=& \sum_{\langle \bf {r,r'}\rangle}\sum_{\alpha,\beta=1,2}\sum_{n_A,n_B,n_A',n_B'}\frac{J^2(t)}{\Delta E_{\alpha}}\Big[Q_{\alpha\beta}(n_A,n_B;n_A',n_B') - Q_{\beta\alpha}(n_A',n_B';n_A,n_B)\Big]
 (1-\delta_{n_A',\bar{n}_A^\alpha}\delta_{n_B',\bar{n}_B^\alpha})\delta_{n_A,\bar{n}_A^\beta} \delta_{n_B,\bar{n}_B^\beta}  
\nonumber\\
P^{(B)}_{b} &=& \sum_{\langle \bf {r,r'}\rangle}\sum_{\alpha,\beta=1,2}\sum_{n_A,n_B,n_A',n_B'}\frac{J^2(t)}{\Delta E_{\alpha}}\Big[Y_{\alpha\beta}(n_A,n_B;n_A',n_B') - Y_{\beta\alpha}(n_A',n_B';n_A,n_B)\Big] (1-\delta_{n_A',\bar{n}_A^\alpha}\delta_{n_B',\bar{n}_B^\alpha})\delta_{n_A,\bar{n}_A^\beta} \delta_{n_B,\bar{n}_B^\beta} \nonumber\\  \label{dynqdef}
\end{eqnarray}   
\end{widetext}
The terms in $P^{(A)[(B)]}_{c}$ can be divided in two parts. The first of these involves a single link and are denoted as  $P^{(A)[(B)]}_{c1}$. It receives contribution from single link parts of $E^{(1)}_c$ and $E^{(2)}_c$ respectively and is given by 
\begin{widetext}
\begin{eqnarray}
P_{c1}^{(A)} &=& \sum_{\langle \bf {r,r'}\rangle}\sum_{\alpha,\beta=1,2}\sum_{n_A,n_B,n_A',n_B'} \frac{J^2(t)}{2\Delta E_{\alpha}} \Big[Q_{\alpha\beta}(n_A,n_B;n_A',n_B') - Q_{\beta\alpha}(n_A',n_B';n_A,n_B)\Big]  \nonumber\\ 
 && \times  (1-\delta_{n_A',\bar{n}_A^\alpha}\delta_{n_B',\bar{n}_B^\alpha}) (1-\delta_{n_A,\bar{n}_A^\beta} \delta_{n_B,\bar{n}_B^\beta}) \nonumber\\
 P_{c1}^{(B)} &=& \sum_{\langle \bf {r,r'}\rangle}\sum_{\alpha,\beta=1,2}\sum_{n_A,n_B,n_A',n_B'} \frac{J^2(t)}{2\Delta E_{\alpha}} \Big[Y_{\alpha\beta}(n_A,n_B;n_A',n_B') - Y_{\beta\alpha}(n_A',n_B';n_A,n_B)\Big] \nonumber\\
 && \times (1-\delta_{n_A',\bar{n}_A^\alpha}\delta_{n_B',\bar{n}_B^\alpha})(1-\delta_{n_A,\bar{n}_A^\beta} \delta_{n_B,\bar{n}_B^\beta})   \label{dynydef}
\end{eqnarray}
\end{widetext}

Finally, we compute the contribution from hopping involving adjacent links. These terms are given by  
\begin{eqnarray} 
P^{(A)[(B)]}_{c2} &=& {\delta E_{c1}}/{\delta f^{n_A \ast[n_B \ast]}_{\bf r}} \nonumber\\
P^{(A)[(B)]}_{c3} &=& {\delta E_{c2}}/{\delta f^{n_A \ast[n_B \ast]}_{\bf r}} \label{pc23eq}
\end{eqnarray}
where only the terms in $E_2$ and $E_3$ that depends on adjacent links have been considered. Each of these yields two terms each for a given $\alpha$ and $\beta$. To express this in a concise manner, we define  
\begin{eqnarray} 
L^{(a)}_{\alpha\beta}(n_A,n_B,n_A',n_B') &=& \delta R^{(a)}_{\alpha\beta}(n_A,n_B,n_A',n_B'))/\delta f^{n_A \ast}_{\bf r} \nonumber\\
Z^{(a)}_{\alpha\beta}(n_A,n_B,n_A',n_B') &=& \delta R^{(a)}_{\alpha\beta}(n_A,n_B,n_A',n_B'))/\delta f^{n_B \ast}_{\bf r} \nonumber\\ \label{lzeq} 
\end{eqnarray} 
where the expressions of $R_{\alpha \beta}^{(a)}$ are given in Appendix \ref{appa}. In terms of these one finds ( see Appendix\ \ref{appb}) 
\begin{widetext}
\begin{eqnarray}
P^{(A)}_{c2} &=& \sum_{a=1,2,3}\sum_{\langle \langle  \bf {r}\rangle\rangle}\sum_{\alpha,\beta=1,2}\sum_{n_A,n_B,n_A',n_B'}\frac{J^2(t)}{\Delta E_{\alpha}}\Big[L^{(a)}_{\alpha\beta}(n_A,n_B;n_A',n_B') - L^{(a)}_{\beta\alpha}(n_A',n_B';n_A,n_B)\Big]
\\
 && \times(1-\delta_{n_A,\bar{n}_A^\alpha}\delta_{n_B,\bar{n}_B^\alpha})\delta_{n_A',\bar{n}_A^\beta} \delta_{n_B',\bar{n}_B^\beta}
\nonumber\\
P^{(B)}_{c2} &=& \sum_{a=1,2,3}\sum_{\langle \langle \bf {r}\rangle\rangle}\sum_{\alpha,\beta=1,2}\sum_{n_A,n_B;n_A',n_B'}\frac{J^2(t)}{\Delta E_{\alpha}}\Big[Z^{(a)}_{\alpha\beta}(n_A,n_B;n_A',n_B') - Z^{(a)}_{\beta\alpha}(n_A',n_B';n_A,n_B)\Big] \\
&& \times (1-\delta_{n_A,\bar{n}_A^\alpha}\delta_{n_B,\bar{n}_B^\alpha}) \delta_{n_A',\bar{n}_A^\beta} \delta_{n_B',\bar{n}_B^\beta} \nonumber \label{dynldef}
\end{eqnarray}
\end{widetext}

\begin{widetext}
\begin{eqnarray}
P^{(A)}_{c3} &=& \sum_{a=1,2,3}\sum_{\langle \langle \bf {r}\rangle\rangle}\sum_{\alpha,\beta=1,2}\sum_{n_e,n_o,n_e',n_o'}\frac{J^2(t)}{\Delta E_{\alpha}}\Big[L^{(a)}_{\alpha\beta}(n_A,n_B;n_A',n_B') - L^{(a)}_{\beta\alpha}(n_A',n_B';n_A,n_B)\Big]
\\
 && \times (1-\delta_{n_A,\bar{n}_A^\alpha}\delta_{n_B,\bar{n}_B^\alpha})(1-\delta_{n_A',\bar{n}_A^\beta} \delta_{n_B',\bar{n}_B^\beta})
\nonumber\\
P^{(B)}_{c3} &=& \sum_{a=1,2,3}\sum_{\langle \langle \bf {r}\rangle\rangle}\sum_{\alpha,\beta=1,2}\sum_{n_e,n_o,n_e',n_o'}\frac{J^2(t)}{\Delta E_{\alpha}}\Big[Z^{(a)}_{\alpha\beta}(n_A,n_B;n_A',n_B') - Z^{(a)}_{\beta\alpha}(n_A',n_B';n_A,n_B)\Big] \\ 
&& \times (1-\delta_{n_A,\bar{n}_A^\alpha}\delta_{n_B,\bar{n}_B^\alpha})(1-\delta_{n_A',\bar{n}_A^\beta} \delta_{n_B',\bar{n}_B^\beta}) \nonumber \label{dynzdef}
\end{eqnarray}
\end{widetext}

Thus the equation governing the dynamics of the bosons can be written as 
\begin{eqnarray}
    i\hbar\frac{\partial}{\partial t}f^{n_{A(B)}}_{\bf r} + S^{(A)[(B)]} &=& P_0^{(A)[(B)])} +\sum_{\alpha=a,b}  P_{\alpha}^{(A)[(B)])} \nonumber\\
    && + \sum_{j=1,2,3}  P_{cj}^{(A)[(B)])}\label{feq1b} 
\end{eqnarray}
In what follows, we obtain numerical solutions to Eq. \ref{feq1b} which allows us to address the dynamics of the bosons for both sudden quench and non-linear ramps.

\subsection{Sudden quench and ramp dynamics}
\label{dynres} 

In this section, we first address the dynamics of bosons after a sudden quench of the hopping amplitude from an initial value $J_i$ to a final value $J_f$. In what follows, we focus on quenches from the CDW insulating phase; the corresponding results for for quenches from the homogeneous MI to the SF phases have been studied earlier \cite{PhysRevLett.106.095702,PhysRevB.86.085140} and discussed in Appendix\ \ref{appd} for completeness. We consider the dynamics of the superfluid order parameter amplitude and density imbalance following the quench; these are given by 
\begin{eqnarray} 
|\Delta_{A(B)}|(t) &=& |\langle\psi(t)|b_{A(B)}|\psi(t)\rangle| \nonumber\\
\delta n (t) &=& \langle\psi(t)|(\hat n_A -\hat n_B)|\psi(t)\rangle. \label{oddef}
\end{eqnarray}

In what follows, we shall first study two separate cases of quench dynamics where the quench starts from the CDW phase with $J_i<J_c$ and proceeds through the tip of the CDW lobe. The first case involves $J_f \simeq J_c$  so that we end up in the supersolid phase through the tip of the CDW lobe. In addition, we shall also study sudden quenches that ends deep inside the superfluid phase $(J_f \simeq J_c)$. Apart from the order parameter dynamics, we shall also study the fidelity and residual energies of the state after such quenches, given by
\begin{eqnarray} 
{\mathcal F} (t) &=& \ln |\langle\psi_0|\psi(t)\rangle| \nonumber\\
Q(t) &=& \langle \psi(t) |\mathcal{H}[J_f]|\psi(t)\rangle - E_G[J_{f}]
\label{fidresdef}
\end{eqnarray} 
where $|\psi_0\rangle$ is the initial state (chosen to be ground-state of the system at $J=J_i$) and $E_G[J_f]$ is the ground state energy at $J= J_f$. We note that $Q(t)$ vanishes for adiabatic dynamics where no excitations are created. 

\begin{widetext}
\begin{figure*}
\rotatebox{0}{\includegraphics[width=0.195\linewidth]{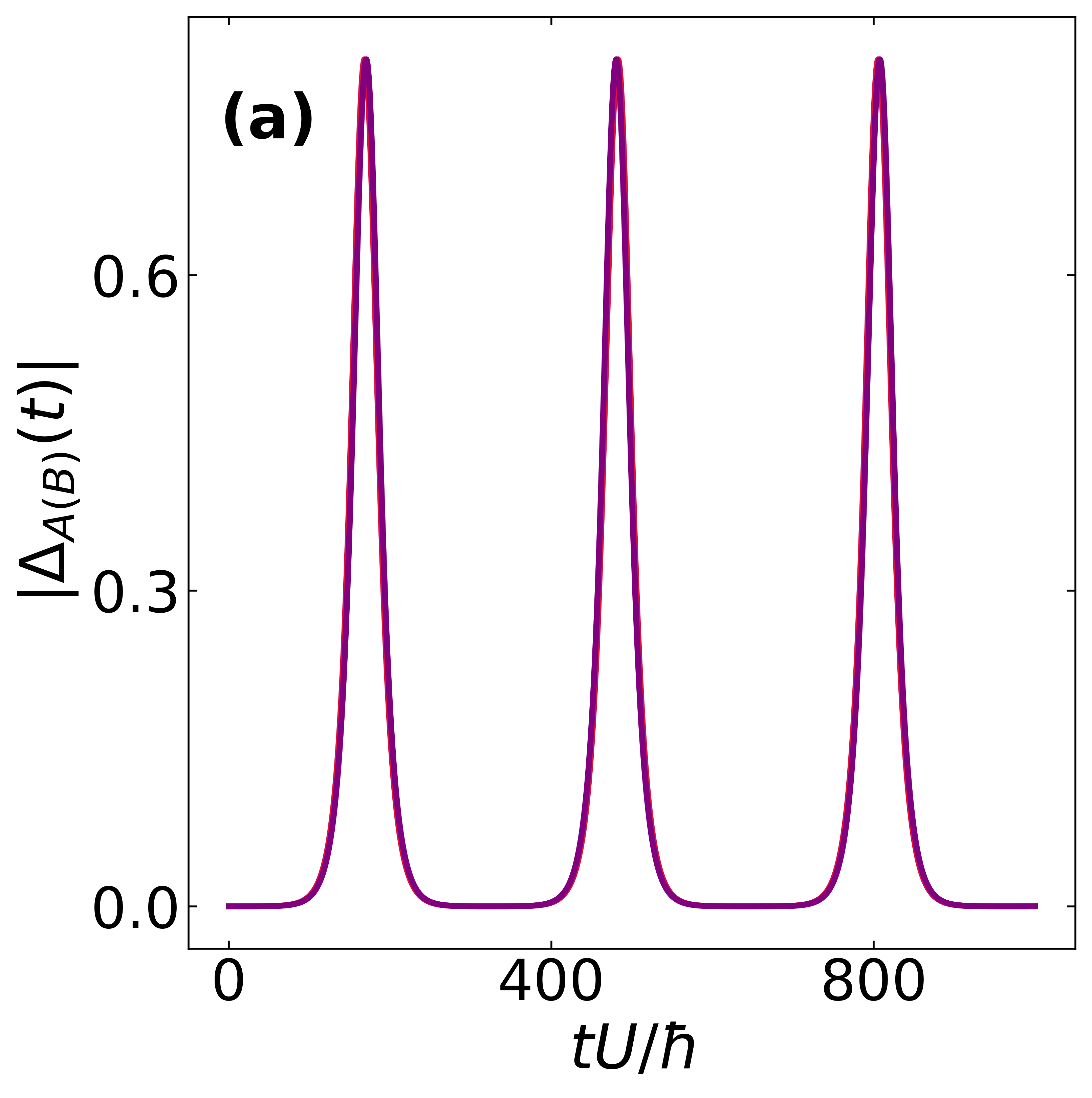}}
\rotatebox{0}{\includegraphics[width=0.195\linewidth]{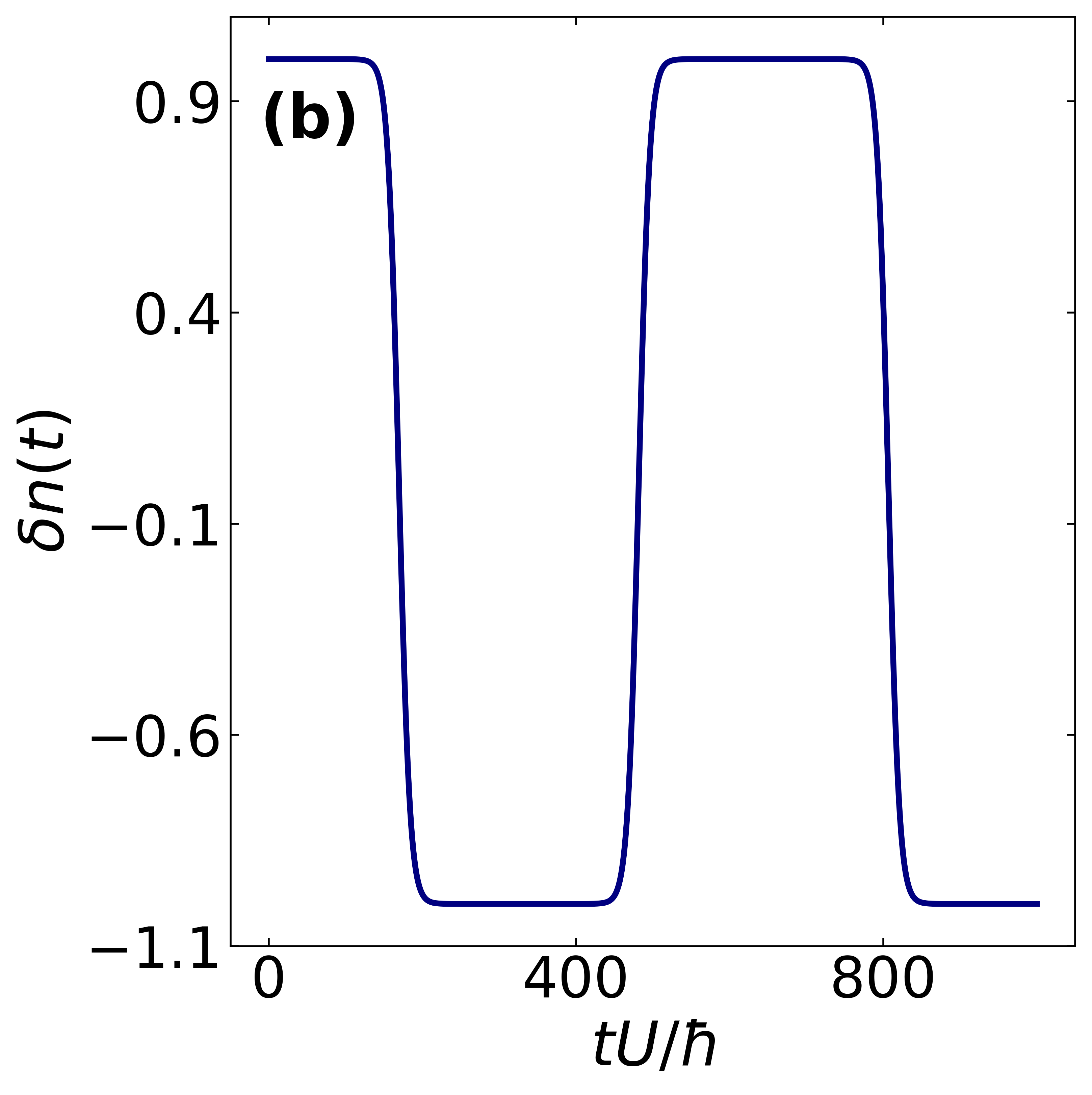}}
\rotatebox{0}{\includegraphics[width=0.195\linewidth]{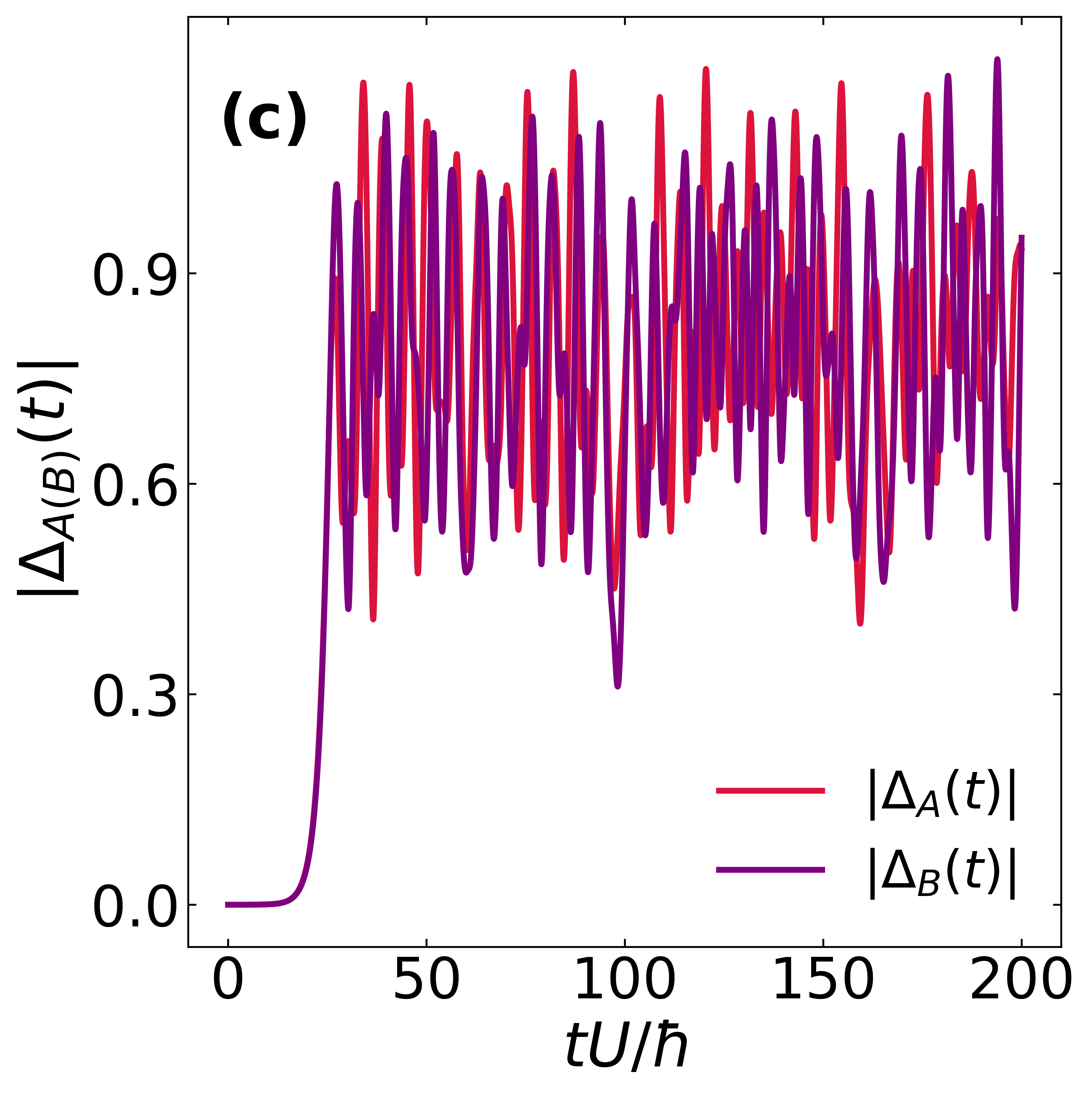}}
\rotatebox{0}{\includegraphics[width=0.195\linewidth]{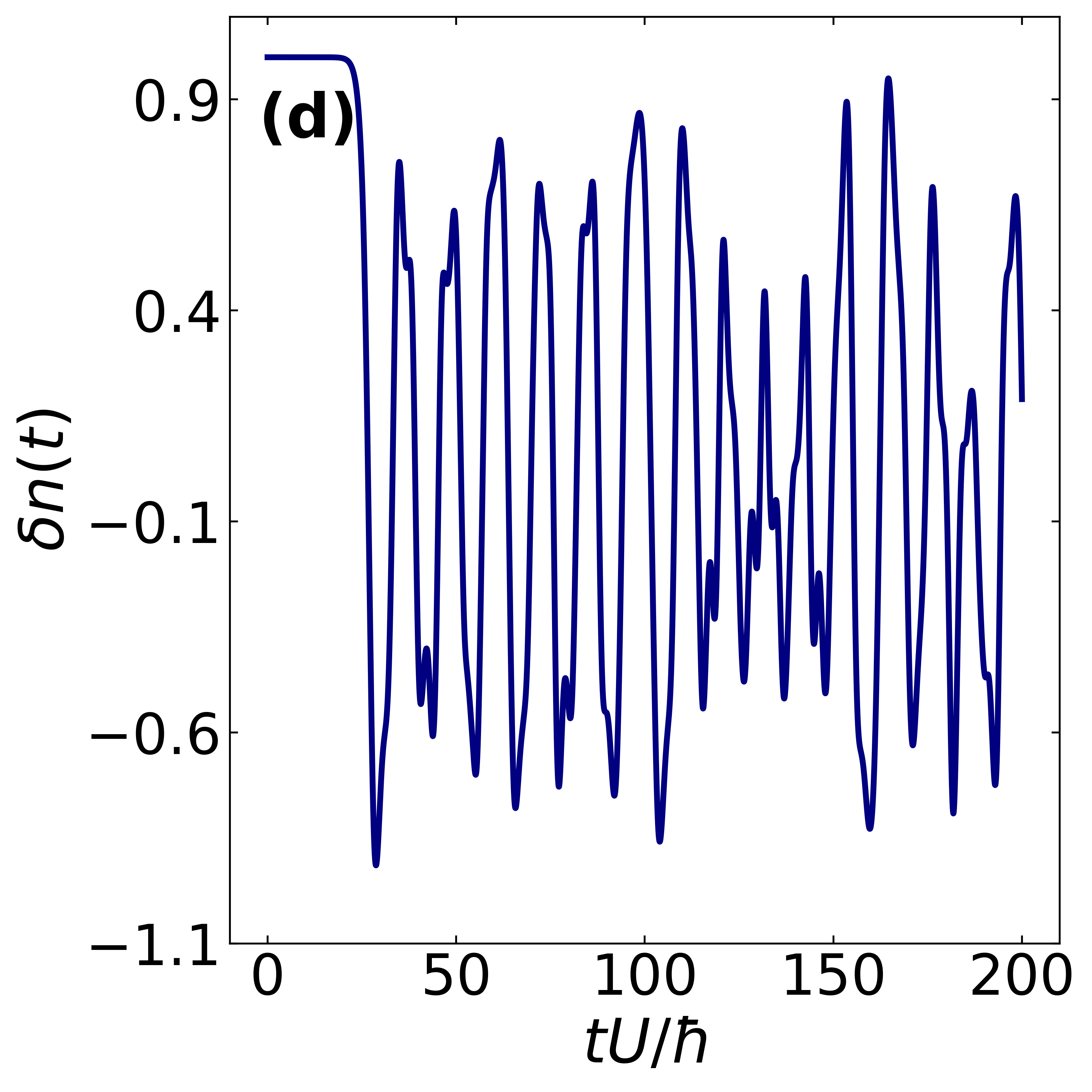}}
\rotatebox{0}{\includegraphics[width=0.195\linewidth]{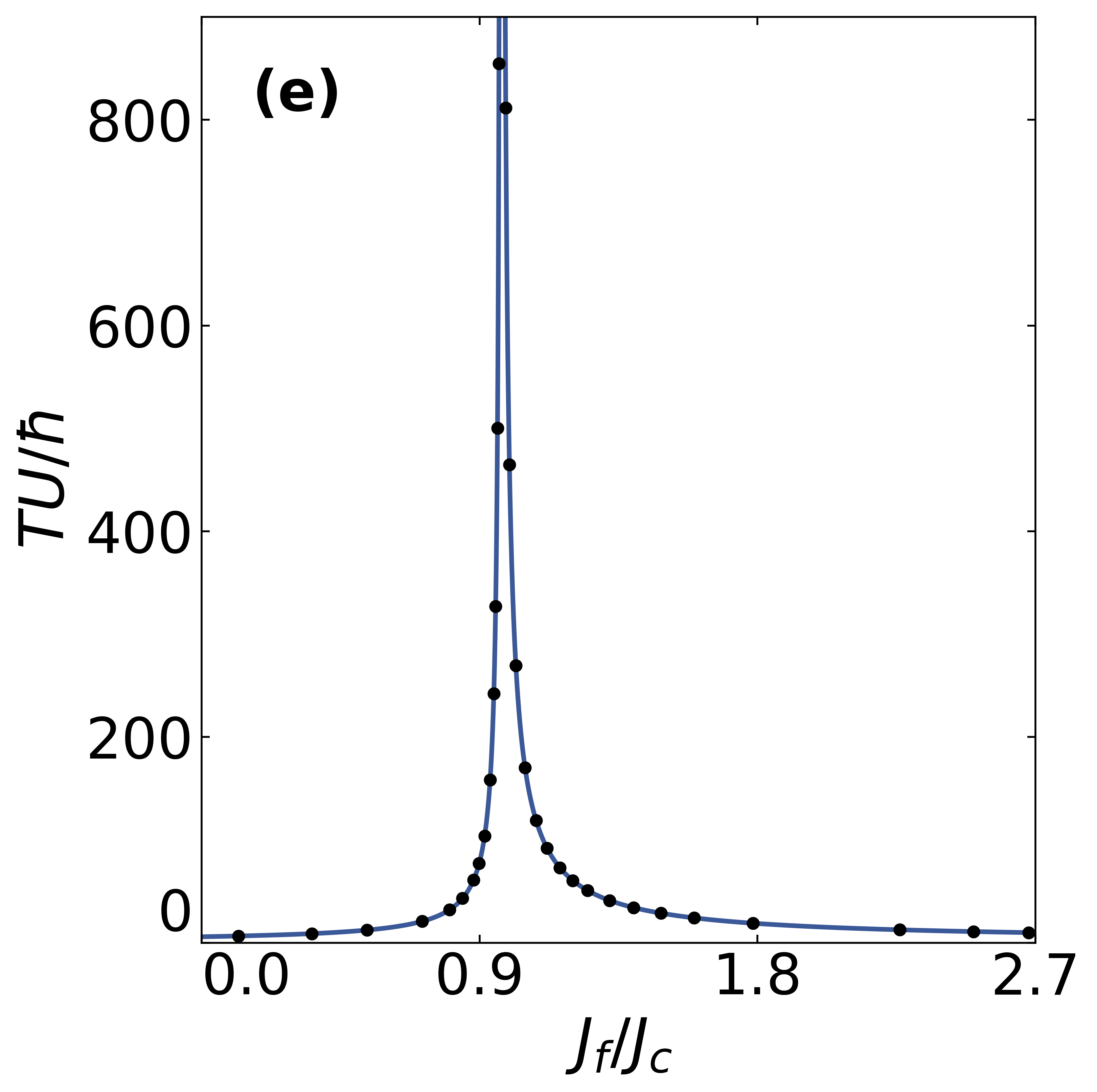}}
    \caption{(a) Plots of the time evolution of the order parameters $|\Delta_{A(B)}(t)|$ as a function of $tU/\hbar$ following a quench to the tip of the CDW lobe with $J_f=1.02 J_c$, $J_i=0.01 J_c$, and $\mu/U=0.98$. (b) Plot of $\delta n(t)$ for the same quench as in (a). (c) Same as in (a) but with $J_f=3.51 J_c$ so that the quench takes the system deep inside the SF phase. (d) Plot of $\delta n(t)$ with all parameters being same as in (c).(e) Plot of the time period of oscillation $TU/\hbar$ as a function of $J_f/J_c$ showing enhancement around $J_c$.}
    \label{fig5}
\end{figure*} 
\end{widetext}

We begin by noting that for a sudden quench, $\partial J / \partial t \sim \delta(t)$, and thus the second term in the  left side of Eq.\ \ref{sch2} does not contribute to the subsequent time evolution of the system for $t>0$. The time evolution of $\Delta_{A(B)}$ and $\delta n$ can be written in terms of the Gutzwiller amplitudes $\{f^{n_A}_{\bf r}(t)f^{n_B}_{r^,}(t)\}$ noting that 
\begin{eqnarray} 
\Delta_{A(B)}(t)=\langle\psi'(t)|e^{iS[J_f]} \hat b_{A(B)} e^{-iS[J_f]}|\psi'(t)\rangle, \nonumber\\
\delta n(t)=\langle\psi'(t)|e^{iS[J_f]} (\hat n_A -\hat n_B) e^{-iS[J_f]}|\psi'(t)\rangle
\label{odexp1} 
\end{eqnarray} 
Using Eq.\ \ref{odexp1}, one can express $\Delta_{A(B)}(t)$ and $\delta n(t)$ in terms of $f_{\bf r}^{n_{A}}(t)$ and $f_{\bf r}^{n_{B}}(t)$. 
We note that for small $J/U$, the exponential factors in Eq.\ \ref{odexp1} can be expanded and the first terms in this expansion represent the mean-field results. The presence of the other terms contributes as quantum fluctuations around this mean-field theory. 

Our results, shown in Fig.\ \ref{fig5}, indicate that quenches near critical boundary
between the CDW and SS phases lead to single-frequency oscillations of both $\delta n$ and $|\Delta|$ (Figs.\ \ref{fig5}(a) and (b)). In this regime, the boson system has large spatial correlation length and weak mixing of the phase and the amplitude collective modes as shown in Ref.\ \onlinecite{PhysRevLett.89.250404}. The contribution to the dynamics here occurs mainly from the amplitude modes leading to single-frequency oscillations. In this regime, the interaction between the phase and the amplitude modes, which will eventually damp out these oscillations in $d=2$, occurs via long-wavelength ($k\simeq 0$) quantum fluctuations. The effect of these fluctuations are not captured by the projection operator formalism. However, we note that the typical timescale for such fluctuations is $T_0 \sim L\hbar/J_c$ and till this timescale, the dynamics is expected to be accurately described by the formalism used here. In contrast, deeper quenches to the SF phase, shown in Fig.\ \ref{fig5}(c) and (d), lead to strong nonlinear coupling among density-wave excitations, amplitude modes, and gapless phase modes resulting in multi-frequency dynamics of both $\Delta_{A(B)}(t)$ and $\delta n$. The time period of these oscillations, $TU/\hbar$, diverges around the critical point reflecting the critical slowing down of dynamics as shown in Fig.\ \ref{fig5}(e). Similar features are seen for quenches from the homogeneous MI to the SF phase as discussed in Appendix\ \ref{appd}.

For $J_f=1.02 J_c$, we note that the oscillations of density (Fig.\ \ref{fig5}(b)) occur between states with $\delta n =1$ to $\delta n \simeq -1$ which corresponds to the values of $\delta n$ for the two CDW ground states. The time period of this oscillation corresponds to $t^{\ast} \simeq 400 \hbar/U$. Concomitantly, as seen from Fig.\ \ref{fig5}(a), the instantaneous SF order parameter amplitudes vanish at these two end points and peaks at an intermediate time $\sim t^{\ast}/2$. Such simple behavior is absent in Fig.\ \ref{fig5}(c) and (d) where multi-frequency dynamics is observed for $J_f=3.51 J_c$. 


\begin{figure}
\rotatebox{0}{\includegraphics[width=0.49\linewidth]{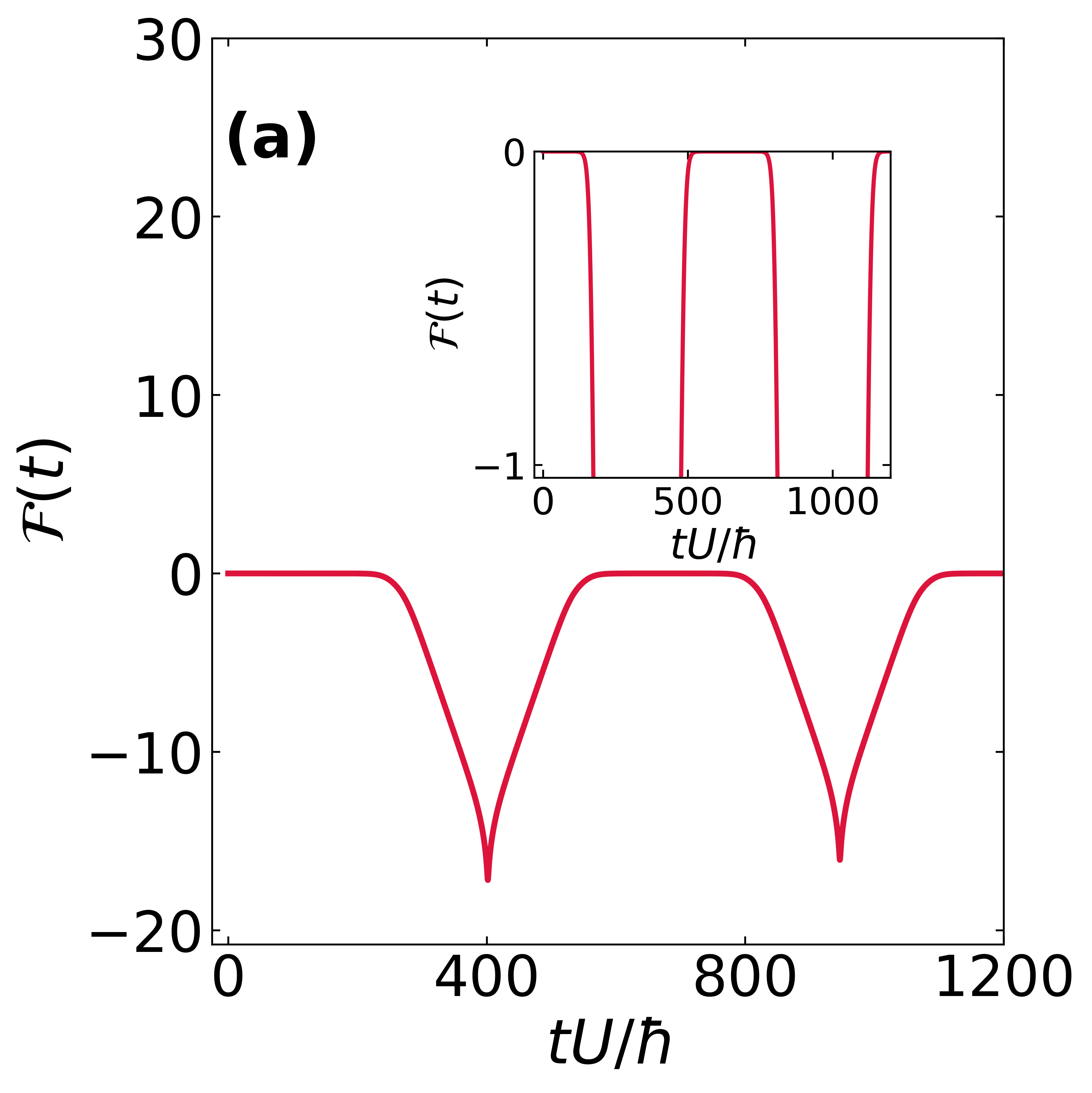}}
\rotatebox{0}{\includegraphics[width=0.49\linewidth]{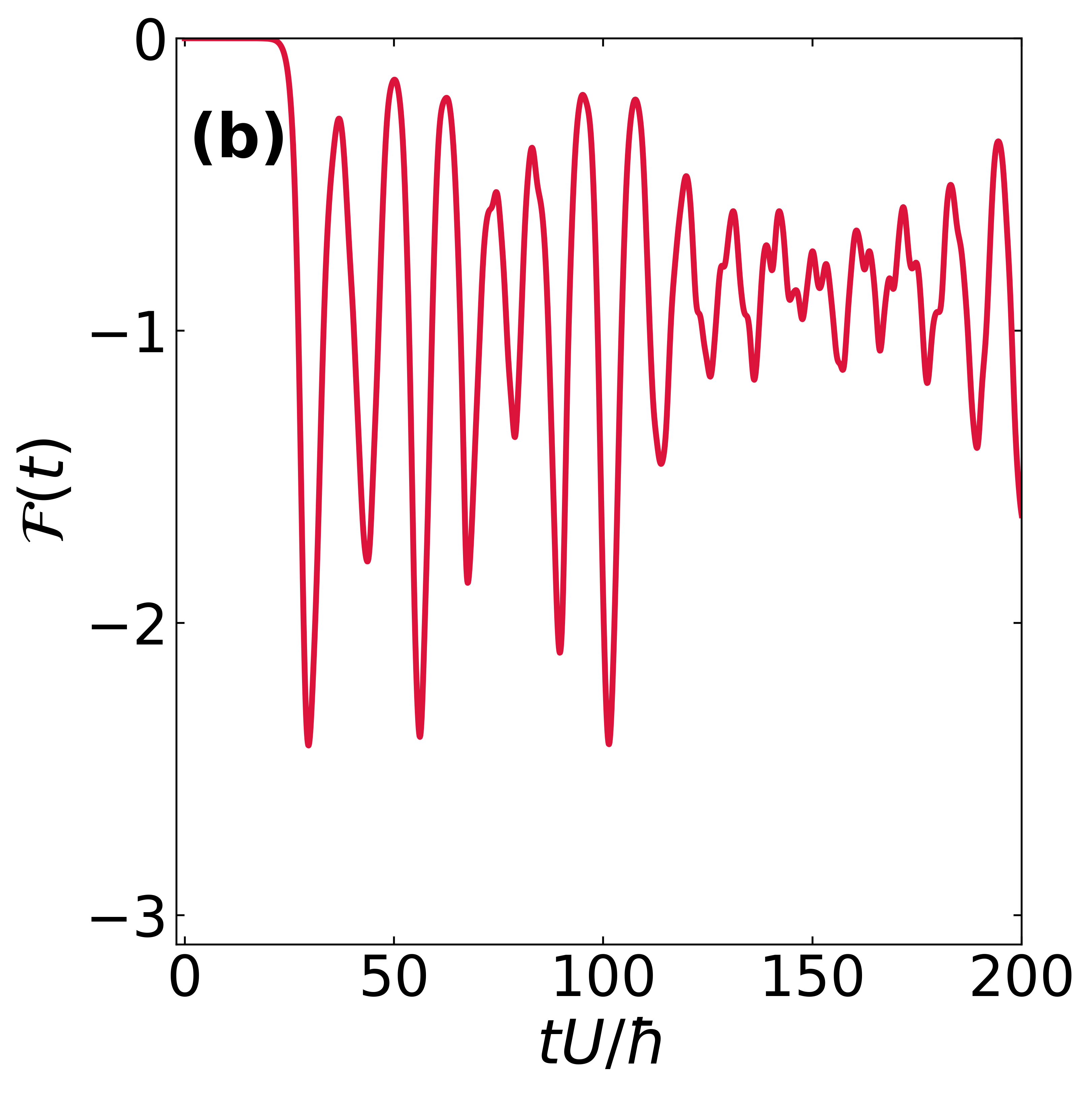}}
\rotatebox{0}{\includegraphics[width=0.49\linewidth]{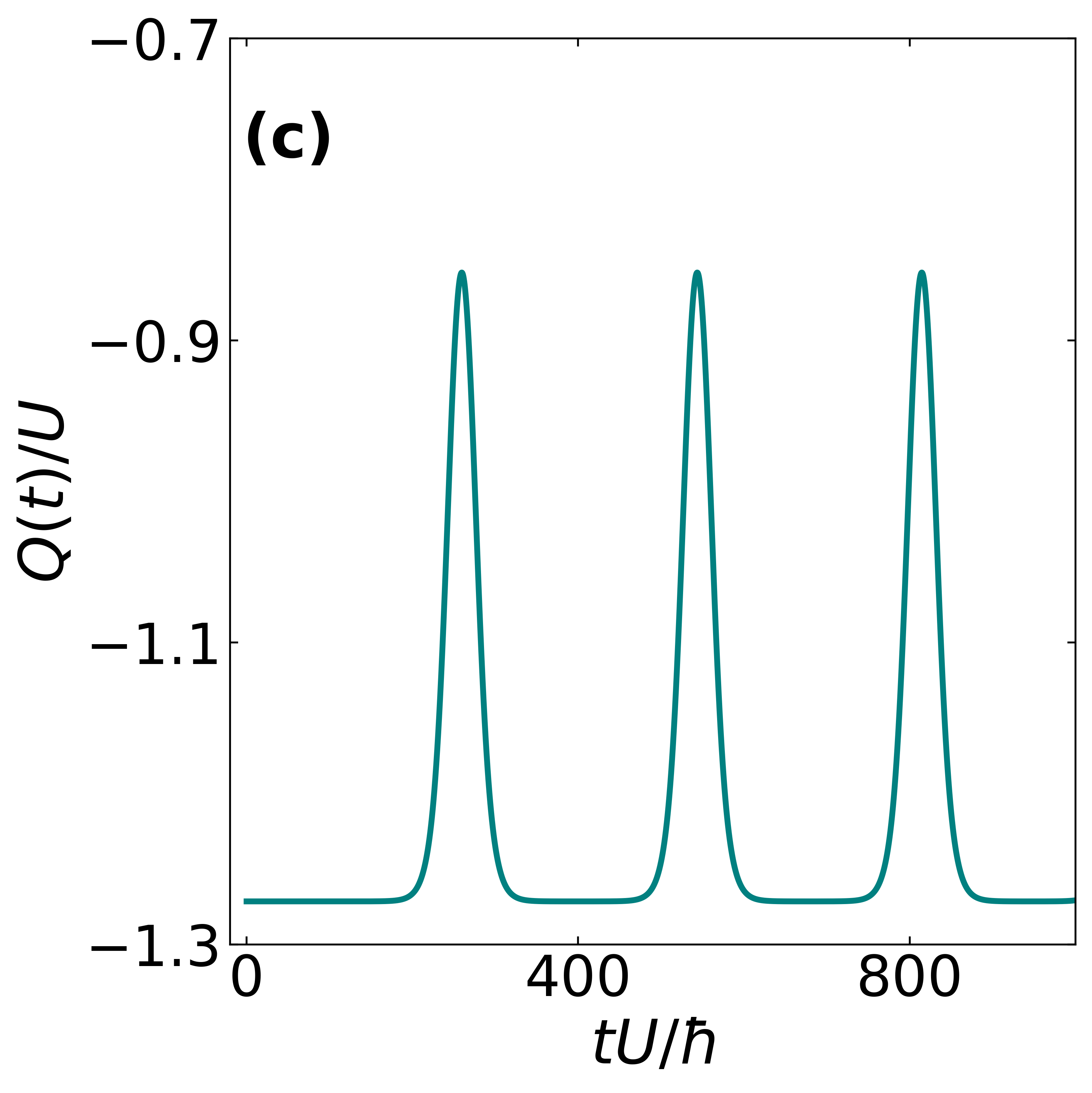}}
\rotatebox{0}{\includegraphics[width=0.49\linewidth]{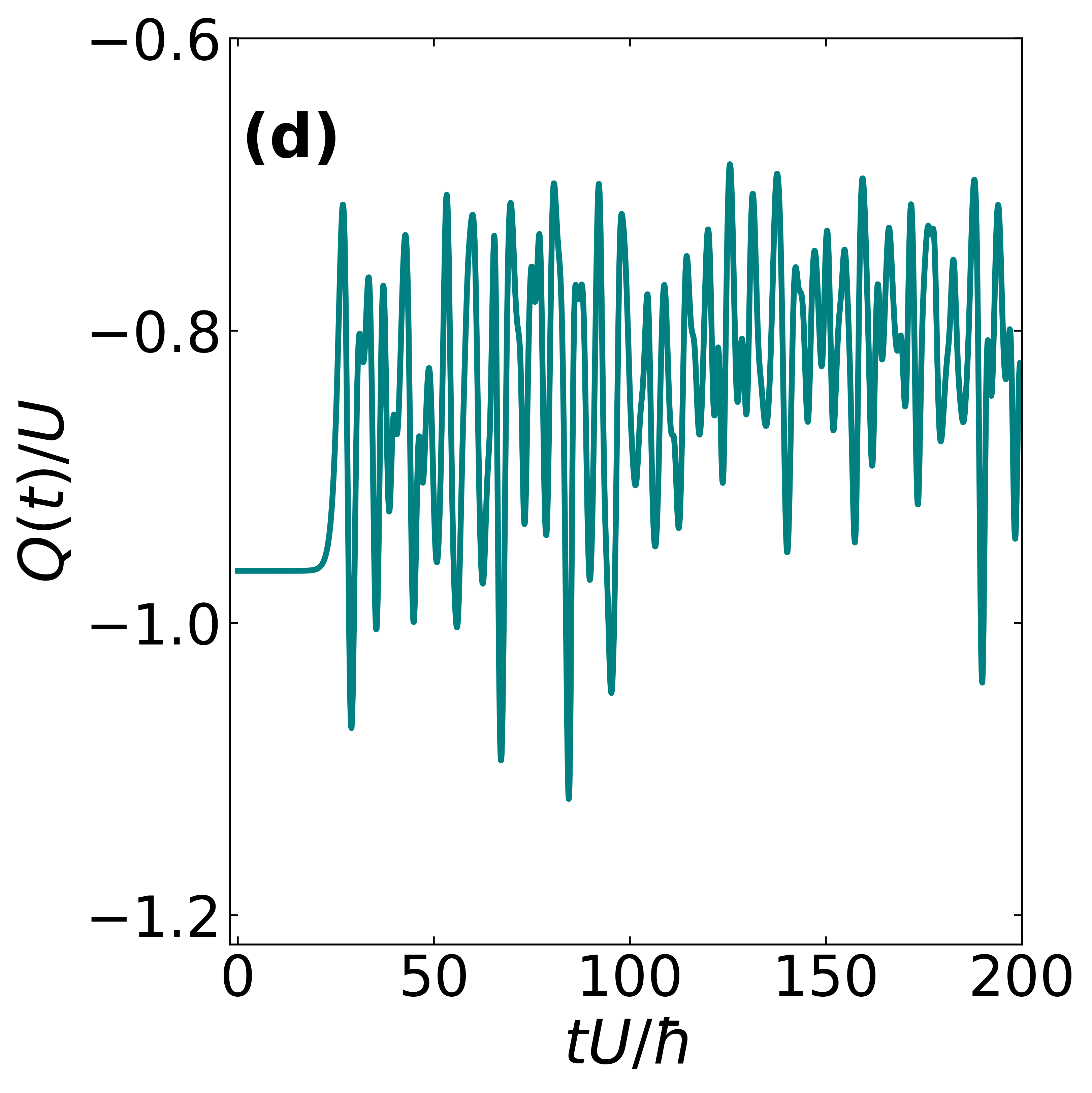}}
    \caption{Plots of the time evolution of ${\mathcal F}(t)$ as a function of $tU/\hbar$. Here $J_i= 0.01 J_c$ for all plots with $\mu=0.98 U$ so that the system is at the tip of the CDW lobe. The left panel (a) correspond to $J_f=1.02 J_c$ while the right panel (b)corresponds to $J_f=3.51 J_c$.  (c) and (d) Plot of the time evolution of $Q(t)$ for same parameters as (a) and (b) respectively. All other parameters are same as in Fig.\ \ref{fig5}. See text for details. }
    \label{fig6}
\end{figure}

Next, we compute the fidelity ${\mathcal F}(t)$ and $Q(t)$ as a function of $t$ after a quench from the ground state of the boson system at $J_i=0.1 J_c$ at the tip of the CDW lobe which corresponds to $\mu=0.98 U$. The results are plotted in Fig.\ \ref{fig6}(a) and (b) for $J_f =1.02 J_c$ and $3.51 J_c$ respectively. We find that ${\mathcal F}(t)$ shows periodic behavior with approximately same periodicity as $\delta n$ when $J_f=1.02 J_c$; in contrast, it exhibits dynamics involving multiple frequencies for $J_f=3.51 J_c$. A similar behavior is found for $Q(t)$, as shown in the bottom panels of Fig.\ \ref{fig6}. 

The dips of ${\mathcal F}(t)$ where ${\mathcal F} \sim 0$, shown in Fig.\ \ref{fig6}(a) for $J_f=1.02 J_c$, can be understood as follows. A comparison of Figs.\ \ref{fig6}(a) and \ref{fig5}(b) shows that these dips occur precisely at $t= n t^{\ast}$ where $n \in Z$; at these points, the wavefunction corresponds $\delta n=-1$. Thus the instantaneous state corresponds to a CDW state with $\langle n_B-n_A \rangle =-1$; consequently, it has almost zero overlap with the initial CDW state for which $\langle n_B-n_A \rangle =1$. This results in a sharp dip of ${\mathcal F}$. We note from Fig.\ \ref{fig6}(b) that $Q$ remains close to its initial value at $t=nt^{\ast}$; this is a consequence of the fact that two CDW states with $\delta n=\pm 1$ have identical instantaneous energies. The peak of $Q(t)$ occurs approximately at $t=t_n^{\ast}/2$. At these times, the wavefunction supports a large instantaneous value of $\Delta(t)$ (Fig.\ \ref{fig5}(a)); consequently, they have a larger instantaneous energy than the initial state. The analogous behavior of ${\mathcal F}(t)$ and $Q(t)$ for $J_f =3.51 J_c$ shown in Fig.\ \ref{fig6}(b) and (d) is more complicated; here $F$ never reaches a value close to zero; also $Q(t)$ shows oscillatory behavior with multiple frequencies similar to $|\Delta|$ and $\delta n$.   
\begin{figure}
\rotatebox{0}{\includegraphics[width=0.49\linewidth]{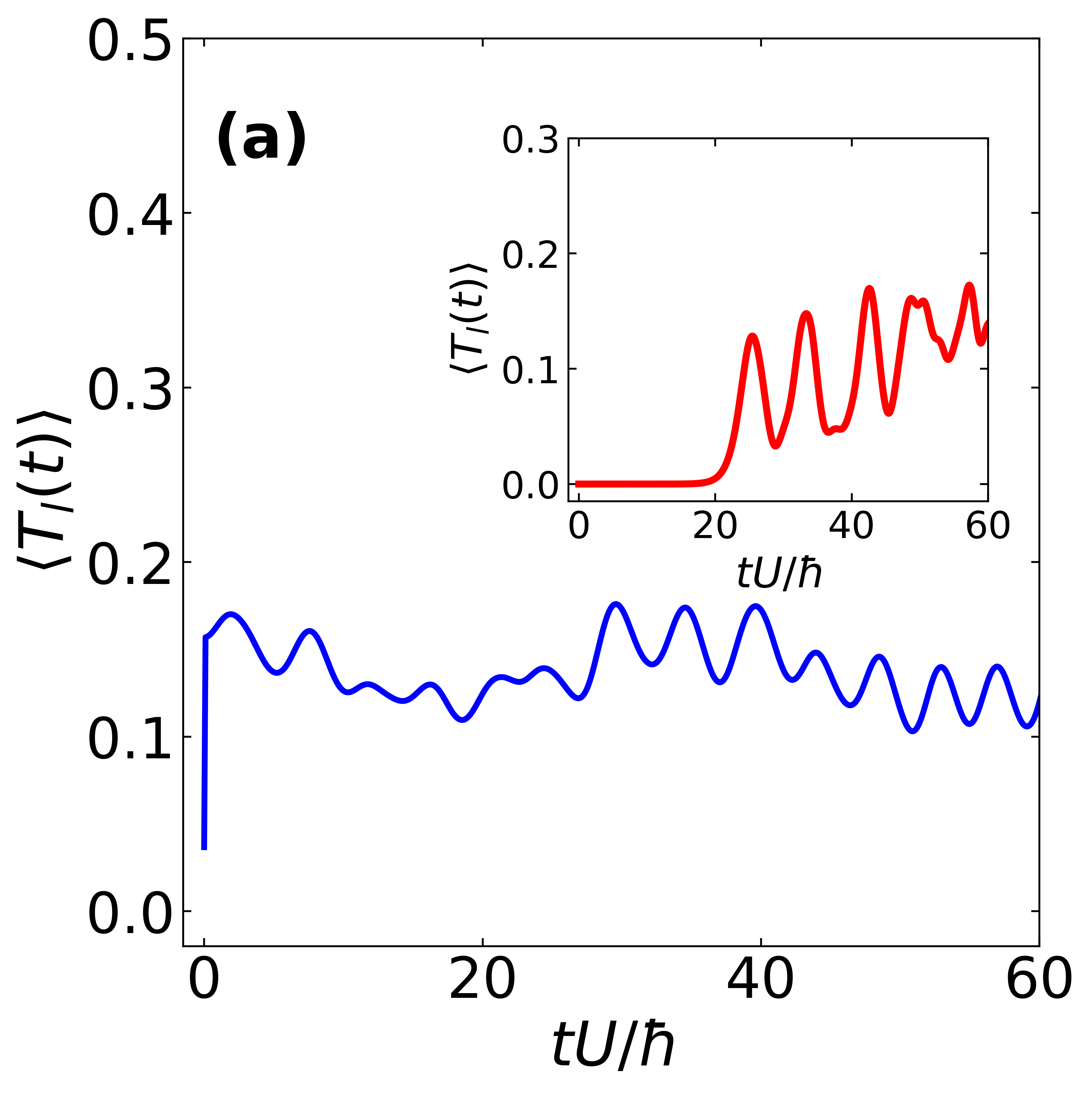}}
\rotatebox{0}{\includegraphics[width=0.49\linewidth]{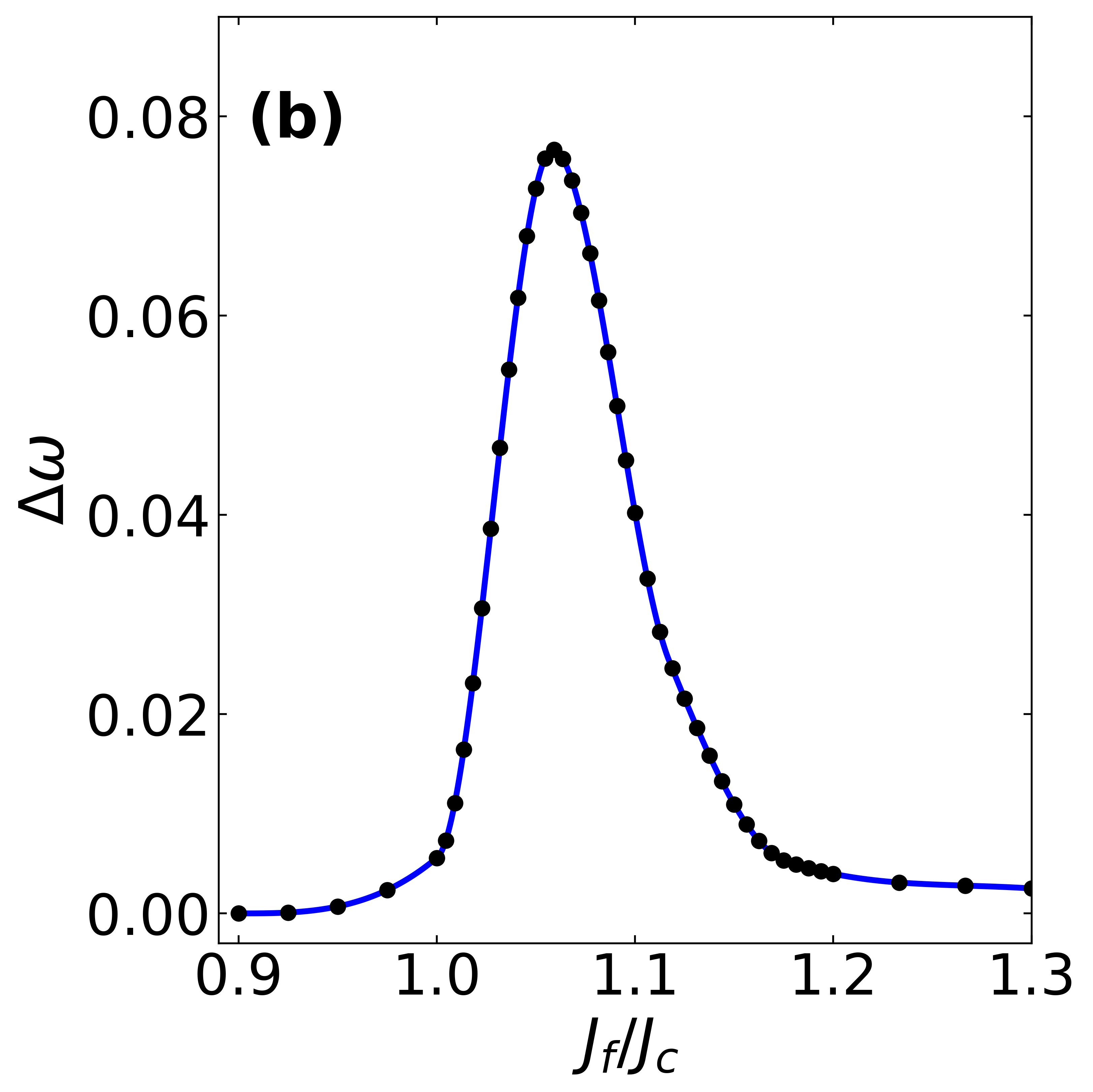}}
    \caption{Plot of the time evolution of the kinetic operator $\langle T_l(t) \rangle$ as a function of $tU/\hbar$ for $\mu=0.98 U$, $J_i=0.98 J_c$ and $J_f=3.51 J_c$ obtained using the projection operator technique.The inset represents the same plot obtained using mean-field theory. A comparison of the two plots shows the crucial role of quantum fluctuations in shaping the short-time dynamics. (b) Plot of $\Delta \omega$ as a function of $J_f/J_c$ for a quench from the CDW to the SS or SF state. The mean-field results become quantitatively more inaccurate near the critical point where fluctuations are more important. See text for details.}
    \label{fig7}
\end{figure}

Next, we provide a comparison of the results obtained using the projection operator approach with that obtained using mean-field theory so as to bring out the crucial role of short-range quantum fluctuations in shaping the post-quench dynamics. To this end, we first consider a quench starting from near the critical point such that $J_i=0.98 J_c$ and quench the hopping amplitude to $J_f= 3.51 J_c$. We study the behavior of $\langle T_{\ell}\rangle(t)$ using projection operator formalism as a function of time, as shown in Fig.\ \ref{fig7}(a). The inset of Fig.\ \ref{fig7}(a) shows the result for an identical quench as obtained from the mean-field theory; for this purpose, we compute $\langle T_{\ell}\rangle_{{\rm MF}} (t)= \langle \psi'(t)|T_{\ell} |\psi'(t)\rangle$. We note that these plots have qualitatively different early time dynamics since $\langle T_{\ell}\rangle_{{\rm MF}}(t=0)=0$ for the mean-field description and finite for the projection operator analysis. A consequence of this different initial condition shows up in the early time dynamics where  $\langle T_{\ell}\rangle_{{\rm MF}}$ remains close to zero at early times and undergoes a subsequent sharp jump in $\langle T_{\ell}\rangle$. This feature is an artifact of the mean-field theory and as can be seen from Fig.\ \ref{fig7}(a). This points out the necessity of incorporation of quantum fluctuations for describing non-equilibrium dynamics of such bosons. 

Apart from this qualitative differences, the two approaches also have disntict quantitative differences. To see this we consider dynamics of the order parameter amplitude following a CDW-SS quench illustrated in Fig.\ \ref{fig5}(a); this shows oscillations with a frequency $\omega_{\rm pr} = 2\pi/T$. We also estimate this frequency by tracking the dynamics for an identical quench using mean-field theory; the corresponding frequency is denoted by $\omega_{\rm mf}$. The difference of these two is defined via $\Delta \omega= |\omega_{\rm pr}-\omega_{\rm mf}|/\omega_{\rm mf}$ which is plotted in Fig.\ \ref{fig7}(b). We note that $\Delta \omega$ peaks close to the critical point which shows the crucial role of short quantum fluctuations in post-quench dynamics when $J_f$ is close to $J_c$.

\begin{figure}
    \centering
    \includegraphics[width=\linewidth]{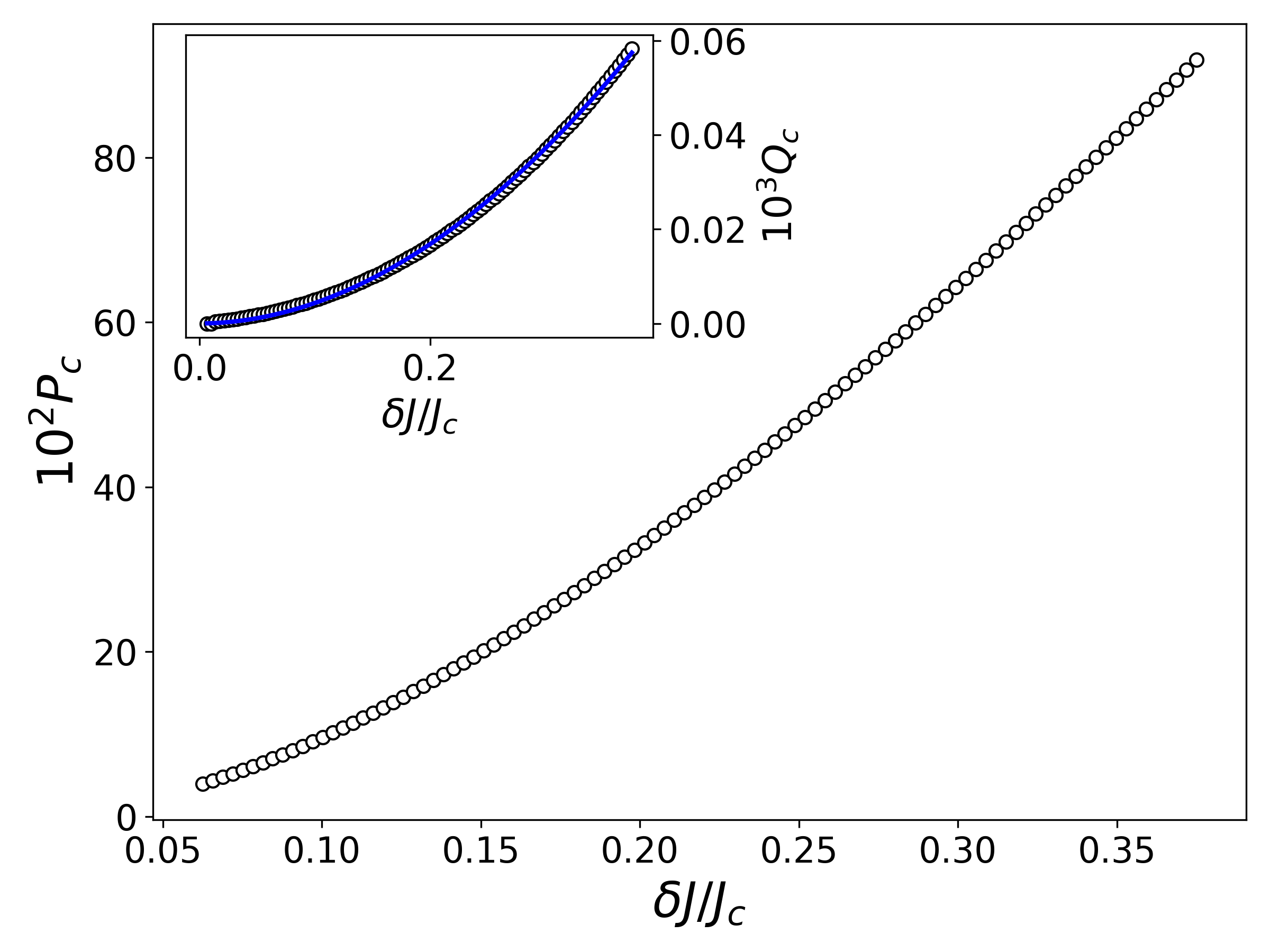}
    \caption{Plots of $P_c$ and $Q_c$ (inset) as function of $\delta J$ for $\delta J/J_c\ll 1$ starting from the tip of the CDW lobe. The solid line corresponds to fits yielding a power law $P_c \sim (\delta J)^{r_1}$ and $Q_c \sim \delta J^{(r_2)}$ with $r_1\simeq 1.6$, $r_2 \simeq 1.96$. See text for details.}
    \label{fig8}
\end{figure}

Next, we consider a quench starting from the critical hopping amplitude $(J_i=J_c)$ to the SS or SF phases with $J_f= J_c+\delta J$. We compute the overlap of the ground-state wave-function, defined as 
\begin{eqnarray}
{\mathcal F}_c &=& |\langle \psi_{f}|\psi_{c}\rangle|^2 = |\langle\psi'_{f}|e^{iS[J_{f}]}e^{-iS[J_{c}]}|\psi'_{c}\rangle|^2
\end{eqnarray} 
for this protocol. Here, $\psi_{f}$ and $\psi_{c}$ denote the ground-state wave-functions at $J = J_{f}$ and $J = J_{c}$, respectively. We also calculate the residual energy for this protocol given by 
\begin{eqnarray} 
Q_c = \langle \psi_f|H[J_c]|\psi_f\rangle - E_{G}[J_{f}].
\end{eqnarray} 

The plot of $P_c= 1-{\mathcal F}_c$ and $Q_c$ (inset) is shown Fig.~\ref{fig8} as functions of $\delta J$. A numerical fit of these data yields the power-law behaviors $P_c \sim \delta J^{r_1}$ with $r_1 \simeq1.6$ and $Q_c \sim \delta J^{r_2}$ with $r_2\simeq 1.96$. The corresponding results in the homogeneous case has been obtained in Ref.\ \cite{PhysRevLett.106.095702} and are known to deviate from the Kibble-Zurek (KZ) behavior which predicts $r_1=d\nu \sim 4/3$ and $r_2=(d+z)\nu \sim 2$ where $d=2$ is the dimension, $z=1$ is the dynamical critical exponent and $\nu \simeq 2/3$ is the correlation length exponent at the tip of the lobe. We find that $r_2$ matches the KZ prediction while $r_1$ shows a much larger deviation. This deviation is due to the fact that the projection operator technique misses the long-wavelength quantum fluctuations which is key to obtaining the correct description of the wavefunctions and hence fidelity \cite{PhysRevLett.106.095702}. In contrast, such fluctuations do not affect the local correlations including residual energies drastically; these quantities are accurately captured by the short-range fluctuations \cite{PhysRevB.86.085140}.

Finally, we investigate the ramp dynamics across the critical point during a ramp of the hopping amplitude $J(t)$. Such a ramp is parameterized by a characteristic rate $\tau^{-1}$ and an exponent $\alpha$ and is given by
\begin{eqnarray} 
J(t) &=& J_{i} + (J_{f} - J_{i})({t}/{\tau})^{\alpha}.
\end{eqnarray} 
Under such a ramp, the hopping amplitude evolves from $J_{i}$ at $t=t_{i} = 0$ to $J_{f}$ at $t= t_{f} = \tau$. As long as the system remains in the strong-coupling regime ($J_{i}/U, J_{f}/U \ll 1$), the perturbative projection method is expected to describe the dynamics accurately, irrespective of the values of $\tau$ and $\alpha$. Consequently, this projection operator framework enables one to treat slow ($\tau\gg\hbar/J$) and fast ($\tau \ll \hbar /J$) ramps on an equal footing for any $\alpha$. 
\begin{figure}
    \centering
    \includegraphics[width=\linewidth]{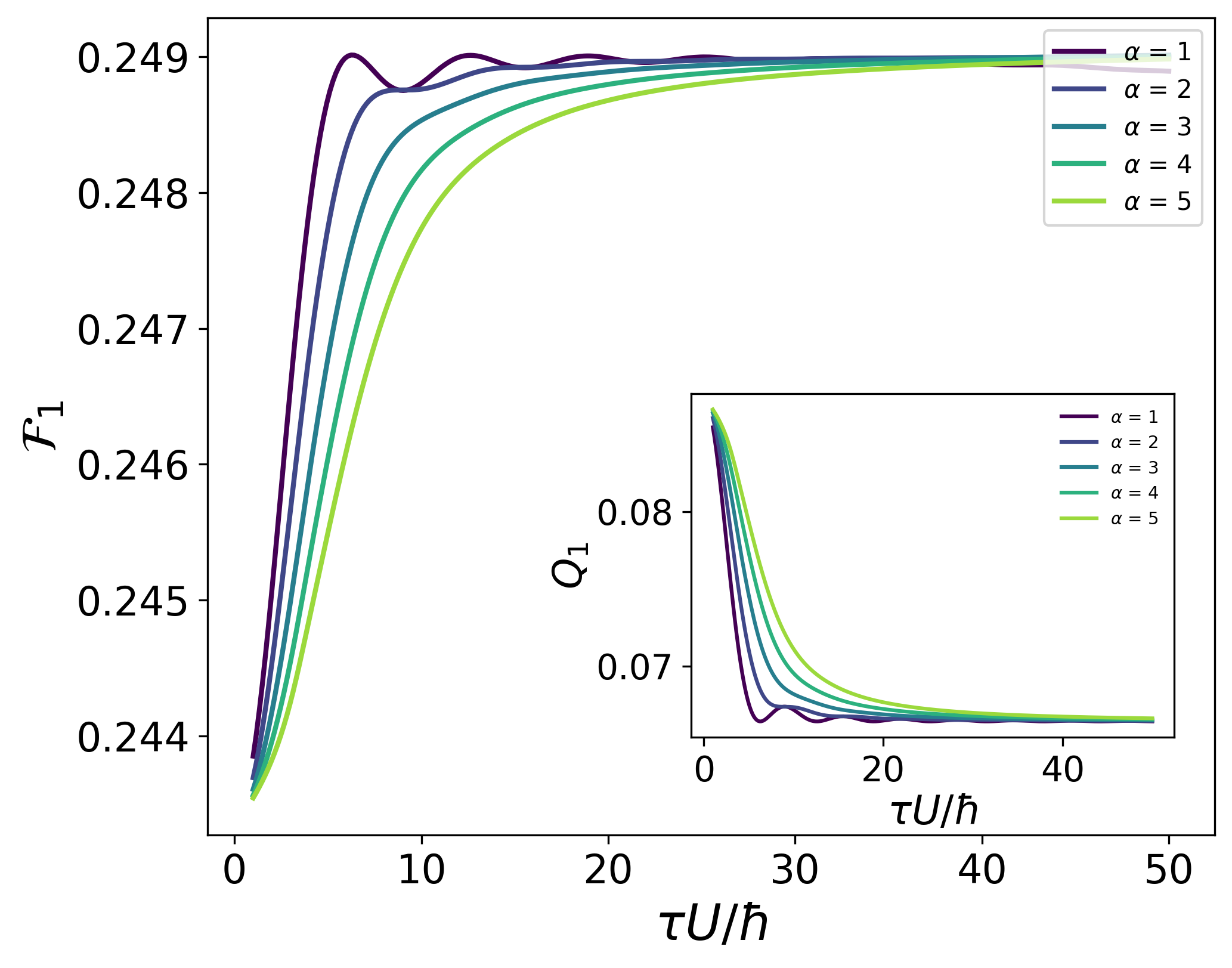}
    \caption{Plot of ${\mathcal F}_1$ as a function of $\tau U$ considering $\hbar=1$ for  $J_i/U=0.02$(SS) and $J_f/U=0.002$(CDW) for different exponents $1\le \alpha \le 5$ showing a plateau-like behavior at large $\tau$. The inset shows $Q_1$ as a function of $\tau$ for same set of $\alpha$. See text for details.}
    \label{fig9}
\end{figure}

To address the dynamics, we solve Eq.\ \ref{feq1} for $f_{\bf r}^{n_A}$ and $f_{r'}^{n_{B}}$ in a system with translational invariance within each sublattice. For all ramp dynamics, we start from an initial state which is the ground state of the system at $t=t_i$ for which $J_i/U=0.02$ so that the ground sttae corresponds to the SS state. We choose $J_f=0.002 U$ so that the it corresponds to the CDW phase. In what follows, we compute the defect formation probability $P_1 = 1 - |\langle \psi_{G}|\psi(t_{f})\rangle|^2 = 1 - {\mathcal F}_1$ and the residual energy $Q_1= \langle \psi_f |H[J_f]|\psi_f\rangle - E_G[J_f]$, where $|\psi_{G}\rangle$ is the final ground state and $|\psi(t_f)\rangle$ is the state evolved after the ramp. 

The resulting behaviors of $P_1$ and $Q_1$ (inset) are displayed in Fig.~\ref{fig9} for several representative values of the ramp exponent $\alpha$. Our results show that both $Q$ and $F$ (and consequently $P$) exhibit a plateau-like behavior in the large-$\tau$ limit. While the scaling of both quantities at small $\tau$ depends on the specific choice of the ramp protocol via $\alpha$, their asymptotic values at large $\tau$ remain entirely independent of $\alpha$. This asymptotic saturation is an artifact of inability of the present approach to capture the effect of long-wavelength fluctuation near the critical point. These fluctuations become important over a timescale $T_1 = L\hbar/J$; so the ramp dynamics is expected to yield accurate result for $\tau \le T_1$ \cite{PhysRevLett.106.095702}. A proper incorporation these long-wavelength fluctuations is expect to reproduce the KZ behavior at slow ramp rates \cite{rotor1}. Thus the dynamics showing approach to the plateau in Fig.\ \ref{fig9} is expected to be accurately described by the present method.

\section{Discussion}
\label{dissc}

In this work we have employed a projection operator formalism to study the ground-state phase diagram and the non-equilibrium quench and ramp dynamics of ultracold bosons coupled to a cavity. Our analysis provides a semi-analytic understanding of the phase diagram and the dynamics while retaining short-range quantum fluctuations over mean-field theory. Such a method has been applied to the standard Bose-Hubbard model in Ref.\ \cite{PhysRevLett.106.095702}; our work constitutes a generalization of that analysis in the presence of translational symmetry-broken phases. 

The equilibrium phase diagram that we obtain clearly shows the importance of the short-range fluctuations. Our results indicate that these fluctuations lead to additional stability of the MI phases; in addition, they also predict a larger regime where SS phases stabilize. The Mott lobe pushes further to higher values of $J_c/U$ and yields numerical values of $J_c/U$ that are closer to the quantum Monte Carlo results than that obtained from mean-field theory.  

We also analyze the non-equilibrium dynamics of these bosons. In this work, we have mainly concentrated on quench and ramp dynamics of bosons in translational symmetry-broken phases. The quench of $J$ starting from the
CDW ground state to the critical point reveals a single-frequency oscillatory post-quench dynamics; in contrast, a quench which takes the system deep inside the SF phase involves multiple frequencies. We also show the key role played by the quantum fluctuations in describing the post-quench dynamics; a plot of $\langle T_{\ell}\rangle$ as a function of time after the quench shows spurious jump in short time dynamics which is smoothened by the presence of quantum fluctuations. In addition, the time period of oscillatory motion following a quench close to $J_c$ shows significant deviation from its mean-field counterpart. Our analysis for the fidelity ${\mathcal F}$ and residual energy $Q$ reproduces the KZ scaling for the latter; this shows that short-range fluctuation accurately describes $Q$. However, for ${\mathcal F}$, our results deviate from the KZ predicted scaling exponent; this is due to the use of a variational wavefunction which misses the contribution from long-wavelength quantum fluctuations. 

We also study the ramp dynamics of these bosons starting from the critical point to the CDW phase. Our analysis indicates that both the fidelity and the residual energy of such ramps reach a plateau at small ramp rate; they do not exhibit the KZ scaling seen in Ref.\ \cite{rotor1}. We provide a qualitative argument for this absence; it stems from the inability of this method to capture long-wavelength quantum fluctuations. However, such fluctuations are expected to become important when the ramp rate either matches or is slower than the timescale $T_0= L\hbar/J$; thus for small $J/U$, there is a regime of intermediate and fast ramp rates where the projection operator approach can describe the ramp dynamics. 

Our results can be verified in standard experimental setup of bosons coupled to the cavity \cite{revcav,RevModPhys.97.011001}. We suggest implementing a quench via a sudden change of optical lattice potential which changes the hopping $J$ from a initial value $J_i$ to a final value $J_f$. Our main prediction is that such a quench starting from the CDW phase to the SS phase near $J_c$ will lead to single-frequency oscillations; these oscillations are expected to have a decay scale controlled by $T_0^{-1}$; consequently for $J/U \ll 1$, they will decay slowly. This makes them experimentally observable in a standard setup. The time period of such oscillations can be matched to our theoretical predictions. Also, an analogous quench from the CDW to the SF phase is expected to lead to a more complicated dynamics involving multiple frequencies. 

In conclusion, we have studied ultracold bosons coupled to a cavity using a projection operator approach which allows systematic inclusion of short-range quantum fluctuations of the bosons. Our analysis shows that these fluctuation have important effects on both the phase boundary and non-equilibrium dynamics of the system. We have also suggested realizable experiments which can test our theory.

\begin{acknowledgments}
KS thanks DST for support through SERB project  JCB/2021/000030.
BD acknowledges support under National Quantum Mission of the Department of Science \& Technology, Govt. of India, in the area of Quantum Sensing and Metrology. 
\end{acknowledgments}

\appendix

\section{Expressions of energy elements for quadratic order}
\label{appa} 
In this section, we first provide explicit expressions of $S_{ij;\ell}(n_A,n_B;n'_A,n'_B) = \langle \psi'|T_{\alpha \ell}^{n_A,n_B} T_{\beta \ell}^{n'_A,n'_B} |\psi'\rangle/J^2$ which is used for computing $E_b$ and $E_c^{(1)}$ in the main text. These terms involve a single link $\ell$ between two neighboring sites. Using Eq.\ \ref{eqvarwav1}, we find, after some algebra
\begin{widetext}
\begin{eqnarray}
S_{11;\ell}(n_A,n_B,n_A',n_B') &=& \sqrt{n_A'(n_A'-1)(n_B'+1)(n_B'+2)}f_{r_1}^{ n_A'-2 \ast}f_{r_1}^{n_A'}f_{r_2}^{n_B'+2 \ast}f_{r_2}^{n_B'}  \delta_{n_A,n_A'-1} \delta_{n_B,n_B'+1}
\\ \nonumber
S_{22;\ell}(n_A,n_B,n_A',n_B') &=& \sqrt{n_B'(n_B'-1)(n_A'+1)(n_A'+2)}f_{r_1}^{ n_A'+2 \ast}f_{r_1}^{n_A'}f_{r_2}^{n_B'-2 \ast}f_{r_2}^{n_B'}  \delta_{n_A,n_A'+1} \delta_{n_B,n_B'-1}
\\ \nonumber
S_{12;\ell}(n_A,n_B,n_A',n_B') &=& (n_A'+1)n_B' |f_{r_1}^{n_A'}|^2|f_{r_2}^{n_B'}|^2  \delta_{n_A,n_A'+1} \delta_{n_B,n_B'-1}
\\ \nonumber
 S_{21;\ell}(n_A,n_B,n_A',n_B') &=& (n_B'+1)n_A' |f_{r_1}^{n_A'}|^2|f_{r_2}^{n_B'}|^2  \delta_{n_A,n_A'-1} \delta_{n_B,n_B'+1}
\nonumber
\end{eqnarray}    
\end{widetext}
where $r_1$ and $r_2$ denote coordinates of the site at the two ends of link $\ell$. 

Next, we consider the terms $R_{ij; \ell_1,\ell_2} = \langle \psi'|T_{\alpha \ell_1}^{n_A,n_B} T_{\beta \ell_2}^{n'_A,n'_B} |\psi'\rangle/J^2$  which involves two links $\ell_1$ and $\ell_2$ connecting three adjacent sites with coordinates $r_1$, $r_2$ and $r_3$. These terms are used in the main text for comupation of $E_2$ and $E_3$. 

The computation of such terms involve two distinct cases. The first of these corresponds to the case where the site which is common to both links belongs to sublattice $A$ and the second corresponds to when it belongs to sublattice $B$. We denote these cases by an index $a=1$ and $a=2$ respectively. We compute the matrix elements where the first hopping occurs on link $\ell_2$ and the second on $\ell_1$. Since we sum over $\ell_1$ and $\ell_2$, we can always choose $\ell_2$ to be to the right of $\ell_1$. With these consideration and after some straighforward, but cumbersome, algebra, one obtains 
\begin{widetext}
\begin{eqnarray}
R^{(1)}_{11;l_1,l_2}(n_A,n_B,n_A',n_B') &=& \sqrt{n_An_A'(n_B'+1)(n_B'+2)}f_{r_1}^{n_A-1 \ast }f_{r_1}^{n_A}f_{r_2}^{n_B'+1 \ast }f_{r_2}^{n_B'}f_{r_3}^{ n_A'-1 \ast}f_{r_3}^{n_A'}\delta_{n_B,n_B'+1}
\\ \nonumber
R^{(2)}_{11;l_1,l_2}(n_A,n_B,n_A',n_B') &=& \sqrt{n_A'(n_A'-1)(n_B'+1)(n_B+1)}f_{r_1}^{ n_B+1 \ast}f_{r_1}^{n_B}f_{r_2}^{ n_A'-2 \ast}f_{r_2}^{n_A'}f_{r_3}^{ n_B'+1 \ast}f_{r_3}^{n_B'}\delta_{n_A,n_A'-1}
\\ \nonumber
R^{(1)}_{22;l_1,l_2}(n_A,n_B,n_A',n_B') &=& \sqrt{n_B'(n_B'-1)(n_A'+1)(n_A+2)}f_{r_1}^{n_A-1 \ast}f_{r_1}^{n_A}f_{r_2}^{n_B'-2 \ast}f_{r_2}^{n_B'}f_{r_3}^{ n_A'-1 \ast}f_{r_3}^{n_A'}\delta_{n_B,n_B'-1}
\\ \nonumber
R^{(2)}_{22;l_1,l_2}(n_A,n_B,n_A',n_B') &=& \sqrt{n_B'n_B(n_A'+1)(n_A'+2)}f_{r_1}^{n_B-1 \ast}f_{r_1}^{n_B}f_{r_2}^{n_A'+2 \ast}f_{r_2}^{n_A'}f_{r_3}^{ n_B'-1 \ast}f_{r_3}^{n_B'}\delta_{n_A,n_A'+1}
\\ \nonumber
R^{(1)}_{12;l_1,l_2}(n_A,n_B,n_A',n_B') &=& n_B'\sqrt{n_A(n_A'+1)}f_{r_1}^{n_A-1 \ast}f_{r_1}^{n_A}|f_{r_2}^{n_B'}|^2 f_{r_3}^{n_A'+1 \ast}f_{r_3}^{n_A'}\delta_{n_B,n_B'-1}
\\ \nonumber
R^{(2)}_{12;l_1,l_2}(n_A,n_B,n_A',n_B') &=& (n_A'+1)\sqrt{n_B'(n_B+1)}f_{r_1}^{n_B+1 \ast}f_{r_1}^{n_B}|f_{r_2}^{n_A'}|^2f_{r_3}^{n_B'-1 \ast }f_{r_3}^{n_B'}\delta_{n_A,n_A'+1}
\\ \nonumber
R^{(1)}_{21;l_1,l_2}(n_A,n_B,n_A',n_B') &=& (n_B'+1)\sqrt{n_A'(n_A+1)}f_{r_1}^{n_A-1 \ast}f_{r_1}^{n_A}|f_{r_2}^{n_B'}|^2 f_{r_3}^{n_A'-1 \ast}f_{r_3}^{n_A'}\delta_{n_B,n_B'+1}
\\ \nonumber
R^{(2)}_{21;l_1,l_2}(n_A,n_B,n_A',n_B') &=& n_A'\sqrt{n_B(n_B'+1)}f_{r_1}^{n_B-1 \ast}f_{r_1}^{n_B}|f_{r_2}^{n_A'}|^2 f_{r_3}^{n_B'+1 \ast}f_{r_3}^{n_B'}\delta_{n_A,n_A'-1}
\end{eqnarray}    
\end{widetext}    
This completes our derivation of $S_{ij;\ell}$ and $R^{(a)}_{i j; \ell_1,\ell_2}$. 

\section{Expressions for dynamics}
\label{appb}
In this section, we chart out the the a class of terms that contribute to the right side of Eq.\ \ref{feq1}. First we consider terms which involve a single link Eq.\ \ref{dynqdef} and \ref{dynydef}. These originate from the variations  
${{\delta E_{b}}/{{\delta f^{n_A \ast(n_B \ast)}_{\bf r}}}}$ and ${{\delta E_{c}^{(1)}}/{{\delta f^{n_A \ast(n_B \ast)}_{\bf r}}}}$ and yield two terms each for a given $\alpha$  and $\beta$. We define 
\begin{eqnarray} 
Q_{\alpha\beta}(n_A,n_B,n_A',n_B') &=& {{\delta }}/{{\delta f^{n_A \ast}_{\bf r}}}(S_{\alpha\beta}(n_A,n_B,n_A',n_B')) \nonumber\\
Y_{\alpha\beta}(n_A,n_B,n_A',n_B') &=&  {{\delta }}/{{\delta f^{n_B \ast}_{\bf r}}}(S_{\alpha\beta}(n_A,n_B,n_A',n_B')) \nonumber\\ \label{yqdef}
\end{eqnarray}
In the notation that we shall follow for this derivation, $n_1 \in n_A$ for $Q_{\alpha \beta}$ and $n_1 \in n_B$ for $Y_{\alpha \beta}$. 

A straightforward but cumbersome calculation using these definitions yield 
\begin{widetext}
\begin{eqnarray}
Q_{11}(n_A,n_B,n_A',n_B') &=& \sqrt{(n_1+2)(n_1+1)(n_B'+2)(n_B'+1)} f_{\bf r}^{n_1+2}f_{r_1}^{ n_B'+2 \ast} f_{r_1}^{n_B'} \delta_{n_A,n_1+1} \delta_{n_B,n_B'+1}\delta_{n_1,n_A'-2}\delta_{r_1,r}
\nonumber \\
Q_{22}(n_A,n_B,n_A',n_B') &=& \sqrt{n_1(n_1-1)n_B'(n_B'-1)} f_{\bf r}^{n_1-2}f_{r1}^{n_B'-2 \ast}f_{r_1}^{n_B'} \delta_{n_A,n_1-1} \delta_{n_B,n_B'-1}\delta_{n_1,n_A'+2}\delta_{r_1,r}
\nonumber \\
Q_{12}(n_A,n_B,n_A',n_B') &=& (n_1+1)n_B' f_{\bf r}^{n_1}|f_{r_1}^{n_B'}|^2 \delta_{n_A,n_1'+1} \delta_{n_A,n_A'-1}\delta_{r_1,r}
\nonumber \\
Q_{21}(n_A,n_B,n_A',n_B') &=& (n_B'+1)n_1 f_{\bf r}^{n_1}|f_{r_1}^{n_B'}|^2 \delta_{n_A,n_1'-1} \delta_{n_B,n_B'+1}\delta_{r_1,r}
\label{qeq12} \\ 
Y_{11}(n_A,n_B,n_A',n_B') &=& \sqrt{(n_1-1)n_1(n_A'-1)n_A'} f_{\bf r}^{n_A'-2 \ast}f_{r}^{n_A'}f_{r_1}^{n_1-2} \delta_{n_A,n_A'-1} \delta_{n_B,n_1-1}\delta_{n_1,n_B'+2}\delta_{r_1,r}
\nonumber \\
Y_{22}(n_A,n_B,n_A',n_B') &=& \sqrt{(n_1+1)(n_1+2)(n_A'+2)(n_A'+1)} f_{\bf r}^{n_A'+2 \ast}f_{r}^{n_A'}f_{r_1}^{n_1+2} \delta_{n_A,n_A'+1} \delta_{n_B,n_1+1}\delta_{n_1,n_B'-2}\delta_{r_1,r}
\nonumber \\
Y_{12}(n_A,n_B,n_A',n_B') &=& n_1(n_A'+1) f_{r_1}^{n_1}|f_{r}^{n_A'}|^2  \delta_{n_A,n_1+1} \delta_{n_B,n_1-1}\delta_{r_1,r}
\nonumber \\
Y_{21}(n_A,n_B,n_A',n_B') &=& n_A'(n_1+1) f_{r_1}^{n_1}|f_{r}^{n_A'}|^2 \delta_{n_A,n_1-1} \delta_{n_B,n_B'+1}\delta_{r_1,r}
 \nonumber
\end{eqnarray}    
\end{widetext}

Next, we consider the terms in the right side of Eq.\ \ref{feq1} whose contribution involves two adjacent links connecting three adjacent sites. These are defined as  
\begin{eqnarray} 
L^{(a)}_{\alpha\beta}(n_A,n_B,n_A',n_B') &=& {{\delta }}/{{\delta f^{n_A \ast}_{\bf r}}}(R^{(a)}_{\alpha\beta}(n_A,n_B,n_A',n_B')) \nonumber\\
Z^{(a)}_{\alpha\beta}(n_A,n_B,n_A',n_B') &=& {{\delta }}/{{\delta f^{n_B \ast}_{\bf r}}}(R^{(a)}_{\alpha\beta}(n_A,n_B,n_A',n_B')) \nonumber\\  \label{lzdef}
\end{eqnarray} 
where $n_1\in n_A$ for each of the terms in $L^{(a)}$ and $n_1\in n_B$ for those in $R^{(a)}$. Here the index $a$ takes values $1,2,3$ for three types of terms. A straightforward calculation yields 
\begin{widetext}
\begin{eqnarray}
L_{11;r}^{(1)}(n_A,n_B,n_A',n_B') &=& \sqrt{(n_1+1)(n_B'+1)(n_B'+2)n_A'} f_{\bf r}^{n_1+1}f_{r_2}^{n_B'+2 \ast} f_{r_2}^{n_B'} f_{r_3}^{n_A'-1 \ast}f_{r_3}^{n_A'} \delta_{n_1,n_A-1}\delta_{r,r_1}\delta_{n_B,n_B'+1}
\label{leqs} \\
L_{11;r}^{(2)}(n_A,n_B,n_A',n_B') &=& \sqrt{(n_1+1)(n_A'+1)(n_A'+2)n_e} f_{r_1}^{n_A-1 \ast}f_{r_1}^{n_A}f_{r_2}^{n_B'+2 \ast} f_{r_2}^{n_B'} f_{r}^{n_1+1} \delta_{n_B,n_B'+1}\delta_{r,r_3}\delta_{n_1,n_A'-1}
\nonumber \\
L_{11;r}^{(3)}(n_A,n_B,n_A',n_B') &=& \sqrt{(n_1+2)(n_B'+1)(n_B'+2)(n_1+1)} f_{r_1}^{n_B-1 \ast}f_{r_1}^{n_B}f_{r}^{n_1+2}f_{r_3}^{n_B'+1 \ast}f_{r_3}^{n_B'} \delta_{n_A,n_A'-1}\delta_{r,r_2}\delta_{n_1,n_A'-2}
 \nonumber \\
L_{22;r}^{(1)}(n_A,n_B,n_A',n_B') &=& \sqrt{(n_1+2)n_B'(n_B'-1)(n_A'+1)} f_{\bf r}^{n_1+1}f_{r_2}^{n_B'-2 \ast} f_{r_2}^{n_B'} f_{r_3}^{n_A'-1 \ast}f_{r_3}^{n_A'} \delta_{n_1,n_A-1}\delta_{r,r_1}\delta_{n_B,n_B'-1}
 \nonumber \\
L_{22;r}^{(2)}(n_A,n_B,n_A',n_B') &=& \sqrt{(n_1+2)(n_B'-1)n_B'(n_A+1)} f_{r_1}^{n_A-1 \ast}f_{r_1}^{n_A}f_{r_2}^{n_B'-2 \ast} f_{r_2}^{n_B'} f_{r}^{n_1+1} \delta_{n_B,n_B'-1}\delta_{r,r_3}\delta_{n_1,n_A'-2}
 \nonumber \\
L_{22;r}^{(3)}(n_A,n_B,n_A',n_B') &=& \sqrt{(n_1-1)n_Bn_B'n_1} f_{r_1}^{n_B-1 \ast}f_{r_1}^{n_B}f_{r}^{n_1-2}f_{r_3}^{n_B'-1 \ast}f_{r_3}^{n_B'} \delta_{n_A,n_A'-1}\delta_{r,r_2}\delta_{n_1,n_A'-2}
 \nonumber \\
L_{12;r}^{(1)}(n_A,n_B,n_A',n_B') &=& n_B'\sqrt{(n_1+1)n_A'} f_{\bf r}^{n_1+1}|f_{r_2}^{n_B'}|^2 f_{r_3}^{n_A'+1 \ast}f_{r_3}^{n_A'} \delta_{n_1,n_A-1}\delta_{r,r_1}\delta_{n_B,n_B'-1}
 \nonumber \\
L_{12;r}^{(2)}(n_A,n_B,n_A',n_B') &=& n_B'\sqrt{(n_1+2)n_A}f_{r_1}^{n_A-1 \ast}f_{r_1}^{n_A}|f_{r_2}^{n_B'}|^2 f_{r}^{n_1-1} \delta_{n_B,n_B'-1}\delta_{r,r_3}\delta_{n_1,n_A+1}
 \nonumber  \\
L_{12;r}^{(3)}(n_A,n_B,n_A',n_B') &=& (n_1+1)\sqrt{(n_B+1)n_B'} f_{r_1}^{ n_B+1 \ast}f_{r_1}^{n_B}f_{r}^{n_1}f_{r_3}^{n_B'-1 \ast}f_{r_3}^{n_B'} \delta_{n_A,n_1+1}\delta_{r,r_2} \nonumber  \\
L_{21;r}^{(1)}(n_A,n_B,n_A',n_B') &=&  (n_B'+1)\sqrt{n_1n_A'} f_{\bf r}^{n_1-1}|f_{r_2}^{n_B'}|^2 f_{r_3}^{n_A'-1 \ast}f_{r_3}^{n_A'} \delta_{n_1,n_A+1}\delta_{r,r_1}\delta_{n_B,n_B'+1}
 \nonumber \\
L_{21;r}^{(2)}(n_A,n_B,n_A',n_B') &=& (n_B'+1)\sqrt{(n_1+1)(n_A+1)} f_{r_1}^{ n_A+1 \ast}f_{r_1}^{n_A}|f_{r_2}^{n_B'}|^2 f_{r}^{n_1+1} \delta_{n_B,n_B'+1}\delta_{r,r_3}\delta_{n_1,n_A'-1}
 \nonumber \\
L_{21;r}^{(3)}(n_A,n_B,n_A',n_B') &=& n_1\sqrt{(n_B'+1)n_B} f_{r_1}^{n_B-1 \ast}f_{r_1}^{n_B}f_{r}^{n_1}f_{r_3}^{ n_B'+1 \ast}f_{r_3}^{n_B'} \delta_{n_A,n_1-1}\delta_{r,r_2} \nonumber  
\end{eqnarray} 
and 
\begin{eqnarray} 
Z_{11;r}^{(1)}(n_A,n_B,n_A',n_B') &=& \sqrt{(n_1-1)n_A'n_1n_A} f_{r_1}^{n_A-1 \ast}f_{r_1}^{n_A}f_{\bf r}^{n_1-2} f_{r_3}^{n_A'+1 \ast}f_{r_3}^{n_A'} \delta_{n_B,n_B'+1}\delta_{r,r_2}\delta_{n_1,n_B'+2}
\\ \label{zeqs} 
Z_{11;r}^{(2)}(n_A,n_B,n_A',n_B') &=& \sqrt{(n_A'-1)(n_B'+1)n_1n_A'} f_{r}^{n_1-1}f_{r_2}^{n_A'-2 \ast} f_{r_2}^{n_A'} f_{r_3}^{n_B'+1 \ast}f_{r_3}^{n_B'} \delta_{n_A,n_A'-1}\delta_{r,r_1}\delta_{n_1,n_B+1}
 \nonumber  \\
Z_{11;r}^{(3)}(n_A,n_B,n_A',n_B') &=& \sqrt{(n_A'-1)(n_B+1)n_1n_A'} f_{r_1}^{ n_B+1 \ast}f_{r_1}^{n_B}f_{r_2}^{ n_A'-2 \ast} f_{r_2}^{n_A'} f_{\bf r}^{n_1-1} \delta_{n_e,n_A'-1}\delta_{r,r_3}\delta_{n_1,n_B'+1}
\nonumber \\
Z_{22;r}^{(1)}(n_A,n_B,n_A',n_B') &=& \sqrt{(n_1+2)(n_A'+1)(n_1+1)(n_A+1)} f_{r_1}^{ n_A-1 \ast}f_{r_1}^{n_A}f_{\bf r}^{n_1+2} f_{r_3}^{n_A'-1 \ast}f_{r_3}^{n_A'} \delta_{n_B,n_B'-1}\delta_{r,r_2}\delta_{n_1,n_B'-2}
 \nonumber  \\
Z_{22;r}^{(2)}(n_A,n_B,n_A',n_B') &=& \sqrt{(n_A'+1)n_B'(n_1+1)(n_A'+2)} f_{r}^{n_1+1}f_{r_2}^{ n_A'+2 \ast} f_{r_2}^{n_A'} f_{r_3}^{n_B'-1 \ast}f_{r_3}^{n_B'} \delta_{n_A,n_A'+1}\delta_{r,r_1}\delta_{n_1,n_B-1}
\nonumber  \\
Z_{22;r}^{(3)}(n_A,n_B,n_A',n_B') &=& \sqrt{(n_A'+1)n_B(n_1+1)(n_A'+2)} f_{r_1}^{ n_B-1 \ast}f_{r_1}^{n_B}f_{r_2}^{ n_A'+2 \ast} f_{r_2}^{n_A'} f_{\bf r}^{n_1+1} \delta_{n_A,n_A'+1}\delta_{r,r_3}\delta_{n_1,n_B'-1}
 \nonumber  \\
Z_{12;r}^{(1)}(n_A,n_B,n_A',n_B') &=& n_1\sqrt{(n_A'+1)n_A} f_{r_1}^{ n_A-1 \ast}f_{r_1}^{n_A}f_{r}^{n_1} f_{r_3}^{ n_A'+1 \ast}f_{r_3}^{n_A'} \delta_{n_B,n_1+1}\delta_{r,r_2}
 \nonumber  \\
Z_{12;r}^{(2)}(n_A,n_B,n_A',n_B') &=& (n_A'+1)\sqrt{n_1 n_B'} f_{r}^{n_1-1}|f_{r_2}^{n_A'}|^2 f_{r_3}^{ n_B'-1 \ast}f_{r_3}^{n_B'} \delta_{n_A,n_A'+1}\delta_{r,r_1}\delta_{n_1,n_B+1}
 \nonumber  \\
Z_{12;r}^{(3)}(n_A,n_B,n_A',n_B') &=& (n_A'+1)\sqrt{(n_1+1)(n_B+1)} f_{r_1}^{ n_B-1 \ast}f_{r_1}^{n_B}|f_{r_2}^{n_A'}|^2 f_{\bf r}^{n_1+1} \delta_{n_A,n_A'+1}\delta_{r,r_3}\delta_{n_1,n_B'-1}
\nonumber  \\
Z_{21;r}^{(1)}(n_A,n_B,n_A',n_B') &=& (n_1+1)\sqrt{(n_A+1)n_A'} f_{r_1}^{ n_A+1 \ast}f_{r_1}^{n_A}f_{r}^{n_1} f_{r_3}^{ n_A'-1 \ast}f_{r_3}^{n_A'} \delta_{n_B,n_1-1}\delta_{r,r_2}
 \nonumber  \\
Z_{21;r}^{(2)}(n_A,n_B,n_A',n_B') &=& n_e'\sqrt{(n_1+1)(n_B'+1)} f_{r}^{n_1+1}|f_{r_2}^{n_A'}|^2 f_{r_3}^{ n_B'+1 \ast}f_{r_3}^{n_B'} \delta_{n_A,n_A'-1}\delta_{r,r_1}\delta_{n_1,n_B-1}
 \nonumber  \\
 Z_{21;r}^{(3)}(n_A,n_B,n_A',n_B') &=& n_A'\sqrt{n_1n_B} f_{r_1}^{ n_B-1 \ast}f_{r_1}^{n_B}|f_{r_2}^{n_A'}|^2 f_{\bf r}^{n_1-1} \delta_{n_A,n_A'-1}\delta_{r,r_3}\delta_{n_1,n_B'+1}
 \nonumber
\end{eqnarray}    
\end{widetext}

\section{Results for quench from homogeneous MI to SF phase}
\label{appd} 

In this appendix, we briefly discuss the results of a quench from the homogeneous Mott phase ( $J=J_i < J_c$) to the superfluid phase through the tip of the Mott lobe. We consider two specific cases; for the first the quench ends near the critical point ($J_f=1.02 J_c$) while for the second, it takes the system deep inside the SF phase ($J_f= 3.51 J_c$). The results of this section reproduces that of earlier works \cite{PhysRevLett.106.095702,PhysRevB.86.085140} and acts a benchmark for our method.

\begin{figure}
\rotatebox{0}{\includegraphics[width=0.49 \linewidth]{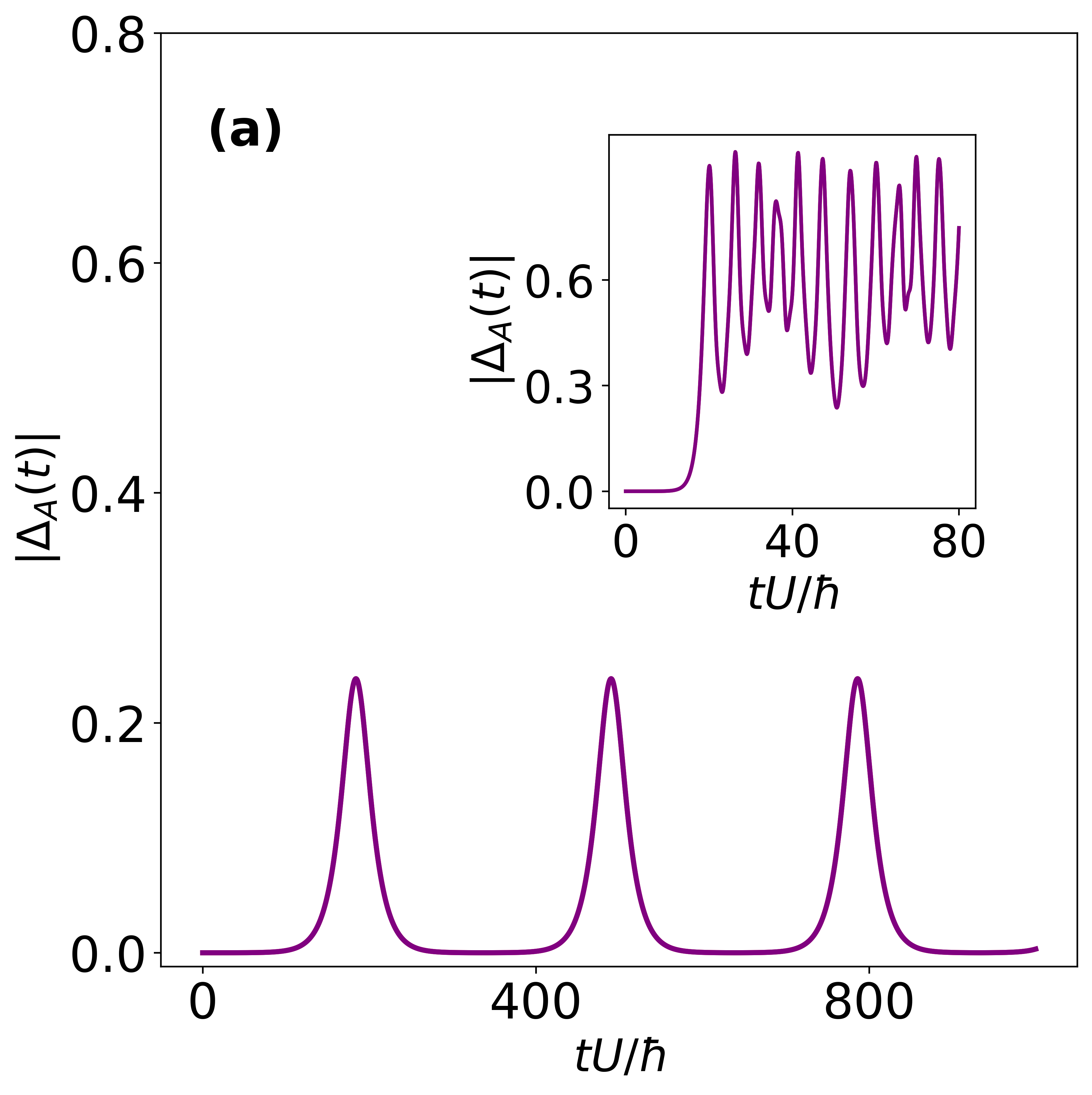}}
\rotatebox{0}{\includegraphics[width=0.49\linewidth]{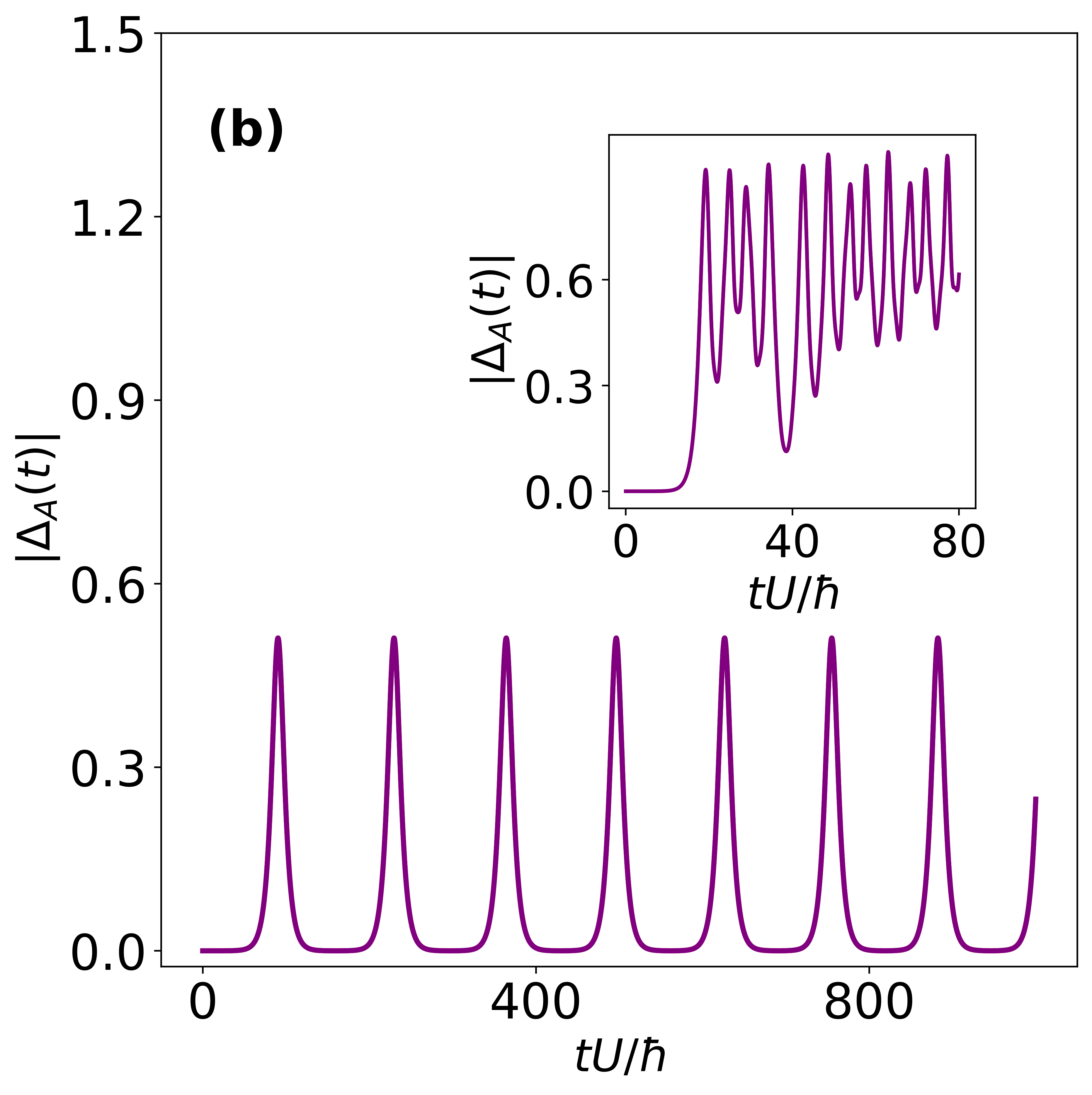}}
    \caption{(a) Plot of $\Delta_{A}(t)$ as a function of $tU/\hbar$ obtained using the projection operator method for $J_i=0.01 J_c$, $J_f=1.02 J_c$ and $\mu=0.41 U$. The inset shows a similar plot but for $J_f= 3.51 J_c$. (b) Similar plot obtained using mean-field theory. See text for details.}
    \label{fig10}
\end{figure}

These results are shown in Fig.\ \ref{fig10}. The left panels of Fig.\ \ref{fig10} shows the dynamics where the quench takes one through the tip of the Mott lobe. The left panel shows the result obtained using the projection operator approach  for the order parameter dynamics, while the right left panel shows the mean-field result. In this regime, $|\Delta_A(t)|= |\Delta_B(t)|$. We find that for $ J_f=1.02 J_c$, when $J_f$ is near the critical $J_c$, the number fluctuations remain small. Moreover, the presence of a small equilibrium superfluid order parameter suppresses coupling between amplitude and phase modes, while a large correlation length blocks short-range spatial fluctuations. These features lead to a single characteristic frequency of oscillation proportional to the equilibrium mass gap at $J= 1.02 J_c$ \cite{PhysRevLett.89.250404, PhysRevLett.97.200601}. We note that this frequency is significantly different in the two approaches; the mean-field theory leads to a much wider peaks with lower frequency in the dynamics of $|\Delta_{A}(t)|$.  

In contrast, quenching deep into the SF regime ($J_f=3.51 J_c$) (insets of Fig.\ \ref{fig10}) allow for larger number fluctuations. In this regime, both the amplitude and the phase modes contribute to the dynamics leading to a multi-frequency response. This response is almost identical for the projection operator and mean-field approaches as can be seen by comparing the insets of Fig.\ \ref{fig10}. This indicates that the mean-field theory leads to accurate results for order parameter dynamics deep inside the SF but not near the critical point. 

These results reproduce those in Refs.\ \onlinecite{PhysRevLett.89.250404, PhysRevLett.97.200601} and provide a benchmark for our formalism
which we apply to quenches starting from the CDW phases in the main text.

\bibliography{Cite}

@article{PhysRevB.86.085140,
  title = {Projection operator approach to the Bose-Hubbard model},
  author = {Dutta, Anirban and Trefzger, C. and Sengupta, K.},
  journal = {Phys. Rev. B},
  volume = {86},
  issue = {8},
  pages = {085140},
  numpages = {10},
  year = {2012},
  month = {Aug},
  publisher = {American Physical Society},
  doi = {10.1103/PhysRevB.86.085140},
  url = {https://link.aps.org/doi/10.1103/PhysRevB.86.085140}
}

@article{PhysRevLett.106.095702,
  title = {Nonequilibrium Dynamics of the Bose-Hubbard Model: A Projection-Operator Approach},
  author = {Trefzger, C. and Sengupta, K.},
  journal = {Phys. Rev. Lett.},
  volume = {106},
  issue = {9},
  pages = {095702},
  numpages = {4},
  year = {2011},
  month = {Feb},
  publisher = {American Physical Society},
  doi = {10.1103/PhysRevLett.106.095702},
  url = {https://link.aps.org/doi/10.1103/PhysRevLett.106.095702}
}

@Article{Chen2020,
author={Chen, Huang-Jie
and Yu, Yan-Qiang
and Zheng, Dong-Chen
and Liao, Renyuan},
title={Extended Bose-Hubbard Model with Cavity-Mediated Infinite-Range Interactions at Finite Temperatures},
journal={Scientific Reports},
year={2020},
month={Jun},
day={03},
volume={10},
number={1},
pages={9076},
issn={2045-2322},
doi={10.1038/s41598-020-66054-1},
url={https://doi.org/10.1038/s41598-020-66054-1}
}

@article{PhysRevA.94.023632,
  title = {Phase transitions in a Bose-Hubbard model with cavity-mediated global-range interactions},
  author = {Dogra, N. and Brennecke, F. and Huber, S. D. and Donner, T.},
  journal = {Phys. Rev. A},
  volume = {94},
  issue = {2},
  pages = {023632},
  numpages = {7},
  year = {2016},
  month = {Aug},
  publisher = {American Physical Society},
  doi = {10.1103/PhysRevA.94.023632},
  url = {https://link.aps.org/doi/10.1103/PhysRevA.94.023632}
}

@article{PhysRevA.99.043633,
  title = {Mean-field phase diagram of the extended Bose-Hubbard model of many-body cavity quantum electrodynamics},
  author = {Himbert, Lukas and Cormick, Cecilia and Kraus, Rebecca and Sharma, Shraddha and Morigi, Giovanna},
  journal = {Phys. Rev. A},
  volume = {99},
  issue = {4},
  pages = {043633},
  numpages = {13},
  year = {2019},
  month = {Apr},
  publisher = {American Physical Society},
  doi = {10.1103/PhysRevA.99.043633},
  url = {https://link.aps.org/doi/10.1103/PhysRevA.99.043633}
}

@Article{Maschler2008,
author={Maschler, C.
and Mekhov, I. B.
and Ritsch, H.},
title={Ultracold atoms in optical lattices generated by quantized light fields},
journal={The European Physical Journal D},
year={2008},
month={Mar},
day={01},
volume={46},
number={3},
pages={545-560},
issn={1434-6079},
doi={10.1140/epjd/e2008-00016-4},
url={https://doi.org/10.1140/epjd/e2008-00016-4}
}

@article{Landig2015QuantumPF,
  title={Quantum phases from competing short- and long-range interactions in an optical lattice},
  author={Renate Landig and Lorenz Hruby and Nishant Dogra and Manuele Landini and Rafael Mottl and Tobias Donner and Tilman Esslinger},
  journal={Nature},
  year={2015},
  volume={532},
  pages={476-479},
  url={https://api.semanticscholar.org/CorpusID:4386279}
}

@article{Li2012LatticesupersolidPO,
  title={Lattice-supersolid phase of strongly correlated bosons in an optical cavity},
  author={Yongqiang Li and Yongqiang Li and Liang He and Walter Hofstetter},
  journal={Physical Review A},
  year={2012},
  volume={87},
  pages={051604},
  url={https://api.semanticscholar.org/CorpusID:54907163}
}

@article{Sundar2016LatticeBW,
  title={Lattice bosons with infinite-range checkerboard interactions},
  author={Bhuvanesh Sundar and Erich J. Mueller},
  journal={Physical Review A},
  year={2016},
  volume={94},
  pages={033631},
  url={https://api.semanticscholar.org/CorpusID:119284718}
}

@article{Chen2016QuantumPT,
  title={Quantum Phase Transitions of the Bose-Hubbard Model inside a Cavity},
  author={Yu Chen and Zhenhua Yu and Hui Zhai},
  journal={Physical Review A},
  year={2016},
  volume={93},
  pages={041601},
  url={https://api.semanticscholar.org/CorpusID:119216901}
}

@article{Bloch2008ManyBody,
  title = {Many-body physics with ultracold gases},
  author = {Bloch, Immanuel and Dalibard, Jean and Zwerger, Wilhelm},
  journal = {Rev. Mod. Phys.},
  volume = {80},
  issue = {3},
  pages = {885--964},
  year = {2008},
  month = {Jul},
  publisher = {American Physical Society},
  doi = {10.1103/RevModPhys.80.885}
}

@article{Lewenstein2007Quantum,
  title={Quantum Gases in Optical Lattices: From Strong Correlations to Macroscopic Quantum Phenomena},
  author={Lewenstein, Maciej and Sanpera, Anna and Ahufinger, Veronica and Damski, Bogdan and Sen De, Aditi and Sen, Ujjwal},
  journal={Advances in Physics},
  volume={56},
  number={2},
  pages={243--379},
  year={2007},
  publisher={Taylor \& Francis},
  doi={10.1080/00018730701223200}
}

@article{Fisher1989Boson,
  title = {Boson localization and the superfluid-insulator transition},
  author = {Fisher, Matthew P. A. and Weichman, Peter B. and Grinstein, G. and Fisher, Daniel S.},
  journal = {Phys. Rev. B},
  volume = {40},
  issue = {1},
  pages = {546--570},
  year = {1989},
  month = {Jul},
  publisher = {American Physical Society},
  doi = {10.1103/PhysRevB.40.546}
}

@article{Jaksch1998Cold,
  title = {Cold Bosonic Atoms in Optical Lattices},
  author = {Jaksch, D. and Bruder, C. and Cirac, J. I. and Gardiner, C. W. and Zoller, P.},
  journal = {Phys. Rev. Lett.},
  volume = {81},
  issue = {15},
  pages = {3108--3111},
  year = {1998},
  month = {Oct},
  publisher = {American Physical Society},
  doi = {10.1103/PhysRevLett.81.3108}
}

@article{Greiner2002Quantum,
  title={Quantum phase transition from a superfluid to a Mott insulator in a gas of ultracold atoms},
  author={Greiner, Markus and Mandel, Olaf0 and Esslinger, Tilman and H{\"a}nsch, Theodor W and Bloch, Immanuel},
  journal={Nature},
  volume={415},
  number={6867},
  pages={39--44},
  year={2002},
  publisher={Nature Publishing Group},
  doi={10.1038/415039a}
}

@article{Lahaye2009Physics,
  title={The physics of dipolar quantum gases},
  author={Lahaye, Thierry and Menotti, C and Santos, L and Lewenstein, M and Pfau, T},
  journal={Reports on Progress in Physics},
  volume={72},
  number={12},
  pages={126401},
  year={2009},
  publisher={IOP Publishing},
  doi={10.1088/0034-4885/72/12/126401}
}

@article{Defenu2023LongRange,
  title={Long-range physics of intercalated, molecular and atomic quantum gases},
  author={Defenu, Nicol{\`o} and Donner, Tobias and Macr{\`i}, Tommaso and Pagano, Guido and Ruffo, Stefano and Trombettoni, Andrea},
  journal={Physics Reports},
  volume={1015},
  pages={1--79},
  year={2023},
  publisher={Elsevier},
  doi={10.1016/j.physrep.2023.03.001}
}

@article{Baumann2010Mott,
  title={Dicke quantum phase transition with a superfluid gas in an optical cavity},
  author={Baumann, Kristian and Guerlin, Christine and Brennecke, Ferdinand and Esslinger, Tilman},
  journal={Nature},
  volume={464},
  number={7293},
  pages={1301--1306},
  year={2010},
  publisher={Nature Publishing Group},
  doi={10.1038/nature09009}
}

@article{Larson2008Mott,
  title = {Mott-Insulator States of Ultracold Atoms in Optical Cavities},
  author = {Larson, J. and Fernandez-Vidal, S. and Morigi, G. and Lewenstein, M.},
  journal = {Phys. Rev. Lett.},
  volume = {100},
  issue = {5},
  pages = {050401},
  year = {2008},
  month = {Feb},
  publisher = {American Physical Society},
  doi = {10.1103/PhysRevLett.100.050401}
}

@article{Leonard2017Supersolid,
  title={Supersolid formation in a quantum gas into an optical lattice},
  author={L{\'e}onard, Julian and Morales, Andrea and Philip, Prosenjit and Langini, Manuele and Esslinger, Tilman and Donner, Tobias},
  journal={Nature},
  volume={543},
  number={7643},
  pages={87--90},
  year={2017},
  publisher={Nature Publishing Group},
  doi={10.1038/nature21067}
}

@article{Li2017A_Supersolid,
  title={A stripe phase with supersolid properties in spin--orbit-coupled Bose--Einstein condensates},
  author={Li, Jun-Ru and Lee, Jeongwon and Huang, Wujie and Burchesky, Sean and Shteynas, Boris and Top, Furkan {\c{C}}agr{\i} and Jamison, Alan O and Ketterle, Wolfgang},
  journal={Nature},
  volume={543},
  number={7643},
  pages={91--94},
  year={2017},
  publisher={Nature Publishing Group},
  doi={10.1038/nature21431}
}

@article{Sheshadri1993Superfluid,
  title={Superfluid and insulating phases in an interacting-boson model: Mean-field theory and the RPA},
  author={Sheshadri, K and Krishnamurthy, HR and Pandit, Rahul and Ramakrishnan, TV},
  journal={EPL (Europhysics Letters)},
  volume={22},
  number={4},
  pages={257},
  year={1993},
  publisher={IOP Publishing},
  doi={10.1209/0295-5075/22/4/002}
}

@article{Schrieffer1966Relation,
  title = {Relation between the Anderson and Kondo Hamiltonians},
  author = {Schrieffer, J. R. and Wolff, P. A.},
  journal = {Phys. Rev.},
  volume = {149},
  issue = {2},
  pages = {491--492},
  year = {1966},
  month = {Sep},
  publisher = {American Physical Society},
  doi = {10.1103/PhysRev.149.491}
}

@article{D.L.Kovrizhin_2005,
doi = {10.1209/epl/i2005-10231-y},
url = {https://doi.org/10.1209/epl/i2005-10231-y},
year = {2005},
month = {sep},
publisher = {},
volume = {72},
number = {2},
pages = {162},
author = {D. L. Kovrizhin and G. Venketeswara Pai and S. Sinha},
title = {Density wave and supersolid phases of correlated bosons  
in an optical lattice},
journal = {Europhysics Letters}
}

@article{Sinha_2025,
doi = {10.1088/1361-648X/adf6fb},
url = {https://doi.org/10.1088/1361-648X/adf6fb},
year = {2025},
month = {aug},
publisher = {IOP Publishing},
volume = {37},
number = {33},
pages = {333001},
author = {Sinha, Sudip and Sinha, Subhasis},
title = {Supersolid phases of bosons},
journal = {Journal of Physics: Condensed Matter}
}

@article{revcav,
  title = {Cold atoms in cavity-generated dynamical optical potentials},
  author = {Ritsch, Helmut and Domokos, Peter and Brennecke, Ferdinand and Esslinger, Tilman},
  journal = {Rev. Mod. Phys.},
  volume = {85},
  issue = {2},
  pages = {553--601},
  numpages = {0},
  year = {2013},
  month = {Apr},
  publisher = {American Physical Society},
  doi = {10.1103/RevModPhys.85.553},
  url = {https://link.aps.org/doi/10.1103/RevModPhys.85.553}
}

@article{PhysRevLett.89.250404,
  title = {Oscillating Superfluidity of Bosons in Optical Lattices},
  author = {Altman, Ehud and Auerbach, Assa},
  journal = {Phys. Rev. Lett.},
  volume = {89},
  issue = {25},
  pages = {250404},
  numpages = {4},
  year = {2002},
  month = {Dec},
  publisher = {American Physical Society},
  doi = {10.1103/PhysRevLett.89.250404},
  url = {https://link.aps.org/doi/10.1103/PhysRevLett.89.250404}
}

@article{PhysRevLett.97.200601,
  title = {Sweeping from the Superfluid to the Mott Phase in the Bose-Hubbard Model},
  author = {Sch\"utzhold, Ralf and Uhlmann, Michael and Xu, Yan and Fischer, Uwe R.},
  journal = {Phys. Rev. Lett.},
  volume = {97},
  issue = {20},
  pages = {200601},
  numpages = {4},
  year = {2006},
  month = {Nov},
  publisher = {American Physical Society},
  doi = {10.1103/PhysRevLett.97.200601},
  url = {https://link.aps.org/doi/10.1103/PhysRevLett.97.200601}
}

@article{Dutta_2015,
  author    = {Dutta, Omjyoti and Gajda, Mariusz and Hauke, Philipp and Lewenstein, Maciej and L\"{u}hmann, Dirk-S\"{o}ren and Malomed, Boris A. and Sowi\'{n}ski, Tomasz and Zakrzewski, Jakub},
  title     = {Non-standard {Hubbard} models in optical lattices: a review},
  journal   = {Reports on Progress in Physics},
  year      = {2015},
  month     = {may},
  volume    = {78},
  number    = {6},
  pages     = {066001},
  doi       = {10.1088/0034-4885/78/6/066001},
  url       = {https://doi.org},
  publisher = {IOP Publishing}
}

@article{RevModPhys.97.011001,
  title = {Colloquium: Synthetic quantum matter in nonstandard geometries},
  author = {Grass, Tobias and Bercioux, Dario and Bhattacharya, Utso and Lewenstein, Maciej and Nguyen, Hai Son and Weitenberg, Christof},
  journal = {Rev. Mod. Phys.},
  volume = {97},
  issue = {1},
  pages = {011001},
  numpages = {31},
  year = {2025},
  month = {Mar},
  publisher = {American Physical Society},
  doi = {10.1103/RevModPhys.97.011001},
  url = {https://link.aps.org/doi/10.1103/RevModPhys.97.011001}
}

@article{Chanda_2025,
  doi = {10.1088/1361-6633/adc3a7},
  url = {https://doi.org},
  year = {2025},
  month = {apr},
  publisher = {IOP Publishing},
  volume = {88},
  number = {4},
  pages = {044501},
  author = {Chanda, Titas and Barbiero, Luca and Lewenstein, Maciej and Mark, Manfred J and Zakrzewski, Jakub},
  title = {Recent progress on quantum simulations of non-standard Bose–Hubbard models},
  journal = {Reports on Progress in Physics}
}

@article{dupuis1,
  title = {Mott-insulator--to--superfluid transition in the Bose-Hubbard model: A strong-coupling approach},
  author = {Sengupta, K. and Dupuis, N.},
  journal = {Phys. Rev. A},
  volume = {71},
  issue = {3},
  pages = {033629},
  numpages = {8},
  year = {2005},
  month = {Mar},
  publisher = {American Physical Society},
  doi = {10.1103/PhysRevA.71.033629},
  url = {https://link.aps.org/doi/10.1103/PhysRevA.71.033629}
}

@article{trivedi1,
author = {Krauth, Werner and Trivedi, Nandini},
year = {2007},
month = {07},
pages = {627},
title = {Mott and Superfluid Transitions in a Strongly Interacting Lattice Boson System},
volume = {14},
journal = {EPL (Europhysics Letters)},
doi = {10.1209/0295-5075/14/7/003}
}

@article{mft1,
  title = {Phase diagram of bosons in a two-dimensional optical lattice with infinite-range cavity-mediated interactions},
  author = {Flottat, T. and de Parny, L. de Forges and H\'ebert, F. and Rousseau, V. G. and Batrouni, G. G.},
  journal = {Phys. Rev. B},
  volume = {95},
  issue = {14},
  pages = {144501},
  numpages = {8},
  year = {2017},
  month = {Apr},
  publisher = {American Physical Society},
  doi = {10.1103/PhysRevB.95.144501},
  url = {https://link.aps.org/doi/10.1103/PhysRevB.95.144501}
}

@article{mft3,
title = {Non-equilibrium dynamics of ultracold lattice bosons inside a cavity},
journal = {Annals of Physics},
volume = {465},
pages = {169667},
year = {2024},
issn = {0003-4916},
doi = {https://doi.org/10.1016/j.aop.2024.169667},
url = {https://www.sciencedirect.com/science/article/pii/S0003491624000757},
author = {Xiayao He and Huan Wang and Min Liu and Hongrong Li and Shuai Li and Bo Liu}
}

@article{mft4,
  title = {Ultracold bosons with cavity-mediated long-range interactions: A local mean-field analysis of the phase diagram},
  author = {Niederle, Astrid E. and Morigi, Giovanna and Rieger, Heiko},
  journal = {Phys. Rev. A},
  volume = {94},
  issue = {3},
  pages = {033607},
  numpages = {9},
  year = {2016},
  month = {Sep},
  publisher = {American Physical Society},
  doi = {10.1103/PhysRevA.94.033607},
  url = {https://link.aps.org/doi/10.1103/PhysRevA.94.033607}
}

@article{mft5,
  title = {Theoretical exploration of competing phases of lattice Bose gases in a cavity},
  author = {Liao, Renyuan and Chen, Huang-Jie and Zheng, Dong-Chen and Huang, Zhi-Gao},
  journal = {Phys. Rev. A},
  volume = {97},
  issue = {1},
  pages = {013624},
  numpages = {6},
  year = {2018},
  month = {Jan},
  publisher = {American Physical Society},
  doi = {10.1103/PhysRevA.97.013624},
  url = {https://link.aps.org/doi/10.1103/PhysRevA.97.013624}
}

@article{dipbose,
  title = {Quantum phases of lattice dipolar bosons coupled to a high-finesse cavity},
  author = {Hebib, Yaghmorassene and Zhang, Chao and Yang, Jin and Capogrosso-Sansone, Barbara},
  journal = {Phys. Rev. A},
  volume = {107},
  issue = {4},
  pages = {043318},
  numpages = {7},
  year = {2023},
  month = {Apr},
  publisher = {American Physical Society},
  doi = {10.1103/PhysRevA.107.043318},
  url = {https://link.aps.org/doi/10.1103/PhysRevA.107.043318}
}

@article{qmc1,
  title = {Quantum Monte Carlo study of systems interacting via long-range interactions mediated by a cavity},
  author = {Dom\'{\i}nguez-Navarro, Marta and Rojo-Franc\`as, Abel and Juli\'a-D\'{\i}az, Bruno and Astrakharchik, Grigori E.},
  journal = {Phys. Rev. A},
  volume = {114},
  issue = {3},
  pages = {033303},
  numpages = {16},
  year = {2026},
  month = {Sep},
  publisher = {American Physical Society},
  doi = {10.1103/6jcl-c1gt},
  url = {https://link.aps.org/doi/10.1103/6jcl-c1gt}
}

@article{bdmft1,
  title = {Nonequilibrium Phase Transition of Interacting Bosons in an Intra-Cavity Optical Lattice},
  author = {Bakhtiari, M. Reza and Hemmerich, A. and Ritsch, H. and Thorwart, M.},
  journal = {Phys. Rev. Lett.},
  volume = {114},
  issue = {12},
  pages = {123601},
  numpages = {5},
  year = {2015},
  month = {Mar},
  publisher = {American Physical Society},
  doi = {10.1103/PhysRevLett.114.123601},
  url = {https://link.aps.org/doi/10.1103/PhysRevLett.114.123601}
}

@article{bdmft2,
  title = {Spectral properties and phase diagram of correlated lattice bosons in an optical cavity within bosonic dynamical mean-field theory},
  author = {Panas, Jaromir and Kauch, Anna and Byczuk, Krzysztof},
  journal = {Phys. Rev. B},
  volume = {95},
  issue = {11},
  pages = {115105},
  numpages = {9},
  year = {2017},
  month = {Mar},
  publisher = {American Physical Society},
  doi = {10.1103/PhysRevB.95.115105},
  url = {https://link.aps.org/doi/10.1103/PhysRevB.95.115105}
}

@article{rotor1,
  title = {Quantum Critical Behavior of Entanglement in Lattice Bosons with Cavity-Mediated Long-Range Interactions},
  author = {Sharma, Shraddha and J\"ager, Simon B. and Kraus, Rebecca and Roscilde, Tommaso and Morigi, Giovanna},
  journal = {Phys. Rev. Lett.},
  volume = {129},
  issue = {14},
  pages = {143001},
  numpages = {6},
  year = {2022},
  month = {Sep},
  publisher = {American Physical Society},
  doi = {10.1103/PhysRevLett.129.143001},
  url = {https://link.aps.org/doi/10.1103/PhysRevLett.129.143001}
}

@article{ergodicity,
  title = {Ergodicity breaking with long-range cavity-induced quasiperiodic interactions},
  author = {Kubala, Piotr and Sierant, Piotr and Morigi, Giovanna and Zakrzewski, Jakub},
  journal = {Phys. Rev. B},
  volume = {103},
  issue = {17},
  pages = {174208},
  numpages = {10},
  year = {2021},
  month = {May},
  publisher = {American Physical Society},
  doi = {10.1103/PhysRevB.103.174208},
  url = {https://link.aps.org/doi/10.1103/PhysRevB.103.174208}
}

@article{PD1,
  author   = {Zhang, Chao and Rieger, Heiko},
  title    = {Phase diagrams of the disordered Bose-Hubbard model with cavity-mediated long-range and nearest-neighbor interactions},
  journal  = {The European Physical Journal B},
  year     = {2020},
  volume   = {93},
  number   = {2},
  pages    = {25},
  issn     = {1434-6036},
  doi      = {10.1140/epjb/e2019-100420-1},
  url      = {https://doi.org/10.1140/epjb/e2019-100420-1}
}

@article{GS1,
  title = {Staggered superfluid phases of dipolar bosons in two-dimensional square lattices},
  author = {Suthar, Kuldeep and Kraus, Rebecca and Sable, Hrushikesh and Angom, Dilip and Morigi, Giovanna and Zakrzewski, Jakub},
  journal = {Phys. Rev. B},
  volume = {102},
  issue = {21},
  pages = {214503},
  numpages = {12},
  year = {2020},
  month = {Dec},
  publisher = {American Physical Society},
  doi = {10.1103/PhysRevB.102.214503},
  url = {https://link.aps.org/doi/10.1103/PhysRevB.102.214503}
}

@article{GS2,
  title = {Staggered ground states in an optical lattice},
  author = {Johnstone, Dean and Westerberg, Niclas and Duncan, Callum W. and \"Ohberg, Patrik},
  journal = {Phys. Rev. A},
  volume = {100},
  issue = {4},
  pages = {043614},
  numpages = {10},
  year = {2019},
  month = {Oct},
  publisher = {American Physical Society},
  doi = {10.1103/PhysRevA.100.043614},
  url = {https://link.aps.org/doi/10.1103/PhysRevA.100.043614}
}

@article{metastable1,
  title = {Ultracold dipolar gas in an optical lattice: The fate of metastable states},
  author = {Trefzger, C. and Menotti, C. and Lewenstein, M.},
  journal = {Phys. Rev. A},
  volume = {78},
  issue = {4},
  pages = {043604},
  numpages = {14},
  year = {2008},
  month = {Oct},
  publisher = {American Physical Society},
  doi = {10.1103/PhysRevA.78.043604},
  url = {https://link.aps.org/doi/10.1103/PhysRevA.78.043604}
}

@article{metastable2,
  author  = {Hruby, Lorenz and Dogra, Nishant and Landini, Manuele and Donner, Tobias and Esslinger, Tilman},
  title   = {Metastability and avalanche dynamics in strongly correlated gases with long-range interactions},
  journal = {Proceedings of the National Academy of Sciences},
  volume  = {115},
  number  = {13},
  pages   = {3279--3284},
  year    = {2018},
  doi     = {10.1073/pnas.1720415115},
  url     = {https://www.pnas.org/doi/abs/10.1073/pnas.1720415115}
}

@article{metastable3,
  title = {Quantum Relaxation and Metastability of Lattice Bosons with Cavity-Induced Long-Range Interactions},
  author = {Bla\ss{}, Benjamin and Rieger, Heiko and Ro\'osz, Gerg\ifmmode \mbox{\H{o}}\else \H{o}\fi{} and Igl\'oi, Ferenc},
  journal = {Phys. Rev. Lett.},
  volume = {121},
  issue = {9},
  pages = {095301},
  numpages = {6},
  year = {2018},
  month = {Aug},
  publisher = {American Physical Society},
  doi = {10.1103/PhysRevLett.121.095301},
  url = {https://link.aps.org/doi/10.1103/PhysRevLett.121.095301}
}

@article{MBL1,
  title = {Many-body localization regime for cavity-induced long-range interacting models},
  author = {Chanda, Titas and Zakrzewski, Jakub},
  journal = {Phys. Rev. B},
  volume = {105},
  issue = {5},
  pages = {054309},
  numpages = {8},
  year = {2022},
  month = {Feb},
  publisher = {American Physical Society},
  doi = {10.1103/PhysRevB.105.054309},
  url = {https://link.aps.org/doi/10.1103/PhysRevB.105.054309}
}

@article{MBL2,
  title = {Coexistence of localized and extended phases: Many-body localization in a harmonic trap},
  author = {Chanda, Titas and Yao, Ruixiao and Zakrzewski, Jakub},
  journal = {Phys. Rev. Res.},
  volume = {2},
  issue = {3},
  pages = {032039(R)},
  numpages = {6},
  year = {2020},
  month = {Aug},
  publisher = {American Physical Society},
  doi = {10.1103/PhysRevResearch.2.032039},
  url = {https://link.aps.org/doi/10.1103/PhysRevResearch.2.032039}
}

@article{haldaneWorm,
  author   = {Sicks, Johannes and Rieger, Heiko},
  title    = {Haldane insulator in the 1D nearest-neighbor extended Bose-Hubbard model with cavity-mediated long-range interactions},
  journal  = {The European Physical Journal B},
  year     = {2020},
  volume   = {93},
  number   = {6},
  pages    = {104},
  issn     = {1434-6036},
  doi      = {10.1140/epjb/e2020-10109-3},
  url      = {https://doi.org/10.1140/epjb/e2020-10109-3}
}

@article{DPT1,
  title = {Observation of a Dissipative Phase Transition in a One-Dimensional Circuit QED Lattice},
  author = {Fitzpatrick, Mattias and Sundaresan, Neereja M. and Li, Andy C. Y. and Koch, Jens and Houck, Andrew A.},
  journal = {Phys. Rev. X},
  volume = {7},
  issue = {1},
  pages = {011016},
  numpages = {8},
  year = {2017},
  month = {Feb},
  publisher = {American Physical Society},
  doi = {10.1103/PhysRevX.7.011016},
  url = {https://link.aps.org/doi/10.1103/PhysRevX.7.011016}
}

\end{document}